\documentclass[twocolumn,longbibliography,nofootinbib,aps,prx,superscriptaddress]{revtex4-2}    
\usepackage{CJK}
\pdfoutput=1 %

\usepackage{tikz}
\usetikzlibrary{positioning, shapes.geometric, arrows.meta,fit}

\usetikzlibrary{shapes.multipart}
\usetikzlibrary{circuits.ee.IEC}
\tikzstyle{bwSpider}=[
       rectangle split,
       rectangle split parts=2,
       rectangle split part fill={black,white},
 minimum size=3.6 mm, inner sep=-2mm, draw=black,scale=0.5,rounded corners=0.8 mm
       ]
 \tikzstyle{wbSpider}=[
       rectangle split,
       rectangle split parts=2,
       rectangle split part fill={white,black},
 minimum size=3.6 mm, inner sep=-2mm, draw=black,scale=0.5,rounded corners=0.8 mm
       ]
\tikzstyle{qWire}=[line width = 2pt, color=violet]
\tikzstyle{hWire}=[line width = 1pt, color=higherorderred]
\tikzstyle{sWire}=[line width = 1pt, color=higherordergreen]
\tikzstyle{cWire}=[color=gray,thin]
\tikzstyle{rWire}=[color=black,thin]

\tikzstyle{env}=[copoint,regular polygon rotate=0,minimum width=0.2cm, fill=black]

\tikzstyle{probs}=[shape=semicircle,fill=white,draw=black,shape border rotate=180,minimum width=1.2cm]

\tikzstyle{every picture}=[baseline=-0.25em,scale=.5]
\tikzstyle{dotpic}=[] 
\tikzstyle{diredges}=[every to/.style={diredge}]
\tikzstyle{math matrix}=[matrix of math nodes,left delimiter=(,right delimiter=),inner sep=2pt,column sep=1em,row sep=0.5em,nodes={inner sep=0pt},text height=1.5ex, text depth=0.25ex]

\tikzstyle{inline text}=[text height=1.5ex, text depth=0.25ex,yshift=0.5mm]
\tikzstyle{label}=[font=\scriptsize,text height=1.5ex, text depth=0.25ex,yshift=0.0mm]
\tikzstyle{left label}=[label,anchor=east,xshift=1.5mm]
\tikzstyle{right label}=[label,anchor=west,xshift=-1mm]

\tikzstyle{braceedge}=[decorate,decoration={brace,amplitude=2mm,raise=-1mm}]
\tikzstyle{small braceedge}=[decorate,decoration={brace,amplitude=1mm,raise=-1mm}]

\tikzstyle{doubled}=[line width=1.6pt] 
\tikzstyle{boldedge}=[doubled,shorten <=-0.17mm,shorten >=-0.17mm]
\tikzstyle{boldedgegray}=[doubled,gray,shorten <=-0.17mm,shorten >=-0.17mm]
\tikzstyle{singleedgegray}=[gray]

\tikzstyle{semidoubled}=[line width=1.4pt] 
\tikzstyle{semiboldedgegray}=[semidoubled,gray,shorten <=-0.17mm,shorten >=-0.17mm]

\tikzstyle{boxedge}=[semiboldedgegray]

\tikzstyle{boldedgedashed}=[very thick,dashed,shorten <=-0.17mm,shorten >=-0.17mm]
\tikzstyle{vboldedgedashed}=[doubled,dashed,shorten <=-0.17mm,shorten >=-0.17mm]
\tikzstyle{left hook arrow}=[left hook-latex]
\tikzstyle{right hook arrow}=[right hook-latex]
\tikzstyle{sembracket}=[line width=0.5pt,shorten <=-0.07mm,shorten >=-0.07mm]

\tikzstyle{causal edge}=[->,thick,gray]
\tikzstyle{causal nondir}=[thick,gray]
\tikzstyle{timeline}=[thick,gray, dashed]

\tikzstyle{cedge}=[<->,thick,gray!70!white]

\tikzstyle{empty diagram}=[draw=gray!40!white,dashed,shape=rectangle,minimum width=1cm,minimum height=1cm]
\tikzstyle{empty circle}=[draw=black,dashed,shape=circle,minimum width=.2cm,minimum height=.2cm, inner sep=1pt]
\tikzstyle{empty diagram small}=[draw=gray!50!white,dashed,shape=rectangle,minimum width=0.4cm,minimum height=0.3cm]

\tikzstyle{dot}=[inner sep=0mm,minimum width=2mm,minimum height=2mm,draw,shape=circle]
\tikzstyle{proj}=[trapezium, trapezium angle=67.5, draw, inner sep=1pt, outer sep=0pt, minimum height=0.5cm, minimum width=0.5cm, rotate=0,trapezium stretches=false]
\tikzstyle{proj2}=[draw=black,rectangle, minimum height=0.5cm, minimum width=1.2cm]
\tikzstyle{inc}=[trapezium, trapezium angle=-67.5, draw, inner sep=1pt, outer sep=0pt, minimum height=6pt, minimum width=2pt,rotate=0,trapezium stretches=true]
\tikzstyle{coproj}=[trapezium, trapezium angle=-67.5, draw, inner sep=1pt, outer sep=0pt, minimum height=6pt, minimum width=2pt,rotate=0,trapezium stretches=true]
\tikzstyle{coinc}=[trapezium, trapezium angle=67.5, draw, inner sep=1pt, outer sep=0pt, minimum height=6pt, minimum width=2pt,rotate=0,trapezium stretches=true]
\tikzstyle{leak}=[white dot, shape=regular polygon, minimum size=3.3 mm, regular polygon sides=3, outer sep=-0.2mm, regular polygon rotate=270]
\tikzstyle{standard_rectangle}=[draw, rectangle, minimum height=1cm, minimum width=1cm]
\tikzstyle{wide proj}=[draw,fill=white,chamfered rectangle,chamfered rectangle angle=30, minimum width=15mm,minimum height=1mm,scale=0.5, outer sep=-0.2mm]
\tikzstyle{very wide proj}=[draw,fill=white,chamfered rectangle,chamfered rectangle angle=30, minimum width=25mm,minimum height=1mm,scale=0.5, outer sep=-0.2mm]
\tikzstyle{very very wide proj}=[draw,fill=white,chamfered rectangle,chamfered rectangle angle=30, minimum width=35mm,minimum height=1mm,scale=0.5, outer sep=-0.2mm]
\tikzstyle{preleak}=[proj]

\tikzstyle{split proj out}=[regular polygon,regular polygon sides=3,draw,scale=0.75,inner sep=-0.5pt,minimum width=3.3mm,fill=white,regular polygon rotate=180]
\tikzstyle{split proj in}=[regular polygon,regular polygon sides=3,draw,scale=0.75,inner sep=-0.5pt,minimum width=3.3mm,fill=white]
\tikzstyle{Vleak}=[white dot, shape=regular polygon, minimum size=3.3 mm, regular polygon sides=3, outer sep=-0.2mm, regular polygon rotate=90]
\tikzstyle{dleak}=[white dot, line width=1.6pt, shape=regular polygon, minimum size=3.3 mm, regular polygon sides=3, outer sep=-0.2mm, regular polygon rotate=270]

\tikzstyle{Wsquare}=[white dot, shape=regular polygon, minimum height=1cm, minimum width=1cm,  regular polygon sides=3, outer sep=-0.2mm]
\tikzstyle{Wsquareadj}=[white dot, shape=regular polygon,  minimum height=1cm, minimum width=1cm, regular polygon sides=3, outer sep=-0.2mm, regular polygon rotate=180]
\tikzstyle{ddot}=[inner sep=0mm, doubled, minimum width=2.5mm,minimum height=2.5mm,draw,shape=circle]

\tikzstyle{black dot}=[dot,fill=black]
\tikzstyle{white dot}=[dot,fill=white,,text depth=-0.2mm]
\tikzstyle{white Wsquare}=[Wsquare,fill=gray,,text depth=-0.2mm]
\tikzstyle{white Wsquareadj}=[Wsquareadj,fill=white,,text depth=-0.2mm]
\tikzstyle{green dot}=[white dot] 
\tikzstyle{gray dot}=[dot,fill=gray!40!white,,text depth=-0.2mm]
\tikzstyle{red dot}=[gray dot] 

\tikzstyle{black ddot}=[ddot,fill=black]
\tikzstyle{white ddot}=[ddot,fill=white]
\tikzstyle{gray ddot}=[ddot,fill=gray!40!white]

\tikzstyle{gray edge}=[gray!60!white]

\tikzstyle{small dot}=[inner sep=0.5mm,minimum width=0pt,minimum height=0pt,draw,shape=circle]

\tikzstyle{small black dot}=[small dot,fill=black]
\tikzstyle{small white dot}=[small dot,fill=white]
\tikzstyle{small gray dot}=[small dot,fill=gray!40!white]

\tikzstyle{causal dot}=[inner sep=0.4mm,minimum width=0pt,minimum height=0pt,draw=white,shape=circle,fill=gray!40!white]

\tikzstyle{phase dimensions}=[minimum size=5mm,font=\footnotesize,rectangle,rounded corners=2.5mm,inner sep=0.2mm,outer sep=-2mm]
\tikzstyle{dphase dimensions}=[minimum size=5mm,font=\footnotesize,rectangle,rounded corners=2.5mm,inner sep=0.2mm,outer sep=-2mm]

\tikzstyle{white phase dot}=[dot,fill=white,phase dimensions]
\tikzstyle{white phase ddot}=[ddot,fill=white,dphase dimensions]

\tikzstyle{white rect ddot}=[draw=black,fill=white,doubled,minimum size=5mm,font=\footnotesize,rectangle,rounded corners=2.5mm,inner sep=0.2mm]
\tikzstyle{gray rect ddot}=[draw=black,fill=gray!40!white,doubled,minimum size=6mm,font=\footnotesize,rectangle,rounded corners=3mm]

\tikzstyle{gray phase dot}=[dot,fill=gray!40!white,phase dimensions]
\tikzstyle{gray phase ddot}=[ddot,fill=gray!40!white,dphase dimensions]
\tikzstyle{grey phase dot}=[gray phase dot]
\tikzstyle{grey phase ddot}=[gray phase ddot]

\tikzstyle{small phase dimensions}=[minimum size=4mm,font=\tiny,rectangle,rounded corners=2mm,inner sep=0.2mm,outer sep=-2mm]
\tikzstyle{small dphase dimensions}=[minimum size=4mm,font=\tiny,rectangle,rounded corners=2mm,inner sep=0.2mm,outer sep=-2mm]

\tikzstyle{small gray phase dot}=[dot,fill=gray!40!white,small phase dimensions]
\tikzstyle{small gray phase ddot}=[ddot,fill=gray!40!white,small dphase dimensions]

\tikzstyle{small map}=[draw,shape=rectangle,minimum height=4mm,minimum width=4mm,fill=white]
\tikzstyle{math map}=[draw,shape=rectangle,minimum height=4mm,minimum width=4mm,fill=black, font=\color{white}]

\tikzstyle{cnot}=[fill=white,shape=circle,inner sep=-1.4pt]

\tikzstyle{asym hadamard}=[fill=white,draw,shape=NEbox,inner sep=0.6mm,font=\footnotesize,minimum height=4mm]
\tikzstyle{asym hadamard conj}=[fill=white,draw,shape=NWbox,inner sep=0.6mm,font=\footnotesize,minimum height=4mm]
\tikzstyle{asym hadamard dag}=[fill=white,draw,shape=SEbox,inner sep=0.6mm,font=\footnotesize,minimum height=4mm]

\tikzstyle{hadamard}=[fill=white,draw,inner sep=0.6mm,font=\footnotesize,minimum height=4mm,minimum width=4mm]
\tikzstyle{small hadamard}=[fill=white,draw,inner sep=0.6mm,minimum height=1.5mm,minimum width=1.5mm]
\tikzstyle{small hadamard rotate}=[small hadamard,rotate=45]
\tikzstyle{dhadamard}=[hadamard,doubled]
\tikzstyle{small dhadamard}=[small hadamard,doubled]
\tikzstyle{small dhadamard rotate}=[small hadamard rotate,doubled]
\tikzstyle{antipode}=[white dot,inner sep=0.3mm,font=\footnotesize]

\tikzstyle{scalar}=[diamond,draw,inner sep=0.5pt,font=\small]
\tikzstyle{dscalar}=[diamond,doubled, draw,inner sep=0.5pt,font=\small]

\tikzstyle{small box}=[rectangle,inline text,fill=white,draw,minimum height=0.5mm,yshift=-0.5mm,minimum width=5mm,font=\small]
\tikzstyle{small gray box}=[small box,fill=gray!30]
\tikzstyle{medium box}=[rectangle,inline text,fill=white,draw,minimum height=5mm,yshift=-0.5mm,minimum width=10mm,font=\small]
\tikzstyle{square box}=[small box] 
\tikzstyle{medium gray box}=[small box,fill=gray!30]
\tikzstyle{semilarge box}=[rectangle,inline text,fill=white,draw,minimum height=5mm,yshift=-0.5mm,minimum width=12.5mm,font=\small]
\tikzstyle{large box}=[rectangle,inline text,fill=white,draw,minimum height=5mm,yshift=-0.5mm,minimum width=15mm,font=\small]
\tikzstyle{large gray box}=[small box,fill=gray!30]

\tikzstyle{Bayes box}=[rectangle,fill=black,draw, minimum height=3mm, minimum width=3mm]

\tikzstyle{gray square point}=[small box,fill=gray!50]

\tikzstyle{dphase box white}=[dhadamard]
\tikzstyle{dphase box gray}=[dhadamard,fill=gray!50!white]
\tikzstyle{phase box white}=[hadamard]
\tikzstyle{phase box gray}=[hadamard,fill=gray!50!white]

\tikzstyle{point}=[regular polygon,regular polygon sides=3,draw,scale=0.75,inner sep=-0.5pt,minimum width=9mm,fill=white,regular polygon rotate=180]
\tikzstyle{point nosep}=[regular polygon,regular polygon sides=3,draw,scale=0.75,inner sep=-2pt,minimum width=9mm,fill=white,regular polygon rotate=180]
\tikzstyle{copoint}=[regular polygon,regular polygon sides=3,draw,scale=0.75,inner sep=-0.5pt,minimum width=9mm,fill=white]
\tikzstyle{dpoint}=[point,doubled]
\tikzstyle{dcopoint}=[copoint,doubled]

\tikzstyle{pointgrow}=[shape=cornerpoint,kpoint common,scale=0.75,inner sep=3pt]
\tikzstyle{pointgrow dag}=[shape=cornercopoint,kpoint common,scale=0.75,inner sep=3pt]

\tikzstyle{wide copoint}=[fill=white,draw,shape=isosceles triangle,shape border rotate=90,isosceles triangle stretches=true,inner sep=0pt,minimum width=1.5cm,minimum height=6.12mm]
\tikzstyle{wide point}=[fill=white,draw,shape=isosceles triangle,shape border rotate=-90,isosceles triangle stretches=true,inner sep=0pt,minimum width=1.5cm,minimum height=6.12mm,yshift=-0.0mm]
\tikzstyle{wide point plus}=[fill=white,draw,shape=isosceles triangle,shape border rotate=-90,isosceles triangle stretches=true,inner sep=0pt,minimum width=1.74cm,minimum height=7mm,yshift=-0.0mm]

\tikzstyle{wide dpoint}=[fill=white,doubled,draw,shape=isosceles triangle,shape border rotate=-90,isosceles triangle stretches=true,inner sep=0pt,minimum width=1.5cm,minimum height=6.12mm,yshift=-0.0mm]

\tikzstyle{tinypoint}=[regular polygon,regular polygon sides=3,draw,scale=0.55,inner sep=-0.15pt,minimum width=6mm,fill=white,regular polygon rotate=180]

\tikzstyle{white point}=[point]
\tikzstyle{white dpoint}=[dpoint]
\tikzstyle{green point}=[white point] 
\tikzstyle{white copoint}=[copoint]
\tikzstyle{gray point}=[point,fill=gray!40!white]
\tikzstyle{gray dpoint}=[gray point,doubled]
\tikzstyle{red point}=[gray point] 
\tikzstyle{gray copoint}=[copoint,fill=gray!40!white]
\tikzstyle{gray dcopoint}=[gray copoint,doubled]

\tikzstyle{white point guide}=[regular polygon,regular polygon sides=3,font=\scriptsize,draw,scale=0.65,inner sep=-0.5pt,minimum width=9mm,fill=white,regular polygon rotate=180]

\tikzstyle{black point}=[point,fill=black,font=\color{white}]
\tikzstyle{black copoint}=[copoint,fill=black,font=\color{white}]

\tikzstyle{tiny gray point}=[tinypoint,fill=gray!40!white]

\tikzstyle{diredge}=[->]
\tikzstyle{ddiredge}=[<->]
\tikzstyle{rdiredge}=[<-]
\tikzstyle{thickdiredge}=[->, very thick]
\tikzstyle{pointer edge}=[->,very thick,gray]
\tikzstyle{pointer edge part}=[very thick,gray]
\tikzstyle{dashed edge}=[dashed]
\tikzstyle{thick dashed edge}=[very thick,dashed]
\tikzstyle{thick gray dashed edge}=[thick dashed edge,gray!40]
\tikzstyle{thick map edge}=[very thick,|->]

\makeatletter
\newcommand{\boxshape}[3]{%
\pgfdeclareshape{#1}{
\inheritsavedanchors[from=rectangle] 
\inheritanchorborder[from=rectangle]
\inheritanchor[from=rectangle]{center}
\inheritanchor[from=rectangle]{north}
\inheritanchor[from=rectangle]{south}
\inheritanchor[from=rectangle]{west}
\inheritanchor[from=rectangle]{east}
\backgroundpath{
\southwest \pgf@xa=\pgf@x \pgf@ya=\pgf@y
\northeast \pgf@xb=\pgf@x \pgf@yb=\pgf@y

\@tempdima=#2
\@tempdimb=#3

\pgfpathmoveto{\pgfpoint{\pgf@xa - 5pt + \@tempdima}{\pgf@ya}}
\pgfpathlineto{\pgfpoint{\pgf@xa - 5pt - \@tempdima}{\pgf@yb}}
\pgfpathlineto{\pgfpoint{\pgf@xb + 5pt + \@tempdimb}{\pgf@yb}}
\pgfpathlineto{\pgfpoint{\pgf@xb + 5pt - \@tempdimb}{\pgf@ya}}
\pgfpathlineto{\pgfpoint{\pgf@xa - 5pt + \@tempdima}{\pgf@ya}}
\pgfpathclose
}
}}

\boxshape{NEbox}{0pt}{3pt}
\boxshape{SEbox}{0pt}{-3pt}
\boxshape{NWbox}{5pt}{0pt}
\boxshape{SWbox}{-5pt}{0pt}
\boxshape{EBox}{-3pt}{3pt}
\boxshape{WBox}{3pt}{-3pt}
\makeatother

\tikzstyle{cloud}=[shape=cloud,draw,minimum width=1.5cm,minimum height=1.5cm]

\tikzstyle{map}=[draw,shape=NEbox,inner sep=1pt,minimum height=4mm,fill=white]
\tikzstyle{dashedmap}=[draw,dashed,shape=NEbox,inner sep=2pt,minimum height=6mm,fill=white]
\tikzstyle{mapdag}=[draw,shape=SEbox,inner sep=1pt,minimum height=4mm,fill=white]
\tikzstyle{mapadj}=[draw,shape=SEbox,inner sep=2pt,minimum height=6mm,fill=white]
\tikzstyle{maptrans}=[draw,shape=SWbox,inner sep=2pt,minimum height=6mm,fill=white]
\tikzstyle{mapconj}=[draw,shape=NWbox,inner sep=2pt,minimum height=6mm,fill=white]

\tikzstyle{medium map}=[draw,shape=NEbox,inner sep=2pt,minimum height=6mm,fill=white,minimum width=7mm]
\tikzstyle{medium map dag}=[draw,shape=SEbox,inner sep=2pt,minimum height=6mm,fill=white,minimum width=7mm]
\tikzstyle{medium map adj}=[draw,shape=SEbox,inner sep=2pt,minimum height=6mm,fill=white,minimum width=7mm]
\tikzstyle{medium map trans}=[draw,shape=SWbox,inner sep=2pt,minimum height=6mm,fill=white,minimum width=7mm]
\tikzstyle{medium map conj}=[draw,shape=NWbox,inner sep=2pt,minimum height=6mm,fill=white,minimum width=7mm]
\tikzstyle{semilarge map}=[draw,shape=NEbox,inner sep=2pt,minimum height=6mm,fill=white,minimum width=9.5mm]
\tikzstyle{semilarge map trans}=[draw,shape=SWbox,inner sep=2pt,minimum height=6mm,fill=white,minimum width=9.5mm]
\tikzstyle{semilarge map adj}=[draw,shape=SEbox,inner sep=2pt,minimum height=6mm,fill=white,minimum width=9.5mm]
\tikzstyle{semilarge map dag}=[draw,shape=SEbox,inner sep=2pt,minimum height=6mm,fill=white,minimum width=9.5mm]
\tikzstyle{semilarge map conj}=[draw,shape=NWbox,inner sep=2pt,minimum height=6mm,fill=white,minimum width=9.5mm]
\tikzstyle{large map}=[draw,shape=NEbox,inner sep=2pt,minimum height=6mm,fill=white,minimum width=12mm]
\tikzstyle{large map conj}=[draw,shape=NWbox,inner sep=2pt,minimum height=6mm,fill=white,minimum width=12mm]
\tikzstyle{very large map}=[draw,shape=NEbox,inner sep=2pt,minimum height=6mm,fill=white,minimum width=17mm]

\tikzstyle{medium dmap}=[draw,doubled,shape=NEbox,inner sep=2pt,minimum height=6mm,fill=white,minimum width=7mm]
\tikzstyle{medium dmap dag}=[draw,doubled,shape=SEbox,inner sep=2pt,minimum height=6mm,fill=white,minimum width=7mm]
\tikzstyle{medium dmap adj}=[draw,doubled,shape=SEbox,inner sep=2pt,minimum height=6mm,fill=white,minimum width=7mm]
\tikzstyle{medium dmap trans}=[draw,doubled,shape=SWbox,inner sep=2pt,minimum height=6mm,fill=white,minimum width=7mm]
\tikzstyle{medium dmap conj}=[draw,doubled,shape=NWbox,inner sep=2pt,minimum height=6mm,fill=white,minimum width=7mm]
\tikzstyle{semilarge dmap}=[draw,doubled,shape=NEbox,inner sep=2pt,minimum height=6mm,fill=white,minimum width=9.5mm]
\tikzstyle{semilarge dmap trans}=[draw,doubled,shape=SWbox,inner sep=2pt,minimum height=6mm,fill=white,minimum width=9.5mm]
\tikzstyle{semilarge dmap adj}=[draw,doubled,shape=SEbox,inner sep=2pt,minimum height=6mm,fill=white,minimum width=9.5mm]
\tikzstyle{semilarge dmap dag}=[draw,doubled,shape=SEbox,inner sep=2pt,minimum height=6mm,fill=white,minimum width=9.5mm]
\tikzstyle{semilarge dmap conj}=[draw,doubled,shape=NWbox,inner sep=2pt,minimum height=6mm,fill=white,minimum width=9.5mm]
\tikzstyle{large dmap}=[draw,doubled,shape=NEbox,inner sep=2pt,minimum height=6mm,fill=white,minimum width=12mm]
\tikzstyle{large dmap conj}=[draw,doubled,shape=NWbox,inner sep=2pt,minimum height=6mm,fill=white,minimum width=12mm]
\tikzstyle{large dmap trans}=[draw,doubled,shape=SWbox,inner sep=2pt,minimum height=6mm,fill=white,minimum width=12mm]
\tikzstyle{large dmap adj}=[draw,doubled,shape=SEbox,inner sep=2pt,minimum height=6mm,fill=white,minimum width=12mm]
\tikzstyle{large dmap dag}=[draw,doubled,shape=SEbox,inner sep=2pt,minimum height=6mm,fill=white,minimum width=12mm]
\tikzstyle{very large dmap}=[draw,doubled,shape=NEbox,inner sep=2pt,minimum height=6mm,fill=white,minimum width=19.5mm]

\tikzstyle{muxbox}=[draw,shape=rectangle,minimum height=3mm,minimum width=3mm,fill=white]
\tikzstyle{dmuxbox}=[muxbox,doubled]

\tikzstyle{box}=[draw,shape=rectangle,inner sep=2pt,minimum height=6mm,minimum width=6mm,fill=white]
\tikzstyle{dbox}=[draw,doubled,shape=rectangle,inner sep=2pt,minimum height=6mm,minimum width=6mm,fill=white]
\tikzstyle{dmap}=[draw,doubled,shape=NEbox,inner sep=2pt,minimum height=6mm,fill=white]
\tikzstyle{dmapdag}=[draw,doubled,shape=SEbox,inner sep=2pt,minimum height=6mm,fill=white]
\tikzstyle{dmapadj}=[draw,doubled,shape=SEbox,inner sep=2pt,minimum height=6mm,fill=white]
\tikzstyle{dmaptrans}=[draw,doubled,shape=SWbox,inner sep=2pt,minimum height=6mm,fill=white]
\tikzstyle{dmapconj}=[draw,doubled,shape=NWbox,inner sep=2pt,minimum height=6mm,fill=white]

\tikzstyle{ddmap}=[draw,doubled,dashed,shape=NEbox,inner sep=2pt,minimum height=6mm,fill=white]
\tikzstyle{ddmapdag}=[draw,doubled,dashed,shape=SEbox,inner sep=2pt,minimum height=6mm,fill=white]
\tikzstyle{ddmapadj}=[draw,doubled,dashed,shape=SEbox,inner sep=2pt,minimum height=6mm,fill=white]
\tikzstyle{ddmaptrans}=[draw,doubled,dashed,shape=SWbox,inner sep=2pt,minimum height=6mm,fill=white]
\tikzstyle{ddmapconj}=[draw,doubled,dashed,shape=NWbox,inner sep=2pt,minimum height=6mm,fill=white]

\boxshape{sNEbox}{0pt}{3pt}
\boxshape{sSEbox}{0pt}{-3pt}
\boxshape{sNWbox}{3pt}{0pt}
\boxshape{sSWbox}{-3pt}{0pt}
\tikzstyle{smap}=[draw,shape=sNEbox,fill=white]
\tikzstyle{smapdag}=[draw,shape=sSEbox,fill=white]
\tikzstyle{smapadj}=[draw,shape=sSEbox,fill=white]
\tikzstyle{smaptrans}=[draw,shape=sSWbox,fill=white]
\tikzstyle{smapconj}=[draw,shape=sNWbox,fill=white]

\tikzstyle{dsmap}=[draw,dashed,shape=sNEbox,fill=white]
\tikzstyle{dsmapdag}=[draw,dashed,shape=sSEbox,fill=white]
\tikzstyle{dsmaptrans}=[draw,dashed,shape=sSWbox,fill=white]
\tikzstyle{dsmapconj}=[draw,dashed,shape=sNWbox,fill=white]

\boxshape{mNEbox}{0pt}{10pt}
\boxshape{mSEbox}{0pt}{-10pt}
\boxshape{mNWbox}{10pt}{0pt}
\boxshape{mSWbox}{-10pt}{0pt}
\tikzstyle{mmap}=[draw,shape=mNEbox]
\tikzstyle{mmapdag}=[draw,shape=mSEbox]
\tikzstyle{mmaptrans}=[draw,shape=mSWbox]
\tikzstyle{mmapconj}=[draw,shape=mNWbox]

\tikzstyle{mmapgray}=[draw,fill=gray!40!white,shape=mNEbox]
\tikzstyle{smapgray}=[draw,fill=gray!40!white,shape=sNEbox]

\makeatletter

\pgfdeclareshape{cornerpoint}{
\inheritsavedanchors[from=rectangle] 
\inheritanchorborder[from=rectangle]
\inheritanchor[from=rectangle]{center}
\inheritanchor[from=rectangle]{north}
\inheritanchor[from=rectangle]{south}
\inheritanchor[from=rectangle]{west}
\inheritanchor[from=rectangle]{east}
\backgroundpath{
\southwest \pgf@xa=\pgf@x \pgf@ya=\pgf@y
\northeast \pgf@xb=\pgf@x \pgf@yb=\pgf@y

\pgfmathsetmacro{\pgf@shorten@left}{\pgfkeysvalueof{/tikz/shorten left}}
\pgfmathsetmacro{\pgf@shorten@right}{\pgfkeysvalueof{/tikz/shorten right}}

\pgfpathmoveto{\pgfpoint{0.5 * (\pgf@xa + \pgf@xb)}{\pgf@ya - 5pt}}
\pgfpathlineto{\pgfpoint{\pgf@xa - 8pt + \pgf@shorten@left}{\pgf@yb - 1.5 * \pgf@shorten@left}}
\pgfpathlineto{\pgfpoint{\pgf@xa - 8pt + \pgf@shorten@left}{\pgf@yb}}
\pgfpathlineto{\pgfpoint{\pgf@xb + 8pt - \pgf@shorten@right}{\pgf@yb}}
\pgfpathlineto{\pgfpoint{\pgf@xb + 8pt - \pgf@shorten@right}{\pgf@yb - 1.5 * \pgf@shorten@right}}
\pgfpathclose
}
}

\pgfdeclareshape{cornercopoint}{
\inheritsavedanchors[from=rectangle] 
\inheritanchorborder[from=rectangle]
\inheritanchor[from=rectangle]{center}
\inheritanchor[from=rectangle]{north}
\inheritanchor[from=rectangle]{south}
\inheritanchor[from=rectangle]{west}
\inheritanchor[from=rectangle]{east}
\backgroundpath{
\southwest \pgf@xa=\pgf@x \pgf@ya=\pgf@y
\northeast \pgf@xb=\pgf@x \pgf@yb=\pgf@y

\pgfmathsetmacro{\pgf@shorten@left}{\pgfkeysvalueof{/tikz/shorten left}}
\pgfmathsetmacro{\pgf@shorten@right}{\pgfkeysvalueof{/tikz/shorten right}}

\pgfpathmoveto{\pgfpoint{0.5 * (\pgf@xa + \pgf@xb)}{\pgf@yb + 5pt}}
\pgfpathlineto{\pgfpoint{\pgf@xa - 8pt + \pgf@shorten@left}{\pgf@ya + 1.5 * \pgf@shorten@left}}
\pgfpathlineto{\pgfpoint{\pgf@xa - 8pt + \pgf@shorten@left}{\pgf@ya}}
\pgfpathlineto{\pgfpoint{\pgf@xb + 8pt - \pgf@shorten@right}{\pgf@ya}}
\pgfpathlineto{\pgfpoint{\pgf@xb + 8pt - \pgf@shorten@right}{\pgf@ya + 1.5 * \pgf@shorten@right}}
\pgfpathclose
}
}

\makeatother

\pgfkeyssetvalue{/tikz/shorten left}{0pt}
\pgfkeyssetvalue{/tikz/shorten right}{0pt}

\tikzstyle{kpoint common}=[draw,fill=white,inner sep=1pt,minimum height=4mm]
\tikzstyle{kpoint sc}=[shape=cornerpoint,kpoint common]
\tikzstyle{kpoint adjoint sc}=[shape=cornercopoint,kpoint common]
\tikzstyle{kpoint}=[shape=cornerpoint,shorten left=5pt,kpoint common]
\tikzstyle{kpoint adjoint}=[shape=cornercopoint,shorten left=5pt,kpoint common]
\tikzstyle{kpoint conjugate}=[shape=cornerpoint,shorten right=5pt,kpoint common]
\tikzstyle{kpoint transpose}=[shape=cornercopoint,shorten right=5pt,kpoint common]
\tikzstyle{kpoint symm}=[shape=cornerpoint,shorten left=5pt,shorten right=5pt,kpoint common]

\tikzstyle{wide kpoint sc}=[shape=cornerpoint,kpoint common, minimum width=1 cm]
\tikzstyle{wide kpointdag sc}=[shape=cornercopoint,kpoint common, minimum width=1 cm]

\tikzstyle{black kpoint}=[shape=cornerpoint,shorten left=5pt,kpoint common,fill=black,font=\color{white}]

\tikzstyle{black kpoint sm}=[shape=cornerpoint,shorten left=5pt,kpoint common,fill=black,font=\color{white},scale=0.75]

\tikzstyle{black kpoint adjoint}=[shape=cornercopoint,shorten left=5pt,kpoint common,fill=black,font=\color{white}]
\tikzstyle{black kpointadj}=[shape=cornercopoint,shorten left=5pt,kpoint common,fill=black,font=\color{white}]

\tikzstyle{black kpointadj sm}=[shape=cornercopoint,shorten left=5pt,kpoint common,fill=black,font=\color{white},scale=0.75]

\tikzstyle{black dkpoint}=[shape=cornerpoint,shorten left=5pt,kpoint common,fill=black, doubled,font=\color{white}]
\tikzstyle{black dkpoint adjoint}=[shape=cornercopoint,shorten left=5pt,kpoint common,fill=black, doubled,font=\color{white}]
\tikzstyle{black dkpointadj}=[shape=cornercopoint,shorten left=5pt,kpoint common,fill=black, doubled,font=\color{white}]

\tikzstyle{black dkpoint sm}=[shape=cornerpoint,shorten left=5pt,kpoint common,fill=black, doubled,font=\color{white},scale=0.75]
\tikzstyle{black dkpointadj sm}=[shape=cornercopoint,shorten left=5pt,kpoint common,fill=black, doubled,font=\color{white},scale=0.75]

\tikzstyle{kpointdag}=[kpoint adjoint]
\tikzstyle{kpointadj}=[kpoint adjoint]
\tikzstyle{kpointconj}=[kpoint conjugate]
\tikzstyle{kpointtrans}=[kpoint transpose]

\tikzstyle{big kpoint}=[kpoint, minimum width=1.2 cm, minimum height=8mm, inner sep=4pt, text depth=3mm]

\tikzstyle{wide kpoint}=[kpoint, minimum width=1 cm, inner sep=2pt]
\tikzstyle{wide kpointdag}=[kpointdag, minimum width=1 cm, inner sep=2pt]
\tikzstyle{wide kpointconj}=[kpointconj, minimum width=1 cm, inner sep=2pt]
\tikzstyle{wide kpointtrans}=[kpointtrans, minimum width=1 cm, inner sep=2pt]

\tikzstyle{wider kpoint}=[kpoint, minimum width=1.25 cm, inner sep=2pt]
\tikzstyle{wider kpointdag}=[kpointdag, minimum width=1.25 cm, inner sep=2pt]
\tikzstyle{wider kpointconj}=[kpointconj, minimum width=1.25 cm, inner sep=2pt]
\tikzstyle{wider kpointtrans}=[kpointtrans, minimum width=1.25 cm, inner sep=2pt]

\tikzstyle{gray kpoint}=[kpoint,fill=gray!50!white]
\tikzstyle{gray kpointdag}=[kpointdag,fill=gray!50!white]
\tikzstyle{gray kpointadj}=[kpointadj,fill=gray!50!white]
\tikzstyle{gray kpointconj}=[kpointconj,fill=gray!50!white]
\tikzstyle{gray kpointtrans}=[kpointtrans,fill=gray!50!white]

\tikzstyle{gray dkpoint}=[kpoint,fill=gray!50!white,doubled]
\tikzstyle{gray dkpointdag}=[kpointdag,fill=gray!50!white,doubled]
\tikzstyle{gray dkpointadj}=[kpointadj,fill=gray!50!white,doubled]
\tikzstyle{gray dkpointconj}=[kpointconj,fill=gray!50!white,doubled]
\tikzstyle{gray dkpointtrans}=[kpointtrans,fill=gray!50!white,doubled]

\tikzstyle{white label}=[draw,fill=white,rectangle,inner sep=0.7 mm]
\tikzstyle{gray label}=[draw,fill=gray!50!white,rectangle,inner sep=0.7 mm]
\tikzstyle{black label}=[draw,fill=black,rectangle,inner sep=0.7 mm]

\tikzstyle{dkpoint}=[kpoint,doubled]
\tikzstyle{wide dkpoint}=[wide kpoint,doubled]
\tikzstyle{dkpointdag}=[kpoint adjoint,doubled]
\tikzstyle{wide dkpointdag}=[wide kpointdag,doubled]
\tikzstyle{dkcopoint}=[kpoint adjoint,doubled]
\tikzstyle{dkpointadj}=[kpoint adjoint,doubled]
\tikzstyle{dkpointconj}=[kpoint conjugate,doubled]
\tikzstyle{dkpointtrans}=[kpoint transpose,doubled]

\tikzstyle{kscalar}=[kpoint common, shape=EBox, inner xsep=-1pt, inner ysep=3pt,font=\small]
\tikzstyle{kscalarconj}=[kpoint common, shape=WBox, inner xsep=-1pt, inner ysep=3pt,font=\small]

\tikzstyle{spekpoint}=[kpoint sc,minimum height=5mm,inner sep=3pt]
\tikzstyle{spekcopoint}=[kpoint adjoint sc,minimum height=5mm,inner sep=3pt]

\tikzstyle{dspekpoint}=[spekpoint,doubled]
\tikzstyle{dspekcopoint}=[spekcopoint,doubled]

 \tikzstyle{upground}=[circuit ee IEC,thick,ground,rotate=90,scale=2.5]
 \tikzstyle{downground}=[circuit ee IEC,thick,ground,rotate=-90,scale=2.5]
 \tikzstyle{bigground}=[regular polygon,regular polygon sides=3,draw=gray,scale=0.50,inner sep=-0.5pt,minimum width=10mm,fill=gray]

\tikzstyle{arrs}=[-latex,font=\small,auto]
\tikzstyle{arrow plain}=[arrs]
\tikzstyle{arrow dashed}=[dashed,arrs]
\tikzstyle{arrow bold}=[very thick,arrs]
\tikzstyle{arrow hide}=[draw=white!0,-]
\tikzstyle{arrow reverse}=[latex-]
\tikzstyle{cdnode}=[]

\tikzstyle{none}=[inner sep=0mm]
\usepackage{mathtools,amssymb,amsfonts,amsthm,physics,subcaption,graphicx,multirow,float,bm,soul}
\usepackage[colorlinks,citecolor=blue,linkcolor=blue,urlcolor=blue, breaklinks=true]{hyperref}
\usepackage[capitalize]{cleveref}
\usepackage[shortlabels]{enumitem}
\counterwithout{figure}{section}
\usepackage[none]{hyphenat}
\usepackage{csquotes}
\usepackage[normalem]{ulem}
\usepackage{eurosym}
\newcommand{\ER}{\text{\euro}}

\usepackage{xcolor}

\newcommand{\blk}{\color{black}}

\setlist{nosep}

\usepackage{thmtools} 
\usepackage{thm-restate}

\newcommand{\nocontentsline}[3]{}
\let\oldaddcontentsline\addcontentsline
\newcommand{\tocless}[2]{%
  \let\addcontentsline=\nocontentsline#1{#2}
  \let\addcontentsline\oldaddcontentsline}
  
\newtheorem{theorem}{Theorem}%
\newtheorem{prop}{Proposition}%
\newtheorem{corollary}{Corollary}%
\newtheorem{lemma}{Lemma}%
\newtheorem{definition}{Definition}%
\newtheorem{proposition}{Proposition}%

\begin{document}
\title{On whether quantum theory needs complex numbers:\\the foil theories perspective}
 
\author{Y{\`i}l{\`e} Y{\=\i}ng$^{*,\dagger}$}%
\affiliation{Perimeter Institute for Theoretical Physics, Waterloo, Ontario, Canada, N2L 2Y5}
\affiliation{Department of Physics and Astronomy, University of Waterloo, Waterloo, Ontario, Canada, N2L 3G1}
\author{Maria Ciudad Alañón$^{*}$}
\affiliation{Perimeter Institute for Theoretical Physics, Waterloo, Ontario, Canada, N2L 2Y5}
\affiliation{Department of Physics and Astronomy, University of Waterloo, Waterloo, Ontario, Canada, N2L 3G1}
\author{Daniel Centeno$^{*}$}
\affiliation{Perimeter Institute for Theoretical Physics, Waterloo, Ontario, Canada, N2L 2Y5}
\affiliation{Department of Physics and Astronomy, University of Waterloo, Waterloo, Ontario, Canada, N2L 3G1}
\author{Jacopo Surace$^{*}$}
\affiliation{Perimeter Institute for Theoretical Physics, Waterloo, Ontario, Canada, N2L 2Y5}
\affiliation{Aix-Marseille University, CNRS, LIS, 13288 Marseille CEDEX 09, France}
\thanks{These authors contributed equally to this work.}
\author{Marina Maciel Ansanelli}
\affiliation{Perimeter Institute for Theoretical Physics, Waterloo, Ontario, Canada, N2L 2Y5}
\affiliation{Department of Physics and Astronomy, University of Waterloo, Waterloo, Ontario, Canada, N2L 3G1}
\author{Ruizhi Liu}
\affiliation{Perimeter Institute for Theoretical Physics, Waterloo, Ontario, Canada, N2L 2Y5}
\affiliation{Department of Mathematics and Statistics, Dalhousie University, Halifax, Nova Scotia, Canada, B3H 4R2}
\author{David Schmid}
\affiliation{Perimeter Institute for Theoretical Physics, Waterloo, Ontario, Canada, N2L 2Y5}
\author{Robert W. Spekkens}
\affiliation{Perimeter Institute for Theoretical Physics, Waterloo, Ontario, Canada, N2L 2Y5}
\affiliation{Department of Physics and Astronomy, University of Waterloo, Waterloo, Ontario, Canada, N2L 3G1}

\begingroup
\renewcommand{\thefootnote}{$\dagger$}
\footnotetext{Corresponding author: yying@gmail.com}
\endgroup

\begin{abstract}
Recent work by Renou {\em et al.} (2021) has led to some controversy concerning the question of whether quantum theory requires complex numbers for its formulation. 
We promote the view that the main result of that work is best understood not as a claim about the relative merits of different {\em representations} of quantum theory, but rather as a claim about the possibility of experimentally adjudicating between standard quantum theory and an alternative theory---a {\em foil theory}---known as real-amplitude quantum theory (RQT).  In particular, the claim is that this adjudication can be achieved given only an assumption about the causal structure of the experiment.
Here, we aim to shed some light on {\em why} this is possible, by reconceptualizing the comparison of the two theories as an instance of a broader class of such theory comparisons. By recasting RQT as the subtheory of quantum theory that arises by symmetrizing with respect to the collective action of a time-reversal symmetry, we can compare it to other subtheories that arise by symmetrization, but for different symmetries. If the symmetry has a unitary representation, the resulting foil theory is termed a {\em twirled quantum world}, and if it does not (as is the case in RQT), the resulting foil theory is termed a {\em swirled quantum world}. We show that, in contrast to RQT, there is \emph{no} possibility of distinguishing any twirled quantum world from quantum theory given only an assumption about causal structure. We also define analogues of twirling and swirling for an arbitrary generalized probabilistic theory and identify certain necessary conditions on a causal structure for it to be able to support a causal compatibility gap between the theory and its symmetrized version. We draw out the implications of these analyses for the question of how a lack of a shared reference frame state features into the possibility of such a gap. 
\end{abstract}

\maketitle

\raggedbottom

Complex numbers play an important role in the standard formulation of quantum theory.  This role has received renewed interest~\cite{hita2025Quantum,hoffreumon2025Quantum,feng2025Locality,weilenmann2025partial,sarkar2025Gap,elliott2025Strict,feng2025Locality,volovich_real_2025} due to the 2021 article by Renou {\em et al.}~\cite{Renou_2021}.  Much recent work focuses on the question of whether or not quantum theory {\em needs} complex numbers---that is, whether or not complex numbers are required for its formulation.  Without stipulating any constraints on the formulation, the answer is that they are not. 
Indeed, {\em any}  operational theory, including quantum theory, can be given a real-valued representation within the framework of generalized probabilistic theories (GPTs)~\cite{hardy2001quantum,barrett2007information,Chiribella2010purification,hardy2013reconstructing,muller2021probabilistic}, as we will detail below and in \cref{app:anyreal}. Such representations are equivalent to quasi-probability representations such as the Wigner representation, which is also real-valued.

Ref.~\cite{Renou_2021} limits its focus to the Hilbert space formulation of quantum theory, specified by four axioms (the standard ``textbook'' axioms).
Relative to these, the field over which the Hilbert space is defined cannot be taken to be real without changing the empirical predictions of the theory. If one relaxes one or more of these axioms in the Hilbert space formulation, real-valued representations \emph{can} be constructed without modifying the predictions,  as shown in, e.g., Refs.~\cite{hita2025Quantum,hoffreumon2025Quantum}\footnote{Note that these works have not considered the representations of transformations, and thus are not yet complete representations of the theory.}. Some motivations for studying nonstandard representations of quantum theory are given in Appendix~\ref{app:repns}. 
We also note and explain there that the standard Hilbert space formulation plays a special role in picking out all and only the processes that are included in quantum theory. 

In this paper, however, we promote an alternative framing of the result of Ref.~\cite{Renou_2021}. We argue that rather than viewing it as being about alternative {\em representations} of quantum theory,  it is better to view it as providing a contrast between quantum theory and an {\em alternative theory}, known as {\em real-amplitude quantum theory} (or {\em real quantum theory}) (RQT)~\cite{Wooter_1990,Renou_2021,caves2002Unknown,Stueckelberg1960,mckague2009Simulating,aleksandrova2013Realvectorspace,myrheim1999Quantum}. As we will explain, their result addresses the question of whether these two distinct theories can be experimentally differentiated given only an assumption about causal structure. 

The study of alternative ways the world could be is an important strategy for better understanding quantum theory.
This is an instance of the {\em methodology of foil theories}~\cite{Chiribella2016,centeno2024twirled}.
A {\em foil} to quantum theory is an alternative theory that is studied not as an empirical competitor, but rather to highlight which aspects of the operational phenomenology of quantum theory are generic and which are distinctive. 
Foil theories are often defined within the framework of GPTs, wherein theories are characterized entirely in terms of their operational predictions. Each system is associated with a real vector space of some dimension,  states and measurement effects\footnote{For example, in quantum theory, any element of a positive operator-valued measure (POVM) is an effect.} on that system each have representations as real-valued vectors in that space, transformations have representations as linear maps on these (therefore as real-valued matrices), and statistics are obtained by the Euclidean inner product. 
Differences in the geometric shapes of the spaces of allowed states, effects, and transformations then lead to different operational predictions and so to different theories.  For instance, when the Hilbert space dimension is two, the GPT state space of quantum theory is a ball, while that of RQT is a disk.

Since different GPTs make different operational predictions, one can evidently adjudicate between them experimentally. A natural method for adjudicating between different GPT hypotheses describing a given system is the technique known as GPT tomography~\cite{mazurek}. (We will later compare this approach to the method in Ref.~\cite{Renou_2021}.) 
This technique requires the experimentalist to implement a set of preparations and a set of measurements on the system, record the statistics obtained in different such pairings,  and deduce the real-valued vector representations of these by fitting to the statistics.
One can then apply standard model selection techniques to adjudicate between different GPT hypotheses. GPT tomography notably does not require any prior knowledge of how experimental processes are represented in the GPT (unlike, e.g., standard state tomography in quantum theory which requires knowing which POVM represents each measurement), and in this sense it is similar to protocols in the device-independent paradigm. 
Applying this technique, the experiment of Ref.~\cite{mazurek} provided strong evidence against the RQT hypothesis when the system is the polarization degree of freedom of a photon.  Note that the validity of the conclusions of such experiments depends on the sets of preparations and measurements being tomographically complete for the degree of freedom being probed (as was discussed at length in Ref.~\cite{mazurek}).\footnote{In fact, for falsifying a GPT hypothesis, it suffices to use sets of preparations and measurements that are {\em relatively tomographic}~\cite{schmid2024Shadows}.} 

In comparison, Ref.~\cite{Renou_2021} adjudicates between two particular GPT hypotheses using a different type of assumption, namely, an assumption about the {\em causal structure} of the experiment.\footnote{One can obtain experimental evidence for both assumptions of tomographic completeness and for assumptions about causal structure, as discussed in~\cref{app:exp}.}
The paradigm example is a Bell experiment, 
with the standard assumption that the only causal connection between the two wings is a common cause acting on the outcome variables. 
Relative to this assumption, different GPTs yield different upper bounds on the degree of violation of Bell inequalities~\cite{henson2014Theoryindependent}: for example, no violation in classical probability theory, the Tsirelson bound for quantum theory, and the algebraic bound for the GPT known as boxworld~\cite{barrett2007information}.

If one seeks to adjudicate between quantum theory and RQT using only an assumption about the causal structure, however, then  a Bell experiment does not suffice. This is because RQT can saturate the Tsirelson bound, even in a multipartite Bell experiment~\cite{mckague2009Simulating}. Nonetheless, Ref.~\cite{Renou_2021} showed that by going to a more complex causal structure, specifically, that of the \emph{bilocality} scenario~\cite{bilocality},  adjudication becomes possible. (The bilocality scenario is a generalization of the Bell scenario which consists of three parties, $A$, $B$ and $C$, with a two-way source shared between $A$ and $B$ and another shared between $B$ and $C$.) Thus, the main result of Ref.~\cite{Renou_2021} can be stated as follows: 
\begin{prop}\label{prop:RenouEtAl}
It is possible to experimentally adjudicate between quantum theory and RQT assuming only the causal structure of the experiment (e.g., assuming the structure of the bilocality scenario).  
\end{prop}

We will refer to this result as the {\em RQT-QT causal compatibility gap} (or {\em RQT-QT gap} for short).  

{\em The foil theories perspective on the gap.---}We would like to understand {\em why} there is such a gap. 
Following the methodology of foil theories, we seek insight by recasting it as an instance of a broader class of related phenomena.

To begin, therefore, we must
choose the broader class of theory comparisons of which the quantum to RQT comparison will be considered an instance.  
Motivated by the fact that RQT can be obtained as the result of applying a symmetrization process to quantum theory, namely, symmetrization with respect to a collective time-reversal (as we will detail below), the broader class of theory comparisons we consider includes comparisons of quantum theory to any symmetrization thereof.
Furthermore, quantum theory is just one instance of a GPT to which such a symmetrization can be applied, and so the broader class also includes comparisons of an arbitrary such GPT to any symmetrization thereof. 

This naturally leads to the question:  
what are the necessary and sufficient conditions on the GPT and the symmetry for there to be a causal compatibility gap? 
This article will take some first steps towards answering this question. 

{\em Types of symmetrization processes within quantum theory.---}We begin with the case where the GPT to be symmetrized is quantum theory. 

Central to our article is the distinction between symmetries that admit of a unitary representation and those that do not, which we term \emph{unitary symmetries} and \emph{nonunitary symmetries} respectively. 
The former type includes rotations, time-translations, and phase-shifts. Symmetrization relative to this type yields a class of foil theories called \emph{twirled quantum worlds}~\cite{centeno2024twirled} (to be defined below),
which can be understood as quantum theory with a certain kind of superselection rule~\cite{centeno2024twirled}. 
RQT is not a twirled quantum world since time-reversal is represented by an antiunitary operator. Theories obtained via symmetrization relative to a \emph{non}unitary symmetry share many features with twirled quantum worlds but differ in some crucial respects. 
We term them \emph{swirled} quantum worlds. 

{\em Twirled worlds and swirled worlds.---}Consider a system $S$ composed of $n$ subsystems, $S_1$, $S_2$, $\dots$, $S_n$. Consider a symmetry that is described by 
a group $G$ and suppose that on the Hilbert space of the subsystem $S_i$, this symmetry has 
a projective unitary representation, $\{U^{S_i}_g\}_{g\in G}$. 
Let $\{ \mathcal{U}^{S_i}_g \}_{g\in G}$ 
denote the associated {\em superoperator} representation of this symmetry group, that is, the representation on the space of Hermitian operators for system $S_i$ such that for any Hermitian operator $O^{S_i}$, the action associated to $g\in G$ is
\begin{align}
\label{eq:calU}
\mathcal{U}^{S_i}_g(O^{S_i}) := U^{S_i}_g O^{S_i} {U^{S_i}_g}^\dagger.
\end{align}
The collective superoperator representation for a 
composite system is 
\begin{align}
\label{eq:collective}
\{\mathcal{U}^{S}_g = \mathcal{U}^{S_1}_g \otimes \mathcal{U}^{S_2}_g\otimes \dots  \otimes \mathcal{U}^{S_n}_g\}_{g\in G}.
\end{align}  
We say that a Hermitian operator $O^{S}$ (state or effect) is $(G,\mathcal{U})$-invariant if and only if 
\begin{align}
\label{eq:inv}
  \mathcal{U}^{S}_g (O^{S}) =O^{S} \quad \forall g\in G.
\end{align}
Similarly, a quantum operation  from $S$ to $S'$, represented by a completely positive trace-nonincreasing map $\mathcal{E}^{S'|S}$, is said to be $(G,\mathcal{U})$-covariant if and only if 
\begin{align}
\label{eq:cov}
  \mathcal{E}^{S'|S} \circ  \mathcal{U}^{S}_g =  \mathcal{U}^{S'}_g\circ \mathcal{E}^{S'|S} \quad \forall g\in G, 
\end{align}
where $\circ$ denotes sequential composition. 

The {\em $(G,\mathcal{U})$-twirled quantum world} is the subtheory of quantum theory where the states and effects are restricted to all and only those that are $(G,\mathcal{U})$-invariant and where the quantum operations are restricted to all and only those that are $(G,\mathcal{U})$-covariant. 
Note that the superoperator $\frac{1}{|G|}\sum_{g\in G} \mathcal{U}_g$ (where $|G|$ denotes the cardinality of $G$)\footnote{Here we consider a discrete group, but all of our analysis can be generalized to any compact Lie group.} takes every Hermitian operator to one that is $(G,\mathcal{U})$-invariant. This is termed the {\em twirling} map~\cite{bartlettReference2007}.  The map 
$\frac{1}{|G|}\sum_{g\in G} \mathcal{U}_g \circ \cdot \circ  \mathcal{U}_g^{\dag}$, meanwhile, takes every superoperator to one that is $(G,\mathcal{U})$-covariant. We refer to the symmetrization process yielding a twirled quantum world as the {\em twirling process}. 

A swirled quantum world is defined analogously, but based on a nonunitary symmetry. Wigner's theorem~\cite{wigner1932Gruppentheorie} implies that, in this case, at least one symmetry transformation must be represented by an antiunitary operator on the Hilbert space. 
Denoting the superoperator representation of such a symmetry on system $S_i$ by $\mathcal{A}^{S_i}_g$
and substituting $\mathcal{A}$ for $\mathcal{U}$ in \cref{eq:calU,eq:collective,eq:inv,eq:cov} and the definition of the twirling process, we obtain the definitions of the action of this symmetry on a composite system, of ($G,\cal A$)-invariance, of ($G,\cal A$)-covariance,  of the swirling process, and of a ($G,\cal A$)-swirled quantum world.\blk

{\em RQT as the time-reversal-swirled quantum world.---}RQT can be defined formally as follows.\footnote{Connections to prior definitions are detailed in \cref{app:chnn}.}  
For each system $S$,  choose a basis of the associated Hilbert space $\mathcal{H}^S$. Let $\overline{O}^S$ denote the operator whose matrix elements are the complex conjugates of those of $O^{S}$ in the chosen basis.  Furthermore, let $\overline{\cal E}^{S'|S}$ denote the superoperator whose Choi matrix elements are the complex conjugates of those of ${\cal E}^{S'|S}$ in the chosen basis. RQT is then defined as the subtheory of quantum theory that includes all and only the $O^S$ and ${\cal E}^{S'|S}$ such that 
   $\overline{O}^S=O^S, \text{ and } \overline{\cal E}^{S'|S} ={\cal E}^{S'|S}$.
It is straightforward to see that these operations are closed under parallel and sequential composition.

We now prove that RQT, defined in the manner just described,   can be obtained from quantum theory by symmetrization  with respect to a collective time-reversal symmetry, i.e., that it is simply the time-reversal-swirled quantum world.

The  time-reversal symmetry is a nonunitary representation of $\mathbb{Z}_2$. In terms of its action on the Hilbert space of a subsystem $S_i$, it can be represented by $\{\mathbb{I}^{S_i}, C^{S_i}\}$ where $\mathbb{I}^{S_i}$ is the identity operator and $C^{S_i}$ is the complex-conjugation operator in a basis depending on the Hamiltonian of the system~\cite{haake2018Time,strasberg2022TimeReversal}. %
The corresponding superoperator representation is then $\{\mathcal{I}^{S_i}, \mathcal{C}^{S_i}\}$ where for a Hermitian operator $O^{S_i}$, ${\cal C}^{S_i}(O^{S_i}) := C^{S_i} O^{S_i} {C^{S_i}}^\dagger.$ As proven in \cref{app:conju}, ${\cal C}^{S_i}(O^{S_i})=\overline{O}^{S_i}$. Thus, for any system $S$ (possibly composite), an operator $O^{S}$ (state or effect) is time-reversal-invariant, i.e., ${\cal C}^{S} (O^{S})=O^{S}$, if and only if $\overline{O}^S=O^S$. Furthermore, a quantum operation $\mathcal{E}^{S'|S}$ is time-reversal-covariant, i.e., $\mathcal{E}^{S'|S} \circ  \mathcal{C}^{S} =  \mathcal{C}^{S'}\circ \mathcal{E}^{S'|S}$,  if and only if $\overline{\cal E}^{S'|S} ={\cal E}^{S'|S} $~\cite{hickey2018Quantifying}. This concludes the proof. 

We can therefore summarize \cref{prop:RenouEtAl} as establishing the existence of a causal compatibility gap between a swirled world (namely, the time-reversal-swirled quantum world) and the nonswirled world (namely, quantum theory).  We refer to such a phenomenon as a {\em swirled-nonswirled causal compatibility \blk gap}. 

In fact, for \emph{any} swirled quantum world, there exists a swirled–nonswirled gap, since any antiunitary map can be expressed as complex conjugation in some basis~\cite{wigner1993Normal}, and the cardinality of $\mathbb{Z}_2$ (the group associated to time-reversal symmetry) is already the smallest possible for a nontrivial group. That is, for any other nonunitary symmetry, the resulting swirled quantum world must be isomorphic to a subtheory of RQT and thus cannot realize more correlations in a given causal structure than those achievable in RQT.

Swirled worlds (including RQT) share many operational features with twirled worlds, such as the failure of tomographic locality~\cite{centeno2024twirled,Wooter_1990}  and the fact that pure state entanglement need not be monogamous
~\cite{caves2001entanglement,bartlett2006entanglement,yingTwirldEntangle}. These analogies suggest that one might expect to also see a \emph{twirled-nontwirled causal compatibility gap}.
As we now show, however, no such gap exists.

{\em Causal-structure-preserving simulations.---}Since both swirled and twirled quantum worlds are subtheories of quantum theory, to determine whether there is a twirled-nontwirled (swirled-nonswirled) gap for a given causal structure, it suffices to determine whether any correlation achievable in quantum theory within this causal structure can be simulated within the twirled (swirled) world in a manner that preserves the causal structure.  

We prove that: 
\begin{theorem}\label{thm:twirledsim}
For all causal structures and all unitary symmetries $(G, \mathcal{U})$, there is no gap between the correlations realizable in quantum theory and those realizable in the $(G, \mathcal{U})$-twirled quantum world.
\end{theorem}
We provide a constructive proof of this theorem by presenting a causal-structure-preserving simulation (see \cref{app:preser} for the formal definition) within any twirled quantum world; we call it the \emph{incoherent simulation} and denote it by $\$$.\footnote{In \cref{app:incosim}, we provide another causal-structure-preserving simulation that we call the \emph{coherent simulation}.} 

Consider an experimental setup and the circuit representing it. For each system $S$ (associated to each wire in the circuit) that carries a nontrivial representation of the symmetry, $\$$ maps it to a composite system consisting of $A$ and a reference frame system $R_{A}$. To achieve a perfect simulation, the Hilbert space dimension of $R_A$ must be at least as large as $|G|$, the cardinality of the group, in which case there exists a reference frame state $\ket{e}^{R_S}$ such that $\langle g| g'\rangle^{R_S} = \delta_{g,g'}$, where $|g\rangle^{R_S} \coloneqq U_g |e\rangle^{R_S}$.

Then, the injection of a state $\rho^A$ is mapped to the injection of the twirling over $R_A A$ of the state $\ketbra{e}^{R_A} \otimes \rho^A$.  Contraction with an effect $E^A$, i.e., the superoperator $\tr(E^A\cdot)$, is mapped to the twirling over $R_A A$ (acting from the right) of the superoperator $\tr(\ketbra{e}^{R_A}\otimes E^A\cdot)$.
Finally, a general quantum operation  $\mathcal{E}^{B|A}$ is mapped to the twirling over $R_AA$ (acting from the right) and an {\em independent} twirling over $R_B B$ (acting from the left) of the quantum operation $\ketbra{e}^{R_B}\otimes \mathcal{E}^{B|A} \otimes \tr (\ketbra{e}^{R_A} \cdot )$. That is, 
\begin{align}
\$(\rho^A) 
\coloneqq &\frac{1}{|G|}\sum_{g}\ketbra{g}^{R_A}  \otimes \mathcal{U}^{A}_{g} \left( \rho^{A} \right),\label{eq:dolrhoA}\\
    \$ (\tr_A (E^A\;\cdot)) \coloneqq &\sum_{g} \tr_{R_A} ( \ketbra{g}^{R_A} \;\cdot) \otimes \tr_A ( E^A \;\cdot)\circ {{\cal U}^A_g}^{\dag}, \label{eq:dolEA} \\
\$( \mathcal{E}^{B|A} ) 
 \coloneqq   &  \left( \frac{1}{|G|} \sum_{{ g'}} \ketbra{g'}^{R_B}\otimes  \mathcal{U}^{ B}_{ g'} \right)  \label{eq:dolEAB} \\
 &\circ \mathcal{E}^{ B| A} 
 \circ \left( \sum_{{ g}} \tr_{R_A} \left(  \ketbra{g}^{R_A}  \cdot \right)  \otimes  {\mathcal{U}^{ A}_{ g}}^{\dag} \right). \nonumber
\end{align}
\cref{eq:dolrhoA,eq:dolEA} are special cases of \cref{eq:dolEAB} where the quantum operation $\mathcal{E}^{ B| A}$ is specialized to injection of a state and contraction with an effect respectively. 

For a \emph{multipartite} process, the $\$$ map associates a reference frame to \emph{each} input or output system that carries a nontrivial representation of the symmetry, and then applies the twirling process \emph{independently} to each $R_{S_i}{S_i}$ pair.
Specifically, when $A=A_1A_2\dots A_n$ and $B=B_1B_2\ldots B_m$:
\begin{align}
\$(\mathcal{E}^{B|A}) \coloneqq &
\left( \frac{1}{|G|^m} 
\bigotimes_{i} \sum_{{g}_i} \ketbra{g_i}^{R_{B_i}}\otimes \mathcal{U}^{B_i }_{g_i}
\right) 
 \circ \mathcal{E}^{ B| A} \nonumber \\
 &\circ
  \left(
  \bigotimes_{j} \sum_{{g}_j} \tr(\ketbra{g_j}^{R_{A_j}}\cdot) \otimes  {\mathcal{U}^{{A}_j}_{g_j}}^{\dag}
  \right).  \label{eq:dolEmul}
\end{align}
When $S_i$ carries a trivial representation, there is no need to add $R_{S_i}$. We adopt the conventional assumption that a measurement setting or outcome is encoded in a classical system carrying a trivial representation. 

In \cref{app:incosim}, we prove that the map $\$$ is a simulation, i.e., it preserves the statistics in any causal structure. Furthermore, we prove that $\$$ is causal-structure-preserving since: (i) it commutes with sequential and parallel compositions and thus preserves the wiring of any circuit; (ii) it also preserves the \emph{internal} causal structure of each operation within the circuit.
This establishes \cref{thm:twirledsim}.  

This simulation strategy does not work for swirled worlds despite their strong analogy to twirled worlds. 
Defining a map, denoted $\$_C$, as the analogue of $\$$ but where the symmetry $(G,\mathcal{U})$ is replaced by time-reversal symmetry, we find that although it {\em does} achieve a valid simulation in a \emph{unipartite prepare-measure} experiment---where it is equivalent to the simulation proposed by Stueckelberg~\cite{Stueckelberg1960}---it fails to define a simulation when one moves to more general scenarios.
To see that, consider a bipartite state $\rho^{AB}$.  Since the swirling processes for $A$ and $B$ are independent, we obtain cross terms where ${\cal I}^{A} \otimes {\cal C}^{B}$ is implemented on $\rho^{AB}$. For certain entangled states, this partial complex conjugation will yield an operator that is nonpositive. Therefore, the image of the $\$_C$ map does not live in RQT, and so the $\$_C$ map fails to define a simulation.
See \cref{app:failure} for more details. 

 \emph{Generalizing to arbitrary GPTs.---}One can define twirling and swirling more broadly---for arbitrary GPTs---as follows. 

First, we define a {\em physical} symmetry in a GPT to be one for which all the symmetry transformations are realizable as physical evolutions within the GPT.  Then, a symmetrization process is defined to be a twirling process if it is with respect to a physical symmetry, and a swirling process if it is with respect to a nonphysical symmetry.  \blk
In quantum theory, the distinction between physical and nonphysical symmetries coincides with the unitary-nonunitary distinction. In \cref{app:GPT}, we show that although an analogue of the $\$$ map can be defined within any GPT, it yields a causal-structure-preserving simulation if \emph{and only if} the symmetry is physical. This result implies that for any GPT, we have an analogue of \cref{thm:twirledsim}, namely, that there is no twirled-nontwirled causal compatibility gap for any causal structure.  

For nonphysical symmetries, however, there may or may not be such a gap for a general GPT. This follows from the fact that in some GPTs there are two different classes of nonphysical symmetries: one wherein adding the symmetry transformations to the GPT would yield a \emph{logical inconsistency} in the sense of obtaining a number outside $[0,1]$ where one expects a probability, which we term {\em strongly nonphysical}, and one wherein this addition does not yield a logical inconsistency, which we term {\em weakly nonphysical}.  (There is only one type of nonphysical symmetry in quantum theory, since all nonunitary symmetries are strongly nonphysical.) For weakly nonphysical symmetries, we show in \cref{app:GPT} that there exist nonclassical GPTs\footnote{Here, nonclassical GPTs refers to GPTs that are not a subtheory of classical probability theory~\cite{GPTembedding}. We restrict our attention to such nonclassical GPTs since, as explained in \cref{app:necessary}, any subtheory of classical probability theory will not exhibit a symmetrized-nonsymmetrized gap regardless of the symmetry (cf. \cref{cor:nonclassical}).}  for which there is no swirled-nonswirled gap, while we conjecture that there is always a gap when one swirls a nonclassical GPT with respect to a strongly nonphysical symmetry.

The absence of any twirled-nontwirled gaps contrasts strongly with the existence of the RQT-QT gap.  To find more examples of theories yielding a symmetrized-nonsymmetrized gap, the considerations above clarify where to look:  GPTs with nonphysical symmetries, and in particular, the ones with strongly nonphysical symmetries.

{\em Conclusions.---}We now return to the question of why there is a RQT-QT causal compatibility gap. 
 
\cref{thm:twirledsim} and its GPT generalization show that in twirled worlds, a simulation can be achieved by encoding into {\em local} degrees of freedom  that are invariant under the symmetry---specifically, relational degrees of freedom between a system and its associated reference frame (cf. \cref{app:rep}), with different systems being associated with {\em independent} reference frames. The failure of this simulation in any swirled world implies that such local invariant encodings are not possible there.\footnote{\label{footnote:yen}Note that the failure of $\$_C$ to yield a simulation 
in some causal structure implies the possibility, but not the necessity, of an RQT-QT gap in that causal structure (for example, $\$_C$ fails to be a simulation in the Bell scenario despite there being no RQT-QT gap there). We can modify $\$_C$ to incorporate a shared reference frame state between systems whenever allowed by the causal structure, and this yields a simulation in certain causal structures where $\$_C$ does not, such as the Bell scenario. 
However, this generalization necessarily fails in some other scenarios. Note that allowing post-selection can facilitate simulations in these scenario.~\cite{ying2025yenmap}.\blk
}

The idea of using  a {\em reference frame} to achieve a simulation of the nonsymmetrized theory by the symmetrized theory, which originated in the context of unitary symmetries~\cite{bartlettReference2007}, can be ported to the context of symmetrization under time-reversal, as is done in~\cite{aleksandrova2013Realvectorspace}. (Ref. \cite{weilenmann2025partial} refers to it as a reference frame for ``complexness''.) 
It has been noted~\cite{weilenmann2025partial} that if the reference frames associated to different systems are prepared in a \emph{shared reference frame} state (cf. \cref{app:shareRF})---something only achievable by going beyond the causal structure of the bilocality scenario\footnote{This is impossible even if classical randomness is shared between the independent sources in the bilocality scenario since, as proven in \cref{app:shareRF}, any state acting as a shared reference frame must be entangled in the symmetrized theory (i.e., not preparable under local operations and shared randomness). In particular, the shared reference frame state used in \cite{weilenmann2025partial} is entangled in RQT. Indeed, Ref.~\cite{Renou_2021} proves that the RQT-QT gap persists even with the classical shared randomness.}---one can achieve a simulation. 
This might suggest that the following intuition explains the RQT-QT gap, or at least makes it unsurprising: a theory can be simulated by a symmetrized version thereof {\em only if} the causal structure allows for the preparation of a shared reference frame state among all the GPT systems.

There are two problems with this intuition: (i) it makes no reference to the nature of the symmetry, and the claim is known to be false for the case of physical symmetries (because, as noted above, no shared reference frame state is needed in this case for any causal structure); (ii) even for strongly nonphysical symmetries (which are known to yield a swirled-nonswirled gap in the bilocality scenario), there are causal structures that do not allow for the preparation of a shared reference frame state among the systems but that still provably do not exhibit a gap. 
In \cref{app:causal}, we provide a number of necessary conditions for a causal structure to admit a gap, and present various examples of causal structures that do not allow for the preparation of a shared reference frame state but nevertheless violate our necessary conditions, and thus cannot admit of a gap.

In short, the interplay between the necessity of a shared reference frame state, the nature of the symmetry, and the causal structure is more subtle than the simple intuition would lead one to believe.  Nonetheless, in seeking to achieve a better understanding of how the nature of a symmetry affects the properties of the corresponding symmetrized worlds, a study of the differences in the phenomenology of reference frames---and consequently of relational degrees of freedom---may ultimately provide the most conceptual insight.

{Acknowledgements.---}We thank John Selby, Thomas D. Galley and Elie Wolfe for their valuable insights. We also thank Mischa Woods, Davide Rolino and Yeong-Cherng Liang for useful discussions. All authors were supported by Perimeter Institute for Theoretical Physics. Research at Perimeter Institute is supported in part by the Government of Canada through the Department of Innovation, Science and Economic Development and by the Province of Ontario through the Ministry of Colleges and Universities. 
YY and MMA are also supported by the Natural Sciences and Engineering Research Council of Canada (Grant No. RGPIN-2024-04419). JS are also supported by the French government under the France 2030 investment plan, as part of the Initiative d'Excellence d'Aix-Marseille Université-A*MIDEX, AMX-22-CEI-01. RL is also supported by the Simons
Collaboration on Global Categorical Symmetries through Simons Foundation grant 888996.

\bibliography{references.bib}

\appendix

\section{The status of assumptions about tomographic completeness and about causal structure}
\label{app:exp}

In the following, we discuss the sense in which there is an opportunity for obtaining experimental evidence about the plausibility of an assumption of tomographic completeness (and then argue that the situation is similar for assumptions about the causal structure).

The cardinality of a minimal tomographically complete set is the dimension of the GPT vector space.  Making a particular assumption about this cardinality, therefore, implies making an assumption about the dimension of the GPT vector space in one's fit. A given such choice of dimension might underfit the data. In this case, one has experimental evidence that weighs against this assumption, particularly if one's statistical confidence is high.

In this way, experiments provide an opportunity to falsify a given hypothesis about an upper bound on the cardinality of the tomographically complete set, thereby yielding a lower bound on this cardinality.  This can serve to confidently rule out certain GPT hypotheses.  For instance, in an experiment on photon polarization, if the best-fit dimension-3 GPT model underfits the data, one can confidently rule out the hypothesis that the GPT governing the experiment is the rebit GPT (i.e., RQT on a 2-dimensional Hilbert space). Consequently, if one is seeking to adjudicate between the hypothesis that photon polarization is described by the qubit GPT and the hypothesis that it is described by the rebit GPT, it is possible to find experimental evidence from GPT tomography that rules in favor of the qubit hypothesis and against the rebit hypothesis.  

Another choice of dimension might {\em overfit} the data.  In this case, one has tentative experimental evidence against that assumption about the dimension.  It is tentative in the sense that in a future experiment involving a greater diversity of preparations and measurements, it might be found that the assumption no longer overfits the data, that the extra expressive power is actually necessary to not underfit the data.  

As noted in the main text, for the purpose of providing evidence {\em against} a given GPT hypothesis, it also suffices to use sets of preparations and measurements that are merely {\em relatively tomographic}~\cite{schmid2024Shadows} according to that GPT hypothesis, since this is a necessary and sufficient condition for obtaining an accurate description of the GPT fragment~\cite{selby2023} describing one's preparations and measurements.

For further details on these assumptions, see the introduction and conclusion sections of Ref.~\cite{mazurek} and Ref.~\cite{schmid2024Shadows}.

The situation with an assumption about the causal structure of an experiment is of a similar type. 

Again, for a given GPT, the best-fit GPT model under a given hypothesis about the causal structure may be seen to underfit the experiment data, in which case one has experimental evidence against this causal hypothesis.
This is what occurs when one takes classical probability theory as the GPT, and one makes the hypothesis that the causal structure of a Bell experiment is the standard one (used for deriving Bell inequalities). If the data violates a Bell inequality, then because one explain the  violation under such a causal assumption, the best-fit model will underfit the data. In other words, one can rule out an assumption about causal structure by hypothesis testing. 

Similarly, for a given hypothesis about the GPT, a given assumption about the causal structure might be found to {\em overfit} the data.  See \cite{daley2022experimentally,daley2025reply} for details.

Consequently, these two types of assumptions---about tomographic completeness and about causal structure---have a similar status regarding experimental testability and are consequently both viable assumptions on which to ground experimental adjudications between different GPT hypotheses.

\section{Why any operational theory admits of real-valued representations}

\label{app:anyreal}

Here we reproduce the important foundational result due to Hardy~\cite{hardy2001quantum}:
\begin{prop}
    Any theory can be given a real-vector-space representation.
\end{prop}
The basic argument is simple. Any theory of the world must make empirical predictions about the probability with which the outcomes of any given measurement occur. Moreover, there is always some minimal set of measurements, called a fiducial set, such that knowing the probabilities that the theory predicts for those measurements is sufficient to deduce the probabilities for {\em all} measurements in the theory. (In the worst case, the fiducial set contains every measurement in the theory, but usually the fiducial set is dramatically smaller. For example, on a qubit, measurements in Pauli $X$, $Y$, and $Z$ bases suffice.) One can then represent the state of the system simply as a vector of the probabilities that are assigned by the theory to the outcomes of these fiducial measurements. This gives a real-valued representation of each state in the theory. It is moreover a {\em complete} representation in the usual sense, as it perfectly encodes the probability assigned by the theory to every outcome of every measurement. The fiducial effects, meanwhile, are represented simply as unit vectors, which are also real-valued. One can then uniquely compute the representation of non-fiducial effects and of transformations in the theory, and it is not hard to show that all of these are also represented by real-valued (although not necessarily positive) matrices~\cite{hardy2001quantum}. Similarly, for a composite system, one can also find fiducial states and measurements and real-valued representations of any state, effect and transformation on the composite system.

Note that, the real vector space used in this representation is also a special case of a Hilbert space, where both pure states and mixed states are represented by vectors in this Hilbert space. In this sense, one might argue that this could be counted as a Hilbert space formulation of quantum theory~\cite{Renou_2021}. However, it violates Axiom 3 in Ref.~\cite{Renou_2021}, since the empirical probabilities are linear, rather than bilinear, in the Hilbert space vectors.

\section{Why representations matter}
\label{app:repns}

We argued in the main text that the result of Ref.~\cite{Renou_2021} is better understood as being about the relationship between two distinct theories, rather than merely about the question of what representations of quantum theory are possible.
But, to be clear, we do think that it is valuable and interesting to study different possible representations of quantum theory.

There are many reasons to do so. A simple first reason is that one or the other representation might
be easier to work with or lead to better intuitions in a particular context. 
Moreover, different representations highlight different features of quantum theory, suggesting different perspectives on the conceptual meaning of the formalism. 

For instance, at the time of the birth of modern quantum theory, there was 
serious debate about whether Schrodinger's wave mechanics or Heisenberg's matrix mechanics was more fundamental or more useful, especially as the two approaches were packaged together with very different philosophical commitments. Ultimately, both representations were found to be useful and indeed both are still in use today. As another example, the Wigner representation is defined on the phase space of classical mechanics, making this representation particularly useful for considering the classical limit of quantum theory. 

Another reason is that different representations of quantum theory suggest different possibilities for constructing foil theories. For example, one can construct subtheories of quantum theory by allowing all and only the processes in quantum theory whose representations are positive in a particular quasiprobability representation of the formalism.
Such foil theories are useful for understanding what features of quantum theory should be deemed genuinely nonclassical~\cite{spekkens2008Negativity}.
RQT itself is an example of this pattern. It is a very natural alternative to standard quantum theory, starting from the Hilbert space formulation of the latter.

Finally, we note an advantage to formulating quantum theory via the textbook postulates (as was done in Renou {\em et al.}), one that we have not previously seen emphasized. In this particular representation, it is straightforward to characterize the set of valid pure states: they are {\em all and only} the normalized rays in the given Hilbert space.  
By contrast, in real-valued vector representations of quantum theory (such as the representation provided by the framework of GPTs, or equivalently, the framework of quasi-probability representations), many of the vectors in the relevant vector space do not correspond to valid quantum states (but merely lie in the linear span of the valid quantum states). So, what such representations leave to be desired is that they must provide an independent specification of precisely {\em which} vectors in real vector space are included in the theory, and it is not clear whether such a specification can be given in any succinct way that does not refer back to the standard Hilbert space formulation. 
However, there are some works that take on this challenge; these works aim to pick out all and only the quantum processes via a set of physical (rather than mathematical) axioms. This is known as the quantum reconstruction program~\cite{muller2021probabilistic,hardy2013reconstructing}.

\section{Properties of the complex conjugation operator
}
\label{app:conju}
Consider $C$, the complex conjugation operator  
relative to a given orthonormal basis $\{\ket{a}\}_a$ of a Hilbert space. That is, when $C$ acts on a vector $\ket{\psi}\coloneqq\sum_{a} \braket{a}{\psi}\ket{a}\coloneqq \sum_a \psi_a \ket{a}$, we have
\begin{align}
C\ket{\psi} = \overline{\ket{\psi}} \coloneqq \sum_a \overline{\psi}_a\ket{a},
\end{align}
where the overbar notation indicates complex conjugation. $C$ is an antilinear operator, meaning that
\begin{align}
  & C (\ket{\psi_1} + \ket{\psi_2}) = C \ket{\psi_1} + C \ket{\psi_2}, \\
  & C (k\ket{\psi}) = \overline{k} \, C \ket{\psi}.
\end{align}
Thus, it cannot be represented as a matrix.  Note that since $C^2=\mathbb{I}$, complex conjugation is its own inverse, 
\begin{align}
    C^{-1} = C.
\end{align}

Now we define the adjoint of $C$. First, note that the usual notion of adjoint is defined only for linear operators. However, that definition can be extended to an antilinear operator such as $C$ as follows~\cite[Eq. 2.2.7]{Weinberg_1995}: for all $\ket{\psi}$ and 
 $\ket{\eta}$, the adjoint of $C$, denoted $C^{\dagger}$, is defined by
\begin{align}
\label{eq:defCadj}
\bra{\eta}\big(C^{\dagger}\ket{\psi} \big)=\bra{\psi} \big(C\ket{\eta} \big).
\end{align}
\cref{eq:defCadj} is only satisfied when
\begin{align}
    C^{\dagger}=C,
\end{align}
 in which case $\bra{\eta}\big(C^{\dagger}\ket{\psi} \big)=\bra{\psi} \big(C\ket{\eta} \big) %
 = \sum_a \overline{\psi_a}\overline{\eta_a}$.

How does $C$ act on operators? For any vector $\ket{\psi}$ and for any operator $O$, one has, writing everything in the $\{\ket{a}\}_a$ basis,
\begin{align}
    CO\ket{\psi}&=C\sum_{i,j,k} O_{j,k} \psi_i \ketbra{j}{k}\ket{i}=C\sum_{i,j,k} \big(O_{j,k} \psi_i \braket{k}{i}\big)\ket{j} \nonumber \\ 
    &=\sum_{i,j,k} \overline{\big(O_{j,k} \psi_i \braket{k}{i}\big)}\ket{j}= \nonumber\sum_{i,j,k} \overline{O}_{j,k} \overline{\psi}_i \overline{\braket{k}{i}} \ket{j} \\ 
    &=\sum_{i,j,k} \overline{O}_{j,k} \overline{\psi}_i \braket{k}{i} \ket{j}=\nonumber  \overline{O}C\ket{\psi},
\end{align}
Thus, in general 
\begin{align}
    CO=\overline{O}C.
\end{align}
Therefore, if the superoperator ${\cal C}$ is defined as
\begin{align}
{\cal C}(O)\coloneqq C O C^{\dagger},
\end{align}
\blk
then 
\begin{align}
\label{eq:ConO}
{\cal C}(O)=\overline{O}CC^{\dagger}=\overline{O},
\end{align}
establishing that the action of ${\cal C}$ on an operator is simply to take the complex conjugate of the elements of the matrix representation of that operator (relative to the chosen basis).

\section{
Connection between the time-reversal-swirled world and definitions of RQT in the literature}
\label{app:chnn}

Recall that we take the conventional definition of RQT to be as follows. For each system $S$, one chooses a basis of the associated Hilbert space $\mathcal{H}^S$. 
RQT is then defined to be the subtheory of quantum theory that includes all and only the density operators and effects whose matrix elements (when expressed relative to the chosen  basis) are all real, and that includes all and only the quantum operations whose Choi matrix (when expressed relative to the chosen  basis) has only real elements. 

Strictly speaking, we do not know of any previous work that explicitly defines RQT as a full theory including a complete specification for its states, effects, and operations. However, as mentioned in the main text, it is straightforward to extend the usual proposals~\cite{Wooter_1990,Renou_2021,caves2002Unknown,Stueckelberg1960,mckague2009Simulating,aleksandrova2013Realvectorspace,myrheim1999Quantum} of RQT to a full theory, and such a full theory is what we refer to as the conventional definition of RQT in the main text.  Note that the usual proposals are distinct from some recent proposals for how to define RQT 
where parallel composition is not represented by the tensor product~\cite{hita2025Quantum,hoffreumon2025Quantum},
or where the states and effects include ones that are not positive semi-definite  operators~\cite{hoffreumon2025Quantum}.  Furthermore, these recent proposals are representations of (the prepare-measure fragment of) quantum theory instead of distinct GPTs. 

Some proposals~\cite{Renou_2021,Stueckelberg1960} only specify which pure states are included within RQT rather providing such a specification for the full scope of density operators. In these proposals, the pure states allowed in RQT are all and only those whose Hilbert-space representation has only real components in a chosen basis.
Similarly, for measurements, some proposals~\cite{Renou_2021,Wooter_1990,caves2002Unknown,caves2001entanglement,aleksandrova2013Realvectorspace,myrheim1999Quantum,Stueckelberg1960} only concern which projection-valued measures (PVMs) (or observables) are included, rather than providing a specification for the full scope of positive-operator values measures (POVMs).  In these proposals, the PVM elements (or observables) allowed in RQT are all and only those whose representation as a matrix acting on Hilbert space is real in a chosen basis. As for the proposals that also concern transformations, the discussion has been limited to reversible transformations~\cite{mckague2009Simulating,aleksandrova2013Realvectorspace,myrheim1999Quantum,Stueckelberg1960}. In these proposals, the valid reversible transformations are all and only the orthogonal transformations (i.e., unitaries whose matrix representations in the chosen basis are real). Note that there is a proposal to include also the {\em isometries} whose matrix representations are real (See the Supplementary Material of \cite{Renou_2021}.)

First, we note that when our specification of the states, measurements and operations that are included in RQT is specialized to pure states, PVMs, and unitaries (or isometries) respectively, we recover the respective specifications made in these prior proposals.

More importantly, it is straightforward to extend these proposals by stipulating that the mixed states, POVMs, and general quantum operations that ought to be included in the theory are all and only those that respectively admit of a purification, Naimark dilation, or Stinespring dilation in terms of pure states, projectors, and unitary channels that are included in the theory. (In fact, the notions of purification and Naimark dilations are subsumed as special cases of a Stinespring dilation.) The result \cite[Theorem 2]{hickey2018Quantifying} establishes that the set of quantum operations that admit of a real-amplitude Stinespring dilation (i.e., one where the pure state of the ancilla is real-amplitude and the reversible transformation on system and ancilla is given by an orthogonal matrix) are all and only the quantum operations with a real Choi matrix.  Consequently, if one extends the conventional proposals on how to define RQT to general quantum operations via the Stinespring dilation, one obtains the definition of RQT that we provided above.

Finally, it is worth noting another result  from  Ref.~\cite{hickey2018Quantifying}, namely, that the set of quantum operations whose Choi matrix is real can also be characterized as the set of quantum operations that are {\em completely real-preserving}, i.e., when acting on a system or any part of a system, they take every real state of the system to a real state.

\begin{widetext}
\section{Simulations in twirled quantum worlds}
\label{app:sim}

Here we first provide the conditions for a map to be a causal-structure-preserving simulation. Then, we provide more details on the $\$$ map and the proof that it is causal-structure-preserving. We further introduce the coherent simulation, denoted $\ER$, and prove that it too is causal-structure-preserving. Finally, we comment on the relational degrees of freedom used by these two maps in \cref{app:rep}.

Note that both $\$$ and $\ER$ are linear and that being causal-structure-preserving implies being  connectivity-preserving  (a.k.a. semi-functorial~\cite{mac2013categories}); as such, each simulation is an instance of a GPT embedding (of quantum theory into twirled quantum world) map in the sense of Ref.~\cite{GPTembedding}.  

\subsection{Causal-structure-preserving simulations}
\label{app:preser}

In a circuit describing some experiment, each wire is associated with a system, and each gate (a.k.a. box) is associated with an operation, such as injection of a state, contraction with an effect, or a transformation, each of which may have multipartite inputs or outputs. 
Given a circuit, facts about its causal structure are of two types:
\begin{enumerate}
    \item the compositional structure of the circuit, that is, the decomposition of the circuit into sequential and parallel compositions of operations; and
    \item the internal causal structure of each operation appearing in the circuit, that is, causal connections between the inputs and the outputs of each operation.
\end{enumerate}
The first type is the most commonly studied in the literature. Here, we also include the second type, which has been studied in recent works on quantum causal models~\cite{allen2017Quantum,barrett2019Quantum,ormrod2024Quantum}.
It is currently unclear how to express the internal causal structure of certain operations purely compositionally, that is, by a decomposition of that operation into more fine-grained components in a circuit where the compositional structure of the components captures the internal causal structure of the operation~\cite{barrett2019Quantum}.

A causal-structure-preserving simulation is a map $\bf M$ that maps a circuit to another circuit such that the statistics and \emph{both} types of causal structures are preserved. Explicitly,  $\bf M$ is a simulation if and only if the statistics are preserved, and $\bf M$ is causal-structure-preserving if and only if
\begin{align}
\label{eq:noin}
   \forall A,B, \qquad A\nrightarrow B \quad \implies \quad {\bf M} (A) \nrightarrow {\bf M} (B).
\end{align}
Here, $A$ and $B$ can be composite systems or subsystems, $\nrightarrow$ denotes a \emph{no-influence} relation, and ${\bf M} (S)$ denotes the image of system $S$. In words, \cref{eq:noin} states that if a system $A$ does not causally influence a system $B$ in the original circuit, either due to the wiring or the internal causal structure of some operations, then their images under ${\bf M}$ also exhibit the same no-influence relation. 

Note that \cref{eq:noin} requires the implication only in the forward direction. That is, it is acceptable for the image of a circuit under a causal-structure-preserving map to have \emph{more} no-influence relations  (hence, \emph{fewer} nontrivial influences) than the circuit did. For example, one may have ${\bf M}(A) \nrightarrow {\bf M}(B)$ even when $A$ and $B$ are causally connected in the original circuit. It is inappropriate to demand that the implication also hold in  the backward direction because a causal connection in a circuit merely indicates the {\em possibility} of a nontrivial causal influence. 

For a map $\bf M$ to preserve the first type of fact about causal structure in a general circuit, it suffices to demand that $\bf M$ commutes with both sequential and parallel compositions. That is, for any operation $\cal E$ in the circuit (such as injection of a state, contraction with an effect, or a transformation), 
\begin{align}
        {\bf M}({\cal E}^{B_1|A_1}) \otimes {\bf M}({\cal E}^{B_2|A_2}) &= {\bf M}({\cal E}^{B_1|A_1} \otimes {\cal E}^{B_2|A_2}), \label{eq:parallel0}\\
     {\bf M}({\cal E}^{C|B}) \circ {\bf M}({\cal E}^{B|A})& = {\bf M}({\cal E}^{C|B} \circ {\cal E}^{B|A}).\label{eq:sequential0}
\end{align}

If the map $\bf M$ is proven to preserve the statistics of a simple prepare-measure scenario; then \cref{eq:parallel0,eq:sequential0} %
guarantees that the statistics can be reproduced in any circuit,
since a circuit with no open quantum wires defines a probability. 

In terms of preserving the second type of fact about causal structure in a given circuit, the specific criteria may vary depending on one’s definition of causal influence internal to an operation. For example, in proposals such as~\cite{barrett2019Quantum}, it is only meaningful to talk about the causal influence relations between systems connected by unitary channels. In such cases, one must first dilate each operation in the circuit, and then consider the preservation of internal causal structure of the unitary operations appearing in these dilations.
Nevertheless, we will later show that both incoherent and coherent simulations (namely, $\$$ and $\ER$) preserve internal causal structures \emph{regardless} of the specific definition used.

Finally, we comment on the motivation for demanding causal-structure-preservation in a simulation. Causal-structure-preservation is a reasonable property to demand of a simulation 
because typically the reasons one can provide in favor of knowing the causal structure when one knows the experimental setup are reasons that are independent of one's hypothesis about which GPT is the correct theory of nature.  For instance, in a Bell experiment, one appeals to space-like separation and the implausibility of superdeterminism or retrocausation to justify the common-cause structure. Consequently,  a willingness to entertain the possibility that some other GPT, distinct from quantum theory, governs one's experiment does not imply that one should be willing to modify one's hypothesis about the causal structure of the experiment. 

For instance, suppose that one implements an experiment of the form of the bilocality scenario and the experimental data exhibits correlations that are only compatible with the causal structure of the bilocality scenario if one assumes quantum theory rather than RQT. To imagine explaining this data by supposing that the causal structure of the experiment is {\em not} in fact that of the bilocality scenario (as one expects) but  rather one wherein there is a quantum common cause shared among all three parties, i.e., a tripartite Bell scenario, imposes a new burden of explanation. 
Moving from the bilocality scenario to the tripartite Bell scenario does not merely provide the possibility for RQT to simulate correlations that quantum theory can realize in the bilocality scenario, it allows for new types of correlations beyond these, such as perfect correlations among the outcomes at the three wings.  
The new burden of explanation, then, is this: why is it that in the experiment we do not see all the correlations that can be realized by RQT in a tripartite Bell scenario? 
A restriction to only the correlations that are realizable by quantum theory in the bilocality scenario can be achieved by a kind of fine-tuning of the parameters of the model, just as it can in accounts of Bell inequality violations that appeal to a classical causal model but posit an exotic departure from the causal structure of the Bell scenario (such as interlab influences, superdeterminism or retrocausation).  In both cases, however, there are reasons to be dissatisfied with such an explanation.  See Ref.~\cite{wood2015Lesson} for the details of the argument.  Furthermore,  classical causal models with exotic causal structures tend to {\em overfit} the data relative to quantum causal models with conservative causal structures~\cite{daley2022experimentally}. 

\subsection{Incoherent simulation}
\label{app:incosim}

We first define $\$$ for unipartite quantum operations from system $A$ to system $B$, where both $A$ and $B$ carry nontrivial representations of the symmetry. Let $\mathcal{T}^{A\rightarrow B}$ denote all the superoperators  (i.e., linear maps) from $\mathcal{L}(\mathcal{H}^A)$ to $\mathcal{L}(\mathcal{H}^B)$. For each superoperator $\mathcal{E}^{B|A}\in \mathcal{T}^{A\rightarrow B}$, we add a reference frame system for the input and for the output, called $R_A$ and $R_B$ respectively, (as mentioned earlier, when a system, e.g., $A$, is trivial, we do not add a reference frame system to it). We define the map
\begin{align}
& \$ : \mathcal{T}^{A\rightarrow B} \rightarrow \mathcal{T}^{R_A A\rightarrow R_B B} \,:: \mathcal{E}^{B| A} \mapsto \mathcal{E}_{\rm inv}^{R_B B| R_A A}
\end{align}
where
\begin{equation}
\label{eq:dollaruni}
\mathcal{E}_{\rm inv}^{R_B B| R_A A} \coloneqq \$( \mathcal{E}^{B|A} )
 \coloneqq     \left( \frac{1}{|G|} \sum_{{ g'}} \ketbra{g'}^{R_B}\otimes  \mathcal{U}^{ B}_{ g'} \right) 
 \circ \mathcal{E}^{ B| A} 
 \circ \left( \sum_{{ g}} \tr \left(  \ketbra{g}^{R_A}  \cdot \right)  \otimes  {\mathcal{U}^{ A}_{ g}}^{\dag} \right), 
\end{equation}
and where we have used the terminological convention that if $E$ is an effect on $\mathcal{L}(\mathcal{H})$, then $\tr(E \cdot)$ denotes the superoperator that takes the Hilbert-Schmidt inner product of its input with $E$, i.e., $\tr(E \cdot) : \mathcal{L}(\mathcal{H}) \rightarrow \mathbb{C} :: O \mapsto \tr(E O) $.  

\pgfdeclarelayer{nodelayer}
\pgfdeclarelayer{edgelayer}
\pgfsetlayers{edgelayer,nodelayer,main}
This is represented in a string diagram (a.k.a. circuit) as:
\begin{equation}
\begin{tikzpicture}
	\begin{pgfonlayer}{nodelayer}		
        \node [style=right label] (6) at (-0.8, -3) {$R_A$};
        \node [style=right label] (7) at (1.1, -3) {$A$};
        \node [style=right label] (1) at (1.4, 1.5) {$\$$};

        \node  [style=proj2] (11) at (-0, 0) {$\mathcal{E}^{B|A}$};
        \node [style=right label] (12) at (0.25, 1.5) {$B$};
        \node [style=right label] (13) at (0.25, -1.5) {$A$};

        \node [style=rectangle, minimum width=2cm, minimum height=2cm, fill=green, opacity=0.3, rounded corners] (17) at (0, 0) { };

        \node [style=right label] (18) at (-0.80, 3) {$R_B$};
        \node [style=right label] (20) at (1.1, 3) {$B$};
        \node [style=rectangle] (21) at (0, 2.25) { };
        \node [style=rectangle] (22) at (0, -2.25) { };
        \node [style=rectangle] (23) at (-0.95, 1.75) { };
         \node [style=rectangle] (24) at (-0.95, 4) { };
         \node [style=rectangle] (25) at (0.95, 1.75) { };
         \node [style=rectangle] (26) at (0.95, 4) { };
         \node [style=rectangle] (27) at (-0.95, -1.75) { };
         \node [style=rectangle] (28) at (-0.95, -4) { };
         \node [style=rectangle] (29) at (0.95, -1.75) { };
         \node [style=rectangle] (30) at (0.95, -4) { };
	\end{pgfonlayer}
	\begin{pgfonlayer}{edgelayer}
        \draw [qWire] (11) to (21);
        \draw [qWire] (11) to (22);
        \draw [qWire] (23) to (24);
        \draw [qWire] (25) to (26);
        \draw [qWire] (27) to (28);
        \draw [qWire] (29) to (30);
	\end{pgfonlayer}
\end{tikzpicture}
 \coloneqq {\hbox{
\begin{tikzpicture}
	\begin{pgfonlayer}{nodelayer}		
        \node [style=right label] (6) at (-2.35, -3.0) {$R_A$};
        \node [style=right label] (7) at (0.15, -3.0) {$A$};
        \node  [style=proj2] (8) at (0, -2.0) {${\mathcal{U}_{g}^A}^\dagger$};
        \node [style=Wsquare] (9) at (-2.5, -2.0) {$g$};
        \node [style=rectangle, minimum width=1.5cm, align=left] (10) at (-4.5, -2.0) {$\sum_{g}$};
        \node  [style=proj2] (11) at (0, 0.0) {$\mathcal{E}^{B|A}$};
        \node [style=right label] (12) at (0.15, 1.0) {$B$};
        \node [style=right label] (13) at (0.15, -0.9) {$A$};
        \node  [style=proj2] (14) at (0, 2.0) {$\mathcal{U}_{g'}^B$};
        \node [style=Wsquareadj] (15) at (-2.5, 2.0) {$g'$};
        \node [style=rectangle, minimum width=1.5cm, align=left] (16) at (-4.5, 2.0) {$\frac{1}{|G|}\sum_{g'}$};
        \node [style=rectangle] (17) at (-2.5, 3.8) { };
        \node [style=right label] (18) at (-2.35, 3.0) {$R_B$};
        \node [style=rectangle] (19) at (0, 3.8) { };
        \node [style=right label] (20) at (0.15, 3.0) {$B$};
        \node [style=rectangle] (22) at (0, -3.8) { };
        \node [style=rectangle] (23) at (-2.5, -3.8) { };
	\end{pgfonlayer}
	\begin{pgfonlayer}{edgelayer}
        \draw [qWire] (8) to (11);
        \draw [qWire] (14) to (11);
        \draw [qWire] (15) to (17);
        \draw [qWire] (14) to (19);
        \draw [qWire] (22) to (8);
        \draw [qWire] (23) to (9);
	\end{pgfonlayer}
\end{tikzpicture}
}}.
\end{equation}

We will use algebraic equations and string diagrams interchangeably.

The $\$$ map preserves the property of being completely positive (CP) and also the property of being trace-preserving (TP).  This follows from the fact that the superoperators that act on the left and right of $\mathcal{E}^{B|A}$ in \cref{eq:dollaruni} are CPTP. (The $\$$ map also preserves the property of unitality since $[\$ (\mathcal{E}^{B|A})]^{\dag} = \$ ([\mathcal{E}^{B|A}]^{\dag})$, where  $\mathcal{E}^{B|A}$ is unital if and only if $[\mathcal{E}^{B|A}]^{\dag}$ is TP.)

Furthermore, $\$({\cal E}^{B|A})$ is a valid process in the $(G,{\cal U})$-twirled quantum world since it is $(G,{\cal U})$-covariant. Its covariance follows from the fact that it is both left-invariant and right-invariant under the group action (hence the subscript ``inv'' in $\mathcal{E}_{\rm inv}^{R_B B| R_A A}$). 
More precisely, $\forall g\in G$, $\$(\mathcal{E}^{B|A})\circ (\mathcal{U}_g^{R_A}\otimes\mathcal{U}_g^{A})= \$(\mathcal{E}^{B|A})$ which in terms of diagrams is
\begin{align}
\label{eq:right}
\vcenter{\hbox{
\begin{tikzpicture}
	\begin{pgfonlayer}{nodelayer}
        \node [style=right label] (6) at (-1.1, -2.5) {$R_A$};
        \node [style=right label] (7) at (1.35, -2.5) {$A$};
        \node [style=right label] (1) at (1.7, 1.5) {$\$$};
        \node  [style=proj2] (11) at (-0, 0) {$\mathcal{E}^{B|A}$};
        \node [style=right label] (12) at (0.25, 1.5) {$B$};
        \node [style=right label] (13) at (0.25, -1.5) {$A$};
        \node [style=rectangle, minimum width=2.5cm, minimum height=2cm, fill=green, opacity=0.3, rounded corners] (17) at (0, 0) { };
        \node [style=right label] (18) at (-1.1, 3) {$R_B$};
        \node [style=right label] (20) at (1.4, 3) {$B$};
        \node [style=rectangle] (21) at (0, 2.25) { };
        \node [style=rectangle] (22) at (0, -2.25) { };
        \node [style=rectangle] (23) at (-1.25, 1.75) { };
        \node [style=rectangle] (24) at (-1.25, 4) { };
         \node [style=rectangle] (25) at (1.25, 1.75) { };
         \node [style=rectangle] (26) at (1.25, 4) { };
         \node [style=rectangle] (27) at (-1.25, -1.75) { };
         \node [style=rectangle] (28) at (-1.25, -3.5) { };
         \node [style=rectangle] (29) at (1.25, -1.75) { };
         \node [style=rectangle] (30) at (1.25, -3.5) { };
        \node  [style=proj2] (31) at (1.25, -4) {${\mathcal{U}_{g}^A}$};
        \node  [style=proj2] (32) at (-1.25, -4) {${\mathcal{U}_{g}^{R_A}}$};
        \node [style=rectangle] (33) at (1.25, -6) { };
        \node [style=right label] (34) at (-1.1, -5.5) {$R_A$};
         \node [style=rectangle] (35) at (-1.25, -6) { };
        \node [style=right label] (36) at (1.4, -5.5) {$A$};
	\end{pgfonlayer}
	\begin{pgfonlayer}{edgelayer}
        \draw [qWire] (11) to (21);
        \draw [qWire] (11) to (22);
        \draw [qWire] (23) to (24);
        \draw [qWire] (25) to (26);
        \draw [qWire] (32) to (35);
        \draw [qWire] (31) to (33);
        \draw [qWire] (32) to (27);
        \draw [qWire] (31) to (29);
	\end{pgfonlayer}
\end{tikzpicture}
}}
=\vcenter{\hbox{
\begin{tikzpicture}
	\begin{pgfonlayer}{nodelayer}		
        \node [style=right label] (6) at (-2.5, 2.5) {$R_A$};
        \node [style=right label] (7) at (0, 2.5) {$A$};
        \node  [style=proj2] (8) at (-0, 3.5) {${\mathcal{U}_{g'}^A}^\dagger$};
        \node [style=Wsquare] (9) at (-2.5, 3.5) {$g'$};
        \node [style=rectangle, minimum width=1.5cm, align=left] (10) at (-4.5, 3.5) {$\sum_{g'}$};
        \node  [style=proj2] (11) at (-0, 5.5) {$\mathcal{E}^{B|A}$};
        \node [style=right label] (12) at (0, 6.5) {$B$};
        \node [style=right label] (13) at (0, 4.6) {$A$};
        \node  [style=proj2] (14) at (-0, 7.5) {$\mathcal{U}_{g''}^B$};
        \node [style=Wsquareadj] (15) at (-2.5, 7.5) {$g''$};
        \node [style=rectangle, minimum width=1.5cm, align=left] (16) at (-4.5, 7.5) {$\frac{1}{|G|}\sum_{g''}$};
        \node [style=rectangle] (17) at (-2.5, 9.3) { };
        \node [style=right label] (18) at (-2.5, 8.5) {$R_B$};
        \node [style=rectangle] (19) at (0, 9.3) { };
        \node [style=right label] (20) at (0, 8.5) {$B$};
        \node  [style=proj2] (22) at (-0, 1.5) {${\mathcal{U}_{g}^A}$};
        \node  [style=proj2] (23) at (-2.5, 1.5) {${\mathcal{U}_{g}^{R_A}}$};
        \node [style=rectangle] (24) at (-2.5, -0.5) { };
        \node [style=right label] (25) at (-2.5, 0.5) {$R_A$};
         \node [style=rectangle] (26) at (0, -0.5) { };
        \node [style=right label] (27) at (0, 0.5) {$A$};
	\end{pgfonlayer}
	\begin{pgfonlayer}{edgelayer}
        \draw [qWire] (8) to (11);
        \draw [qWire] (14) to (11);
        \draw [qWire] (15) to (17);
        \draw [qWire] (14) to (19);
        \draw [qWire] (22) to (8);
        \draw [qWire] (23) to (9);
        \draw [qWire] (23) to (24);
        \draw [qWire] (22) to (26);
	\end{pgfonlayer}
\end{tikzpicture}
}}
&=\vcenter{\hbox{
\begin{tikzpicture}
	\begin{pgfonlayer}{nodelayer}		
        \node [style=right label] (6) at (-2.5, 2.7) {$R_A$};
        \node [style=right label] (7) at (0, 0.4) {$A$};
        \node  [style=proj2] (8) at (-0, 1.5) {${\mathcal{U}_{g'}^A}^\dagger$};
        \node [style=Wsquare] (9) at (-2.5, 3.5) {$e$};
        \node [style=rectangle, minimum width=1.5cm, align=left] (10) at (-4.5, 1.5) {$\sum_{g'}$};
        \node  [style=proj2] (11) at (-0, 5.5) {$\mathcal{E}^{B|A}$};
        \node [style=right label] (12) at (0, 6.5) {$B$};
        \node [style=right label] (13) at (0, 3.5) {$A$};
        \node  [style=proj2] (14) at (-0, 7.5) {$\mathcal{U}_{g''}^B$};
        \node [style=Wsquareadj] (15) at (-2.5, 7.5) {$g''$};
        \node [style=rectangle, minimum width=1.5cm, align=left] (16) at (-4.5, 7.5) {$\frac{1}{|G|}\sum_{g''}$};
        \node [style=rectangle] (17) at (-2.5, 9.3) { };
        \node [style=right label] (18) at (-2.5, 8.5) {$R_B$};
        \node [style=rectangle] (19) at (0, 9.3) { };
        \node [style=right label] (20) at (0, 8.5) {$B$};
        \node  [style=proj2] (22) at (-0, -0.7) {${\mathcal{U}_{g}^A}$};
        \node  [style=proj2] (23) at (-2.5, -0.7) {${\mathcal{U}_{g}^{R_A}}$};
        \node [style=rectangle] (24) at (-2.5, -3) { };
        \node [style=right label] (25) at (-2.5, -2) {$R_A$};
         \node [style=rectangle] (26) at (0, -3) { };
        \node [style=right label] (27) at (0, -2) {$A$};
        \node  [style=proj2] (28) at (-2.5, 1.5) {${\mathcal{U}_{g'}^{R_A}}^\dagger$};
        \node [style=right label] (29) at (-2.5, 0.4) {$R_A$};
	\end{pgfonlayer}
	\begin{pgfonlayer}{edgelayer}
        \draw [qWire] (8) to (11);
        \draw [qWire] (14) to (11);
        \draw [qWire] (15) to (17);
        \draw [qWire] (14) to (19);
        \draw [qWire] (22) to (8);
        \draw [qWire] (28) to (9);
        \draw [qWire] (23) to (28);
        \draw [qWire] (23) to (24);
        \draw [qWire] (22) to (26);
	\end{pgfonlayer}
\end{tikzpicture}
}} \\ \nonumber & = \vcenter{\hbox{
\begin{tikzpicture}
	\begin{pgfonlayer}{nodelayer}		
        \node [style=right label] (6) at (-2.5, 2.7) {$R_A$};
        \node [style=right label] (7) at (0.3, 0.4) {$A$};
        \node  [style=proj2] (8) at (0.3, 1.5) {${\mathcal{U}_{g'g^{-1}}^A}^\dagger$};
        \node [style=Wsquare] (9) at (-2.5, 3.5) {$e$};
        \node [style=rectangle, minimum width=1.5cm, align=left] (10) at (-4.5, 1.5) {$\sum_{g'}$};
        \node  [style=proj2] (11) at (0.3, 5.5) {$\mathcal{E}^{B|A}$};
        \node [style=right label] (12) at (0.3, 6.5) {$B$};
        \node [style=right label] (13) at (0.3, 3.5) {$A$};
        \node  [style=proj2] (14) at (0.3, 7.5) {$\mathcal{U}_{g''}^B$};
        \node [style=Wsquareadj] (15) at (-2.5, 7.5) {$g''$};
        \node [style=rectangle, minimum width=1.5cm, align=left] (16) at (-4.5, 7.5) {$\frac{1}{|G|}\sum_{g''}$};
        \node [style=rectangle] (17) at (-2.5, 9.3) { };
        \node [style=right label] (18) at (-2.5, 8.5) {$R_B$};
        \node [style=rectangle] (19) at (0.3, 9.3) { };
        \node [style=right label] (20) at (0.3, 8.5) {$B$};
        \node  [style=rectangle] (22) at (0.3, -0.7) { };
        \node  [style=rectangle] (23) at (-2.5, -0.7) {};
        \node  [style=proj2] (28) at (-2.5, 1.5) {${\mathcal{U}_{g'g^{-1}}^{R_A}}^\dagger$};
        \node [style=right label] (29) at (-2.5, 0.4) {$R_A$};
	\end{pgfonlayer}
	\begin{pgfonlayer}{edgelayer}
        \draw [qWire] (8) to (11);
        \draw [qWire] (14) to (11);
        \draw [qWire] (15) to (17);
        \draw [qWire] (14) to (19);
        \draw [qWire] (22) to (8);
        \draw [qWire] (28) to (9);
        \draw [qWire] (23) to (28);
	\end{pgfonlayer}
\end{tikzpicture}
}} = \vcenter{\hbox{
\begin{tikzpicture}
	\begin{pgfonlayer}{nodelayer}		
        \node [style=right label] (6) at (-0.8, -3) {$R_A$};
        \node [style=right label] (7) at (1.1, -3) {$A$};
        \node [style=right label] (1) at (1.4, 1.5) {$\$$};
        \node  [style=proj2] (11) at (-0, 0) {$\mathcal{E}^{B|A}$};
        \node [style=right label] (12) at (0.25, 1.5) {$B$};
        \node [style=right label] (13) at (0.25, -1.5) {$A$};
        \node [style=rectangle, minimum width=2cm, minimum height=2cm, fill=green, opacity=0.3, rounded corners] (17) at (0, 0) { };
        \node [style=right label] (18) at (-0.80, 3) {$R_B$};
        \node [style=right label] (20) at (1.1, 3) {$B$};
        \node [style=rectangle] (21) at (0, 2.25) { };
        \node [style=rectangle] (22) at (0, -2.25) { };
        \node [style=rectangle] (23) at (-0.95, 1.75) { };
         \node [style=rectangle] (24) at (-0.95, 4) { };
         \node [style=rectangle] (25) at (0.95, 1.75) { };
         \node [style=rectangle] (26) at (0.95, 4) { };
         \node [style=rectangle] (27) at (-0.95, -1.75) { };
         \node [style=rectangle] (28) at (-0.95, -4) { };
         \node [style=rectangle] (29) at (0.95, -1.75) { };
         \node [style=rectangle] (30) at (0.95, -4) { };
	\end{pgfonlayer}
	\begin{pgfonlayer}{edgelayer}
        \draw [qWire] (11) to (21);
        \draw [qWire] (11) to (22);
        \draw [qWire] (23) to (24);
        \draw [qWire] (25) to (26);
        \draw [qWire] (27) to (28);
        \draw [qWire] (29) to (30);
	\end{pgfonlayer}
\end{tikzpicture}
}}.
\end{align}
Similarly, 
\begin{align}
\label{eq:left}
    &(\mathcal{U}^{R_B}_g \otimes \mathcal{U}^{B}_g)  \circ \$({\cal E}^{B|A}) =
 \$({\cal E}^{B|A}).
\end{align}
So 
\begin{align}
\label{eq:covdol}
     &(\mathcal{U}^{R_B}_g \otimes \mathcal{U}^{B}_g)  \circ \$({\cal E}^{B|A})  = \$({\cal E}^{B|A}) \circ (\mathcal{U}^{R_A}_g \otimes \mathcal{U}^{A}_g) .
\end{align}
That is, $\$({\cal E}^{B|A})$ is $(G,{\cal U})$-covariant.

If we specialize $\mathcal{E}^{B|A}$ to the case of the superoperator that injects the state $\rho_{B}$, then because this superoperator has a trivial input space (which carries a trivial representation of the symmetry and thus does not get associated with a reference frame system under $\$$), the $\$$ map yields: 
\begin{align}
\label{eq:dolrho}
\$(\rho^B) 
= &\frac{1}{|G|}\sum_{g}\ketbra{g}^{R_B}  \otimes \left( \mathcal{U}^{B}_{g} \circ \rho^{B} \right).
\end{align}
\cref{eq:dolrho} is equivalent to \cref{eq:dolrhoA} in the main text.

If we specialize $\mathcal{E}^{B|A}$ to the case of the superoperator describing contraction with the effect $E$, i.e., $\tr(E^A\cdot)$, then because this superoperator has a trivial output space  (which carries a trivial representation of the symmetry and thus does not get associated with a reference frame system under $\$$), the $\$$ map yields:
\begin{align}
\label{eq:dolE}
\$\left(\tr_A (E^A \; \cdot)\right) 
=& \tr_{R_A A} \left[\sum_{g}\ketbra{g}^{R_A}  \otimes ( E^{A} \circ {\mathcal{U}^{A}_{g}}^{\dag} ) \;\cdot  \right].
\end{align}
\cref{eq:dolE} is equivalent in content to \cref{eq:dolEA} in the main text

Note, in particular, that the superoperator associated to the unit effect on $A$, $\tr_A[\mathbb{I}^A \cdot]$, which is equivalent to the trace on $A$, is mapped to the superoperator associated to the unit effect on $R_A A$, which is equivalent  to the trace over $R_A A$, 
\begin{equation}
    \$(\tr[\mathbb{I}^A \cdot])  = \tr \left[ 
 \sum_{{ g}}   \ketbra{g}^{R_A} \otimes  (\mathbb{I}^A \circ {\mathcal{U}^{ A}_{ g}}^{\dag})  \cdot \right]
= \tr \left(\mathbb{I}^{R_AA}\cdot \right),
\end{equation} 
where we used the fact that $\mathbb{I}^A \circ {\mathcal{U}^{ A}_{ g}}^{\dag}=\mathbb{I}^A$ since $\mathbb{I}^A=  \sum_{{ g'}}   |{ g'} \rangle^{A} \langle { g'}|$.  It follows that a POVM on $A$ is mapped to a POVM on $R_A A$.

The statistics of any prepare-transform-measure scenario are reproduced by the $\$$ map, that is:
\begin{align}
 \label{eq:repro}
 \$(\tr_B (E^B\;\cdot ))\circ \$ (\mathcal{E}^{B|A})\circ \$ (\rho^A) = \tr_B \left(E^B \mathcal{E}^{B|A}(\rho^A)\right).
\end{align}

The proof is that:
\begin{align}
 \begin{tikzpicture}
	\begin{pgfonlayer}{nodelayer}		
        \node [style=right label] (6) at (-1.1, -2.7) {$R_A$};
        \node [style=right label] (7) at (1.4, -2.7) {$A$};
        \node [style=right label] (1) at (1.4, 1.5) {$\$$};
        \node [style=right label] (50) at (1.4, 7) {$\$$};
        \node [style=right label] (51) at (1.4, -4) {$\$$};
        \node  [style=proj2] (11) at (-0, 0) {$\mathcal{E}^{B|A}$};
        \node [style=right label] (12) at (0.15, 1.5) {$B$};
        \node [style=right label] (13) at (0.15, -1.5) {$A$};
        \node [style=rectangle, minimum width=2cm, minimum height=2cm, fill=green, opacity=0.3, rounded corners] (17) at (0, 0) { };       
        \node [style=right label] (18) at (-1.1, 3) {$R_B$};
        \node [style=rectangle] (19) at (0, 9.3) { };
        \node [style=right label] (20) at (1.4, 3) {$B$};
        \node [style=rectangle] (21) at (0, 2.25) { };
        \node [style=rectangle] (22) at (0, -2.25) { };
        \node [style=rectangle] (23) at (-1.25, 1.75) { };
        \node [style=rectangle] (24) at (-1.25, 3.75) { };
         \node [style=rectangle] (25) at (1.25, 1.75) { };
         \node [style=rectangle] (26) at (1.25, 3.75) { };
         \node [style=rectangle] (27) at (-1.25, -1.75) { };
         \node [style=rectangle] (28) at (-1.25, -3.75) { };
         \node [style=rectangle] (29) at (1.25, -1.75) { };
         \node [style=rectangle] (30) at (1.25, -3.75) { };
         \node [style=rectangle] (46) at (0, -3.25) { };
         \node [style=Wsquareadj] (32) at (0, -6) {$\rho^A$};
		\node [style=right label] (33) at (0.15, -4.5) {$A$};
        \node [style=rectangle, minimum width=2cm, minimum height=2cm, fill=green, opacity=0.3, rounded corners] (45) at (0, -5.5) { }; 
        \node [style=Wsquare] (43) at (0, 5.5) {$E^B$};
        \node [style=right label] (44) at (0.15, 4.2) {$B$};
        \node [style=rectangle, minimum width=2cm, minimum height=2cm, fill=green, opacity=0.3, rounded corners] (47) at (0, 5.5) { };
        \node [style=rectangle] (48) at (0, 3.25) { };
	\end{pgfonlayer}
	\begin{pgfonlayer}{edgelayer}
        \draw [qWire] (11) to (21);
        \draw [qWire] (11) to (22);
        \draw [qWire] (23) to (24);
        \draw [qWire] (25) to (26);
        \draw [qWire] (27) to (28);
        \draw [qWire] (29) to (30);
        \draw [qWire] (32) to (46);
        \draw [qWire] (48) to (43);
	\end{pgfonlayer}
\end{tikzpicture} =\vcenter{\hbox{
\begin{tikzpicture}
	\begin{pgfonlayer}{nodelayer}
		\node  [style=proj2] (1) at (-0, 1.5) {$\mathcal{U}_{g''}^A$};
		\node [style=Wsquareadj] (2) at (-0, -0.5) {$\rho^A$};
		\node [style=right label] (3) at (-0, 0.5) {$A$};
        \node [style=Wsquareadj] (4) at (-2.5, 1.5) {$g''$};
        \node [style=rectangle, minimum width=1.5cm, align=left] (5) at (-4.5, 1.5) {$\frac{1}{|G|}\sum_{g''}$};
        \node [style=right label] (6) at (-2.5, 2.5) {$R_A$};
        \node [style=right label] (7) at (0, 2.5) {$A$};
        \node  [style=proj2] (8) at (-0, 3.5) {${\mathcal{U}_{g}^A}^\dagger$};
        \node [style=Wsquare] (9) at (-2.5, 3.5) {$g$};
        \node [style=rectangle, minimum width=1.5cm, align=left] (10) at (-4.5, 3.5) {$\sum_{g}$};
        \node  [style=proj2] (11) at (-0, 5.5) {$\mathcal{E}^{B|A}$};
        \node [style=right label] (12) at (0, 6.5) {$B$};
        \node [style=right label] (13) at (0, 4.6) {$A$};
        \node  [style=proj2] (14) at (-0, 7.5) {$\mathcal{U}_{g'}^B$};
        \node [style=Wsquareadj] (15) at (-2.5, 7.5) {$g'$};
        \node [style=rectangle, minimum width=1.5cm, align=left] (16) at (-4.5, 7.5) {$\frac{1}{|G|}\sum_{g'}$};
        \node [style=rectangle] (17) at (-2.5, 9.3) { };
        \node [style=right label] (18) at (-2.5, 8.5) {$R_B$};
        \node [style=proj2] (19) at (0, 9.6) {${\mathcal{U}_{g'''}^B}^\dagger$};
        \node [style=right label] (20) at (0, 8.5) {$B$};
        \node [style=Wsquare] (21) at (-2.5, 9.6) {\small{$g'''$}};
        \node [style=rectangle, minimum width=1.5cm, align=left] (22) at (-4.5, 9.6) {$\sum_{g'''}$};
        \node [style=Wsquare] (23) at (0, 11.7) {$E^B$};
        \node [style=right label] (20) at (0, 10.7) {$B$};
	\end{pgfonlayer}
	\begin{pgfonlayer}{edgelayer}
		\draw [qWire] (1) to (2.center);
        \draw [qWire] (1) to (8);
        \draw [qWire] (4) to (9);
        \draw [qWire] (8) to (11);
        \draw [qWire] (14) to (11);
        \draw [qWire] (15) to (17);
        \draw [qWire] (14) to (19);
        \draw [qWire] (23) to (19);
	\end{pgfonlayer}
\end{tikzpicture}
}}= \vcenter{\hbox{
\begin{tikzpicture}
	\begin{pgfonlayer}{nodelayer}
		\node [style=Wsquareadj] (2) at (-0, -0.5) {$\rho^A$};
		\node [style=right label] (3) at (-0, 2.5) {$A$};
        \node  [style=proj2] (11) at (-0, 5.5) {$\mathcal{E}^{B|A}$};
        \node [style=right label] (12) at (0, 8.5) {$B$};
        \node [style=Wsquare] (23) at (0, 11.7) {$E^B$};
	\end{pgfonlayer}
	\begin{pgfonlayer}{edgelayer}
		\draw [qWire] (2) to (11);
        \draw [qWire] (11) to (23);
	\end{pgfonlayer}
\end{tikzpicture}
}}.
\end{align}

Now we define the map for a quantum operation $\mathcal{E}^{{B}  | {A} }$ on composite systems where ${A}=A_1A_2\ldots A_n$ and ${B}=B_1B_2\ldots B_m$.  Let the $i$-tuple of group elements be denoted ${g} = (g_1, \dots, g_i)$. For each multipartite quantum operation $\mathcal{E}^{{B}  | {A} }$, its image under $\$$ is
\begin{equation}
\$(\mathcal{E}^{B|A}) \coloneqq 
\left( \frac{1}{|G|^m} 
\bigotimes_{i} \sum_{{g}_i} \ketbra{g_i}^{R_{B_i}}\otimes \mathcal{U}^{B_i }_{g_i}
\right) 
 \circ \mathcal{E}^{B|A} \circ
 \left(
  \bigotimes_{j} \sum_{{g}_j} \tr(\ketbra{g_j}^{R_{A_j}}\cdot) \otimes  {\mathcal{U}^{{A}_j}_{g_j}}^{\dag}
  \right).  
\end{equation}
When an input or an output system $S_i$ carries a trivial representation (e.g., when $S_i$ is classical), we can drop the reference frame system $R_{S_i}$. Note that although the following proofs are all presented for cases where all systems carry nontrivial representations, they can be straightforwardly generalized to cases where some systems carry the trivial representation, as the latter is simply a special case of the former.

Similar to the case for unipartite quantum operations, $\$( \mathcal{E}^{ B|A})$ is a CPTP map whenever $\mathcal{E}^{ B|A}$ is CPTP and is trace-nonincreasing whenever $\mathcal{E}^{ B|A}$ is trace-nonincreasing.
Following the analogous steps of \cref{eq:right,eq:left}, we can also prove that $\$(\mathcal{E}^{B|A})$ is a valid process in the ($G,\cal U$)-twirled quantum world. Also, analogous to \cref{eq:dolrho,eq:dolE}, we can also specialize this to the case of a state on the multipartite system ${B}$ or an effect on the multipartite system ${A}$.

The definition of $\$(\mathcal{E}^{B|A})$ when $A$ and $B$ are multipartite systems ensures that the $\$$ map commutes with parallel composition, which can be expressed diagrammatically as
\begin{equation}    \begin{tikzpicture}
	\begin{pgfonlayer}{nodelayer}		
        \node [style=right label] (6) at (-0.8, -3) {$R_{A_1}$};
        \node [style=right label] (7) at (1.1, -3) {$A_1$};     
        \node  [style=proj2] (11) at (-0, 0) {$\mathcal{E}^{B_1|A_1}$};
        \node [style=right label] (12) at (0.25, 1.5) {$B_1$};
        \node [style=right label] (13) at (0.25, -1.5) {$A_1$};
        \node [style=rectangle, minimum width=5cm, minimum height=2cm, fill=green, opacity=0.3, rounded corners] (17) at (2.75, 0) { };
        \node [style=right label] (18) at (-0.80, 3) {$R_{B_1}$};
        \node [style=right label] (20) at (1.1, 3) {$B_1$};
        \node [style=rectangle] (21) at (0, 2.25) { };
        \node [style=rectangle] (22) at (0, -2.25) { };
        \node [style=rectangle] (23) at (-0.95, 1.75) { };
         \node [style=rectangle] (24) at (-0.95, 4) { };
         \node [style=rectangle] (25) at (0.95, 1.75) { };
         \node [style=rectangle] (26) at (0.95, 4) { };
         \node [style=rectangle] (27) at (-0.95, -1.75) { };
         \node [style=rectangle] (28) at (-0.95, -4) { };
         \node [style=rectangle] (29) at (0.95, -1.75) { };
         \node [style=rectangle] (30) at (0.95, -4) { };
         
        \begin{scope}[shift={(3,0)}] 
        \node [style=right label] (31) at (1.2, -3) {$R_{A_2}$};
        \node [style=right label] (32) at (3.1, -3) {$A_2$};
        \node [style=right label] (33) at (4.1, 1.5) {$\$$};       
        \node  [style=proj2] (34) at (2, 0) {$\mathcal{E}^{B_2|A_2}$};
        \node [style=right label] (35) at (2.25, 1.5) {$B_2$};
        \node [style=right label] (36) at (2.25, -1.5) {$A_2$};
        \node [style=right label] (38) at (1.2, 3) {$R_{B_2}$};
        \node [style=right label] (40) at (3.1, 3) {$B_2$};
        \node [style=rectangle] (41) at (2, 2.25) { };
        \node [style=rectangle] (42) at (2, -2.25) { };
        \node [style=rectangle] (43) at (1.05, 1.75) { };
         \node [style=rectangle] (44) at (1.05, 4) { };
         \node [style=rectangle] (45) at (2.95, 1.75) { };
         \node [style=rectangle] (46) at (2.95, 4) { };
         \node [style=rectangle] (47) at (1.05, -1.75) { };
         \node [style=rectangle] (48) at (1.05, -4) { };
         \node [style=rectangle] (49) at (2.95, -1.75) { };
         \node [style=rectangle] (50) at (2.95, -4) { };
         \end{scope}
	\end{pgfonlayer}
	\begin{pgfonlayer}{edgelayer}
        \draw [qWire] (11) to (21);
        \draw [qWire] (11) to (22);
        \draw [qWire] (23) to (24);
        \draw [qWire] (25) to (26);
        \draw [qWire] (27) to (28);
        \draw [qWire] (29) to (30);
        
        \draw [qWire] (34) to (41);
        \draw [qWire] (34) to (42);
        \draw [qWire] (43) to (44);
        \draw [qWire] (45) to (46);
        \draw [qWire] (47) to (48);
        \draw [qWire] (49) to (50);
	\end{pgfonlayer}
\end{tikzpicture}=
    \begin{tikzpicture}
	\begin{pgfonlayer}{nodelayer}		
        \node [style=right label] (6) at (-0.8, -3) {$R_{A_1}$};
        \node [style=right label] (7) at (1.1, -3) {$A_1$};
        \node [style=right label] (1) at (1.4, 1.5) {$\$$};       
        \node  [style=proj2] (11) at (-0, 0) {$\mathcal{E}^{B_1|A_1}$};
        \node [style=right label] (12) at (0.25, 1.5) {$B_1$};
        \node [style=right label] (13) at (0.25, -1.5) {$A_1$};
        \node [style=rectangle, minimum width=2cm, minimum height=2cm, fill=green, opacity=0.3, rounded corners] (17) at (0, 0) { };
        \node [style=right label] (18) at (-0.80, 3) {$R_{B_1}$};
        \node [style=right label] (20) at (1.1, 3) {$B_1$};
        \node [style=rectangle] (21) at (0, 2.25) { };
        \node [style=rectangle] (22) at (0, -2.25) { };
        \node [style=rectangle] (23) at (-0.95, 1.75) { };
         \node [style=rectangle] (24) at (-0.95, 4) { };
         \node [style=rectangle] (25) at (0.95, 1.75) { };
         \node [style=rectangle] (26) at (0.95, 4) { };
         \node [style=rectangle] (27) at (-0.95, -1.75) { };
         \node [style=rectangle] (28) at (-0.95, -4) { };
         \node [style=rectangle] (29) at (0.95, -1.75) { };
         \node [style=rectangle] (30) at (0.95, -4) { };
         
        \begin{scope}[shift={(3,0)}] 
        \node [style=right label] (31) at (1.2, -3) {$R_{A_2}$};
        \node [style=right label] (32) at (3.1, -3) {$A_2$};
        \node [style=right label] (33) at (3.4, 1.5) {$\$$};       
        \node  [style=proj2] (34) at (2, 0) {$\mathcal{E}^{B_2|A_2}$};
        \node [style=right label] (35) at (2.25, 1.5) {$B_2$};
        \node [style=right label] (36) at (2.25, -1.5) {$A_2$};
        \node [style=rectangle, minimum width=2cm, minimum height=2cm, fill=green, opacity=0.3, rounded corners] (37) at (2, 0) { };
        \node [style=right label] (38) at (1.2, 3) {$R_{B_2}$};
        \node [style=right label] (40) at (3.1, 3) {$B_2$};
        \node [style=rectangle] (41) at (2, 2.25) { };
        \node [style=rectangle] (42) at (2, -2.25) { };
        \node [style=rectangle] (43) at (1.05, 1.75) { };
         \node [style=rectangle] (44) at (1.05, 4) { };
         \node [style=rectangle] (45) at (2.95, 1.75) { };
         \node [style=rectangle] (46) at (2.95, 4) { };
         \node [style=rectangle] (47) at (1.05, -1.75) { };
         \node [style=rectangle] (48) at (1.05, -4) { };
         \node [style=rectangle] (49) at (2.95, -1.75) { };
         \node [style=rectangle] (50) at (2.95, -4) { };
         \end{scope}
	\end{pgfonlayer}
	\begin{pgfonlayer}{edgelayer}
        \draw [qWire] (11) to (21);
        \draw [qWire] (11) to (22);
        \draw [qWire] (23) to (24);
        \draw [qWire] (25) to (26);
        \draw [qWire] (27) to (28);
        \draw [qWire] (29) to (30);
        
        \draw [qWire] (34) to (41);
        \draw [qWire] (34) to (42);
        \draw [qWire] (43) to (44);
        \draw [qWire] (45) to (46);
        \draw [qWire] (47) to (48);
        \draw [qWire] (49) to (50);
	\end{pgfonlayer}
\end{tikzpicture}
\label{eq:para}
\end{equation}
The proof is that both the left-hand-side and the right-hand-side are equal to 
\begin{equation}
\vcenter{\hbox{
\begin{tikzpicture}
	\begin{pgfonlayer}{nodelayer}		
        \node [style=right label] (6) at (-2.5, 2.5) {$R_{A_1}$};
        \node [style=right label] (7) at (0, 2.3) {$A_1$};
        \node  [style=proj2] (8) at (-0, 3.5) {$\left(\mathcal{U}_{g_1}^{A_1}\right)^\dagger$};
        \node [style=Wsquare] (9) at (-2.5, 3.5) {$g_1$};
        \node [style=rectangle, minimum width=1.5cm, align=left] (10) at (-4.5, 3.5) {$\sum_{g_1}$};
        \node  [style=proj2] (11) at (-0, 5.5) {$\mathcal{E}^{B_1|A_1}$};
        \node [style=right label] (12) at (0, 6.5) {$B_1$};
        \node [style=right label] (13) at (0, 4.6) {$A_1$};
        \node  [style=proj2] (14) at (-0, 7.5) {$\mathcal{U}_{g_1'}^{B_1}$};
        \node [style=Wsquareadj] (15) at (-2.5, 7.5) {$g_1'$};
        \node [style=rectangle, minimum width=1.5cm, align=left] (16) at (-4.5, 7.5) {$\frac{1}{|G|}\sum_{g_1'}$};
        \node [style=rectangle] (17) at (-2.5, 9.3) { };
        \node [style=right label] (18) at (-2.5, 8.7) {$R_{B_1}$};
        \node [style=rectangle] (19) at (0, 9.3) { };
        \node [style=right label] (20) at (0, 8.7) {$B_1$};
        \node [style=rectangle] (22) at (0, 1.7) { };
        \node [style=rectangle] (23) at (-2.5, 1.7) { };
        \node [style=right label] (26) at (5, 2.5) {$R_{A_2}$};
        \node [style=right label] (27) at (8, 2.3) {$A_2$};
        \node  [style=proj2] (28) at (8, 3.5) {$\left(\mathcal{U}_{g_2}^{A_2}\right)^\dagger$};
        \node [style=Wsquare] (29) at (5, 3.5) {$g_2$};
        \node [style=rectangle, minimum width=1.5cm, align=left] (30) at (3, 3.5) {$\sum_{g_2}$};
        \node  [style=proj2] (31) at (8, 5.5) {$\mathcal{E}^{B_2|A_2}$};
        \node [style=right label] (32) at (8, 6.5) {$B_2$};
        \node [style=right label] (33) at (8, 4.6) {$A_2$};
        \node  [style=proj2] (34) at (8, 7.5) {$\mathcal{U}_{g_2'}^{B_2}$};
        \node [style=Wsquareadj] (35) at (5, 7.5) {$g_2'$};
        \node [style=rectangle, minimum width=1.5cm, align=left] (36) at (3, 7.5) {$\frac{1}{|G|}\sum_{g_2'}$};
        \node [style=rectangle] (37) at (5, 9.3) { };
        \node [style=right label] (38) at (5, 8.7) {$R_{B_2}$};
        \node [style=rectangle] (39) at (8, 9.3) { };
        \node [style=right label] (40) at (8, 8.7) {$B_2$};
        \node [style=rectangle] (42) at (8, 1.7) { };
        \node [style=rectangle] (43) at (5, 1.7) { };
	\end{pgfonlayer}
	\begin{pgfonlayer}{edgelayer}
          \draw [qWire] (8) to (11);
        \draw [qWire] (14) to (11);
        \draw [qWire] (15) to (17);
        \draw [qWire] (14) to (19);
        \draw [qWire] (22) to (8);
        \draw [qWire] (23) to (9);
        \draw [qWire] (28) to (31);
        \draw [qWire] (34) to (31);
        \draw [qWire] (35) to (37);
        \draw [qWire] (34) to (39);
        \draw [qWire] (42) to (28);
        \draw [qWire] (43) to (29);
	\end{pgfonlayer}
\end{tikzpicture}
}}
\end{equation}

The incoherent simulation map $\$$ also commutes with sequential composition, which can be expressed diagrammatically as
\begin{equation}
   \vcenter{\hbox{\begin{tikzpicture}
	\begin{pgfonlayer}{nodelayer}		
        \node [style=right label] (6) at (-0.8, -3) {$R_A$};
        \node [style=right label] (7) at (1.1, -3) {$A$};
        \node [style=right label] (1) at (1.4, 4.5) {$\$$};
        \node  [style=proj2] (11) at (0, -0.5) {$\mathcal{E}^{B|A}$};
        \node [style=right label] (12) at (0.15, 1.5) {$B$};
        \node [style=right label] (13) at (0.15, -1.5) {$A$};

        \begin{scope}[shift={(0,3.0)}]
            \node  [style=proj2] (31) at (0, 0.5) {$\mathcal{E}^{C|B}$};
            \node [style=right label] (32) at (0.15, 1.5) {$C$};
            \node [style=rectangle] (41) at (0, 2.25) { };
        \end{scope}

        \node [style=rectangle, minimum width=2cm, minimum height=3.52cm, fill=green, opacity=0.3, rounded corners] (17) at (0, 1.48) { };

        \begin{scope}[shift={(0,3.0)}]
            \node [style=right label] (18) at (-0.80, 3) {$R_C$};
            \node [style=right label] (20) at (1.1, 3) {$C$};
            \node [style=rectangle] (23) at (-0.95, 1.75) { };
            \node [style=rectangle] (24) at (-0.95, 4.25) { };
            \node [style=rectangle] (25) at (0.95, 1.75) { };
            \node [style=rectangle] (26) at (0.95, 4.25) { };
        \end{scope}

        \node [style=rectangle] (22) at (0, -2.25) { };
        \node [style=rectangle] (27) at (-0.95, -1.75) { };
        \node [style=rectangle] (28) at (-0.95, -4) { };
        \node [style=rectangle] (29) at (0.95, -1.75) { };
        \node [style=rectangle] (30) at (0.95, -4) { };
	\end{pgfonlayer}

	\begin{pgfonlayer}{edgelayer}
        \draw [qWire] (11) to (31);
        \draw [qWire] (11) to (22);
        \draw [qWire] (23) to (24);
        \draw [qWire] (25) to (26);
        \draw [qWire] (27) to (28);
        \draw [qWire] (29) to (30);
        \draw [qWire] (31) to (41);
	\end{pgfonlayer}
\end{tikzpicture} }}= 
\vcenter{\hbox{\begin{tikzpicture}
	\begin{pgfonlayer}{nodelayer}		
        \node [style=right label] (6) at (-0.8, -3) {$R_A$};
        \node [style=right label] (7) at (1.1, -3) {$A$};
        \node [style=right label] (1) at (1.4, 4.5) {$\$$};
        \node  [style=proj2] (11) at (0, -0.5) {$\mathcal{E}^{B|A}$};
        \node [style=right label] (12) at (0.15, 0.5) {$B$};
        \node [style=right label] (13) at (0.15, -1.5) {$A$};

        \node [style=rectangle] (50) at (0, 1.0) { };
        \node [style=rectangle] (51) at (0, 2) { };
        \node [style=rectangle] (52) at (-0.8, 2) { };
        \node [style=rectangle] (53) at (-0.8, 1.0) { };
        \node [style=rectangle] (54) at (1.1, 2) { };
        \node [style=rectangle] (55) at (1.1, 1.0) { };

        \node [style=right label] (56) at (1.1, 1.5) {$B$};
        \node [style=right label] (57) at (-0.8, 1.5) {$R_B$};
        \node [style=right label] (1) at (1.4, 0.5) {$\$$};
        \begin{scope}[shift={(0,3.0)}]
            \node  [style=proj2] (31) at (0, 0.5) {$\mathcal{E}^{C|B}$};
            \node [style=right label] (32) at (0.15, 1.5) {$C$};
            \node [style=right label] (32) at (0.15, -0.5) {$B$};
            \node [style=rectangle] (41) at (0, 2.25) { };
            \node [style=right label] (18) at (-0.80, 3) {$R_C$};
            \node [style=right label] (20) at (1.1, 3) {$C$};
            \node [style=rectangle] (23) at (-0.95, 1.75) { };
            \node [style=rectangle] (24) at (-0.95, 4.25) { };
            \node [style=rectangle] (25) at (0.95, 1.75) { };
            \node [style=rectangle] (26) at (0.95, 4.25) { };
        \end{scope}

        \node [style=rectangle, minimum width=2cm, minimum height=1.5cm, fill=green, opacity=0.3, rounded corners] (topbox) at (0, 3.5) { };
        \node [style=rectangle, minimum width=2cm, minimum height=1.5cm, fill=green, opacity=0.3, rounded corners] (botbox) at (0, -0.5) { };

        \node [style=rectangle] (22) at (0, -2.25) { };
        \node [style=rectangle] (27) at (-0.95, -1.75) { };
        \node [style=rectangle] (28) at (-0.95, -4) { };
        \node [style=rectangle] (29) at (0.95, -1.75) { };
        \node [style=rectangle] (30) at (0.95, -4) { };
	\end{pgfonlayer}

	\begin{pgfonlayer}{edgelayer}
        \draw [qWire] (11) to (50.center);
        \draw [qWire] (11) to (22);
        \draw [qWire] (23) to (24);
        \draw [qWire] (25) to (26);
        \draw [qWire] (27) to (28);
        \draw [qWire] (29) to (30);
        \draw [qWire] (31) to (41);
        \draw [qWire] (31) to (51.center);
        \draw [qWire] (52.center) to (53.center);
        \draw [qWire] (54.center) to (55.center);
	\end{pgfonlayer}
\end{tikzpicture}}}
\label{eq:seq}
\end{equation}
since 
    \begin{equation}
\vcenter{\hbox{\begin{tikzpicture}
	\begin{pgfonlayer}{nodelayer}		
        \node [style=right label] (6) at (-0.8, -3) {$R_A$};
        \node [style=right label] (7) at (1.1, -3) {$A$};
        \node [style=right label] (1) at (1.4, 4.5) {$\$$};
        \node  [style=proj2] (11) at (0, -0.5) {$\mathcal{E}^{B|A}$};
        \node [style=right label] (12) at (0.15, 0.5) {$B$};
        \node [style=right label] (13) at (0.15, -1.5) {$A$};

        \node [style=rectangle] (50) at (0, 1.0) { };
        \node [style=rectangle] (51) at (0, 2) { };
        \node [style=rectangle] (52) at (-0.8, 2) { };
        \node [style=rectangle] (53) at (-0.8, 1.0) { };
        \node [style=rectangle] (54) at (1.1, 2) { };
        \node [style=rectangle] (55) at (1.1, 1.0) { };

        \node [style=right label] (56) at (1.1, 1.5) {$B$};
        \node [style=right label] (57) at (-0.8, 1.5) {$R_B$};
        \node [style=right label] (1) at (1.4, 0.5) {$\$$};
        \begin{scope}[shift={(0,3.0)}]
            \node  [style=proj2] (31) at (0, 0.5) {$\mathcal{E}^{C|B}$};
            \node [style=right label] (32) at (0.15, 1.5) {$C$};
            \node [style=right label] (32) at (0.15, -0.5) {$B$};
            \node [style=rectangle] (41) at (0, 2.25) { };
            \node [style=right label] (18) at (-0.80, 3) {$R_C$};
            \node [style=right label] (20) at (1.1, 3) {$C$};
            \node [style=rectangle] (23) at (-0.95, 1.75) { };
            \node [style=rectangle] (24) at (-0.95, 4.25) { };
            \node [style=rectangle] (25) at (0.95, 1.75) { };
            \node [style=rectangle] (26) at (0.95, 4.25) { };
        \end{scope}

        \node [style=rectangle, minimum width=2cm, minimum height=1.5cm, fill=green, opacity=0.3, rounded corners] (topbox) at (0, 3.5) { };
        \node [style=rectangle, minimum width=2cm, minimum height=1.5cm, fill=green, opacity=0.3, rounded corners] (botbox) at (0, -0.5) { };

        \node [style=rectangle] (22) at (0, -2.25) { };
        \node [style=rectangle] (27) at (-0.95, -1.75) { };
        \node [style=rectangle] (28) at (-0.95, -4) { };
        \node [style=rectangle] (29) at (0.95, -1.75) { };
        \node [style=rectangle] (30) at (0.95, -4) { };
	\end{pgfonlayer}

	\begin{pgfonlayer}{edgelayer}
        \draw [qWire] (11) to (50.center);
        \draw [qWire] (11) to (22);
        \draw [qWire] (23) to (24);
        \draw [qWire] (25) to (26);
        \draw [qWire] (27) to (28);
        \draw [qWire] (29) to (30);
        \draw [qWire] (31) to (41);
        \draw [qWire] (31) to (51.center);
        \draw [qWire] (52.center) to (53.center);
        \draw [qWire] (54.center) to (55.center);
	\end{pgfonlayer}
\end{tikzpicture}}} = \vcenter{\hbox{
\begin{tikzpicture}
	\begin{pgfonlayer}{nodelayer}
        \node [style=rectangle] (4) at (-2.5, 2) {};
        \node [style=rectangle] (5) at (0, 2) {};
        \node [style=right label] (6) at (-2.5, 2.5) {$R_A$};
        \node [style=right label] (7) at (0, 2.5) {$A$};
        \node  [style=proj2] (8) at (-0, 3.5) {${\mathcal{U}_{g}^A}^\dagger$};
        \node [style=Wsquare] (9) at (-2.5, 3.5) {$g$};
        \node [style=rectangle, minimum width=1.5cm, align=left] (10) at (-4.5, 3.5) {$\sum_{g}$};
        \node  [style=proj2] (11) at (-0, 5.5) {$\mathcal{E}^{B|A}$};
        \node [style=right label] (12) at (0, 6.5) {$B$};
        \node [style=right label] (13) at (0, 4.6) {$A$};
        \node  [style=proj2] (14) at (-0, 7.5) {$\mathcal{U}_{g'}^B$};
        \node [style=Wsquareadj] (15) at (-2.5, 7.5) {$g'$};
        \node [style=rectangle, minimum width=1.5cm, align=left] (16) at (-4.5, 7.5) {$\frac{1}{|G|}\sum_{g'}$};
        \node [style=rectangle] (17) at (-2.5, 9.3) { };
        \node [style=right label] (18) at (-2.5, 8.5) {$R_B$};
        \node [style=proj2] (19) at (0, 9.6) {${\mathcal{U}_{g''}^B}^\dagger$};
        \node [style=right label] (20) at (0, 8.5) {$B$};
        \node [style=Wsquare] (21) at (-2.5, 9.6) {\small{$g''$}};
        \node [style=rectangle, minimum width=1.5cm, align=left] (22) at (-4.5, 9.6) {$\sum_{g''}$};
        \node [style=proj2] (23) at (0, 11.7) {$\mathcal{E}^{C|B}$};
        \node [style=right label] (24) at (0, 10.7) {$B$};
        \node  [style=proj2] (25) at (-0, 13.7) {${\mathcal{U}_{g'''}^C}$};
        \node [style=right label] (26) at (0, 12.7) {$C$};
        \node [style=Wsquareadj] (27) at (-2.5, 13.7) {\small{$g'''$}};
        \node [style=rectangle, minimum width=1.5cm, align=left] (28) at (-4.5, 13.7) {$\frac{1}{|G|} \sum_{g'''}$};
        \node [style=rectangle] (29) at (-2.5, 15.3) {};
        \node [style=rectangle] (30) at (0, 15.3) {};
         \node [style=right label] (31) at (-2.35, 14.8) {$R_C$};
         \node [style=right label] (32) at (0.15, 14.8) {$C$};
         
	\end{pgfonlayer}
	\begin{pgfonlayer}{edgelayer}
          \draw [qWire] (5) to (8);
        \draw [qWire] (4) to (9);
        \draw [qWire] (8) to (11);
        \draw [qWire] (14) to (11);
        \draw [qWire] (15) to (17);
        \draw [qWire] (14) to (19);
        \draw [qWire] (23) to (19);
        \draw [qWire] (23) to (25);
        \draw [qWire] (27) to (29);
        \draw [qWire] (25) to (30);
	\end{pgfonlayer}
\end{tikzpicture} 
}} = \vcenter{\hbox{
\begin{tikzpicture}
	\begin{pgfonlayer}{nodelayer}
        \node [style=rectangle] (4) at (-2.5, 2) {};
        \node [style=rectangle] (5) at (0, 2) {};
        \node [style=right label] (6) at (-2.5, 2.5) {$R_A$};
        \node [style=right label] (7) at (0, 2.5) {$A$};
        \node  [style=proj2] (8) at (-0, 3.5) {${\mathcal{U}_{g}^A}^\dagger$};
        \node [style=Wsquare] (9) at (-2.5, 3.5) {$g$};
        \node [style=rectangle, minimum width=1.5cm, align=left] (10) at (-4.5, 3.5) {$\sum_{g}$};
        \node  [style=proj2] (11) at (-0, 5.5) {$\mathcal{E}^{B|A}$};
        \node [style=right label] (13) at (0, 4.6) {$A$};
        \node [style=proj2] (23) at (0, 11.7) {$\mathcal{E}^{C|B}$};
        \node [style=right label] (24) at (0, 8.5) {$B$};
        \node  [style=proj2] (25) at (-0, 13.7) {${\mathcal{U}_{g'''}^C}$};
        \node [style=right label] (26) at (0, 12.7) {$C$};
        \node [style=Wsquareadj] (27) at (-2.5, 13.7) {\small{$g'''$}};
        \node [style=rectangle, minimum width=1.5cm, align=left] (28) at (-4.5, 13.7) {$\frac{1}{|G|}\sum_{g'''}$};
        \node [style=rectangle] (29) at (-2.5, 15.3) {};
        \node [style=rectangle] (30) at (0, 15.3) {};
        \node [style=right label] (31) at (-2.35, 14.8) {$R_C$};
         \node [style=right label] (32) at (0.15, 14.8) {$C$};
	\end{pgfonlayer}
	\begin{pgfonlayer}{edgelayer}
          \draw [qWire] (5) to (8);
        \draw [qWire] (4) to (9);
        \draw [qWire] (8) to (11);
        \draw [qWire] (23) to (25);
        \draw [qWire] (27) to (29);
        \draw [qWire] (25) to (30);
        \draw [qWire] (11) to (23);
	\end{pgfonlayer}
\end{tikzpicture} 
}} = \vcenter{\hbox{\begin{tikzpicture}
	\begin{pgfonlayer}{nodelayer}		
        \node [style=right label] (6) at (-0.8, -3) {$R_A$};
        \node [style=right label] (7) at (1.1, -3) {$A$};
        \node [style=right label] (1) at (1.4, 4.5) {$\$$};
        \node  [style=proj2] (11) at (0, -0.5) {$\mathcal{E}^{B|A}$};
        \node [style=right label] (12) at (0.15, 1.5) {$B$};
        \node [style=right label] (13) at (0.15, -1.5) {$A$};

        \begin{scope}[shift={(0,3.0)}]
            \node  [style=proj2] (31) at (0, 0.5) {$\mathcal{E}^{C|B}$};
            \node [style=right label] (32) at (0.15, 1.5) {$C$};
            \node [style=rectangle] (41) at (0, 2.25) { };
        \end{scope}

        \node [style=rectangle, minimum width=2cm, minimum height=3.52cm, fill=green, opacity=0.3, rounded corners] (17) at (0, 1.48) { };

        \begin{scope}[shift={(0,3.0)}]
            \node [style=right label] (18) at (-0.80, 3) {$R_C$};
            \node [style=right label] (20) at (1.1, 3) {$C$};
            \node [style=rectangle] (23) at (-0.95, 1.75) { };
            \node [style=rectangle] (24) at (-0.95, 4.25) { };
            \node [style=rectangle] (25) at (0.95, 1.75) { };
            \node [style=rectangle] (26) at (0.95, 4.25) { };
        \end{scope}

        \node [style=rectangle] (22) at (0, -2.25) { };
        \node [style=rectangle] (27) at (-0.95, -1.75) { };
        \node [style=rectangle] (28) at (-0.95, -4) { };
        \node [style=rectangle] (29) at (0.95, -1.75) { };
        \node [style=rectangle] (30) at (0.95, -4) { };
	\end{pgfonlayer}

	\begin{pgfonlayer}{edgelayer}
        \draw [qWire] (11) to (31);
        \draw [qWire] (11) to (22);
        \draw [qWire] (23) to (24);
        \draw [qWire] (25) to (26);
        \draw [qWire] (27) to (28);
        \draw [qWire] (29) to (30);
        \draw [qWire] (31) to (41);
	\end{pgfonlayer}
\end{tikzpicture} }}.
    \end{equation}

Applying these identities to the bilocality scenario, we have
\begin{equation}    \vcenter{\hbox{\begin{tikzpicture}
	\begin{pgfonlayer}{nodelayer}
        \node [style=Wsquareadj] (1) at (-3, -6) {$\rho^{AB_1}$};
        \node [style=Wsquareadj] (2) at (3, -6) {$\rho^{B_2C}$};
        \node  [style=proj2] (3) at (-0, 0) {$E^{B_1B_2}_b$}; 
        \node  [style=proj2] (4) at (-6, 0) {$E^A_{a|x}$};
        \node  [style=proj2] (5) at (6, 0) {$E^C_{c|z}$};
        \node [style=rectangle, minimum width=8cm, minimum height=4.8cm, fill=green, opacity=0.3, rounded corners] (6) at (0, -3.8) { };
        \node [style=right label] (7) at (-4.3, -3) {$A$};
        \node [style=right label] (8) at (-1.3, -3) {$B_1$};
        \node [style=right label] (9) at (1.8, -3) {$B_2$};
        \node [style=right label] (10) at (4.8, -3) {$C$};
        \node [style=right label] (11) at (7.4, 0.7) {$\$$};    
        \node [style=rectangle] (12) at (-6, 2.25) { };
        \node [style=rectangle] (13) at (0, 2.25) { };
        \node [style=rectangle] (14) at (6, 2.25) { };
        \node [style=rectangle] (15) at (-6.5, -10) { };
        \node [style=rectangle] (16) at (6.5, -10) { };
        \node [style=right label,font=\normalsize ] (17) at (-6.25, -9) {$x$};
        \node [style=right label,font=\normalsize ] (18) at (6.65, -9) {$z$};
        \node [style=right label,font=\normalsize ] (19) at (-5.85, 1.5) {$a$};
        \node [style=right label,font=\normalsize ] (20) at (0.15, 1.5) {$b$};
        \node [style=right label,font=\normalsize ] (21) at (6.15, 1.5) {$c$};
	\end{pgfonlayer}
	\begin{pgfonlayer}{edgelayer}
        \draw [qWire] (1) to (3);
        \draw [qWire] (2) to (3);
        \draw [qWire] (1) to (4);
        \draw [qWire] (2) to (5);
        \draw[] (4) to (12);
        \draw[] (3) to (13);
        \draw[] (5) to (14);
        \draw[] (4) to (15);
        \draw[] (5) to (16);
	\end{pgfonlayer}
\end{tikzpicture}}}=\vcenter{\hbox{\begin{tikzpicture}
	\begin{pgfonlayer}{nodelayer}
        \node [style=Wsquareadj] (1) at (-3, -6) {$\rho^{AB_1}$};
        \node [style=Wsquareadj] (2) at (3, -6) {$\rho^{B_2C}$};
        \node  [style=proj2] (3) at (-0, 0) {$E^{B_1B_2}_b$}; 
        \node  [style=proj2] (4) at (-6, 0) {$E^A_{a|x}$};
        \node  [style=proj2] (5) at (6, 0) {$E^C_{c|z}$};
        \node [style=rectangle, minimum width=1.8cm, minimum height=1.2cm, fill=green, opacity=0.3, rounded corners] (6) at (-6, 0) { };
        \node [style=right label] (7) at (-3.9, -3) {$A$};
        \node [style=right label] (8) at (-1, -3) {$B_1$};
        \node [style=right label] (9) at (2, -3) {$B_2$};
        \node [style=right label] (10) at (5, -3) {$C$};
        \node [style=right label] (11) at (7.3, 0.8) {$\$$};    
        \node [style=rectangle] (12) at (-6, 2.25) { };
        \node [style=rectangle] (13) at (0, 2.25) { };
        \node [style=rectangle] (14) at (6, 2.25) { };
        \node [style=rectangle] (15) at (-6.5, -10) { };
        \node [style=rectangle] (16) at (6.5, -10) { };
        \node [style=right label,font=\normalsize ] (17) at (-6.25, -9) {$x$};
        \node [style=right label,font=\normalsize ] (18) at (6.65, -9) {$z$};
        \node [style=right label,font=\normalsize ] (19) at (-5.85, 1.5) {$a$};
        \node [style=right label,font=\normalsize ] (20) at (0.15, 1.5) {$b$};
        \node [style=right label,font=\normalsize ] (21) at (6.15, 1.5) {$c$};
        \node [style=rectangle, minimum width=1.8cm, minimum height=1.2cm, fill=green, opacity=0.3, rounded corners] (22) at (0, 0) { };
        \node [style=rectangle, minimum width=1.8cm, minimum height=1.2cm, fill=green, opacity=0.3, rounded corners] (23) at (6, 0) { };
        \node [style=rectangle, minimum width=2.25cm, minimum height=2.25cm, fill=green, opacity=0.3, rounded corners] (24) at (-3, -6.25) { };
        \node [style=rectangle, minimum width=2.25cm, minimum height=2.25cm, fill=green, opacity=0.3, rounded corners] (25) at (3, -6.25) { };
        \node [style=rectangle] (26) at (-4.3, -4.25) { };
        \node [style=rectangle] (27) at (-5.9, -0.95) { };
        \node [style=right label] (28) at (-5.8, -3) {$R_A$};
        \node [style=rectangle] (29) at (1.7, -4.25) { };
        \node [style=rectangle] (30) at (0.1, -0.95) { };
         \node [style=right label] (31) at (0, -3) {$R_{B_2}$};
          \node [style=rectangle] (32) at (-2.5, -4.25) { };
        \node [style=rectangle] (33) at (-0.9, -0.95) { };
        \node [style=right label] (34) at (-3.3, -3) {$R_{B_1}$};
        \node [style=rectangle] (35) at (3.45, -4.25) { };
        \node [style=rectangle] (36) at (5.10, -0.95) { };
        \node [style=right label] (37) at (3, -3) {$R_C$};
        \node [style=right label] (38) at (1.3, 0.8) {$\$$};
        \node [style=right label] (39) at (-4.7, 0.8) {$\$$};
        \node [style=right label] (40) at (-1.3, -4.4) {$\$$};
        \node [style=right label] (41) at (4.7, -4.4) {$\$$};
         \node [style=rectangle] (42) at (-3.5, -4.25) { };
        \node [style=rectangle] (43) at (-5.1, -0.95) { };
        \node [style=rectangle] (44) at (-4.15, -3.75) { };
        \node [style=rectangle] (45) at (-5.25, -1.45) { };
        \node [style=right label] (46) at (-5.35, -0.95) {$A$};
        \node [style=right label] (47) at (-3.55, -4.5) {$A$};
        \node [style=rectangle] (48) at (-1.7, -4.25) { };
        \node [style=rectangle] (49) at (-0.1, -0.95) { };
        \node [style=right label] (50) at (-2.2, -4.5) {$B_1$};
        \node [style=right label] (50) at (-1.4, -0.9) {$B_1$};
         \node [style=rectangle] (51) at (-1.85, -3.75) { };
        \node [style=rectangle] (52) at (-0.75, -1.45) { };
        \node [style=rectangle] (53) at (2.5, -4.25) { };
        \node [style=rectangle] (54) at (0.9, -0.95) { };
        \node [style=rectangle] (55) at (1.85, -3.75) { };
        \node [style=rectangle] (56) at (0.75, -1.45) { };
        \node [style=right label] (57) at (0.7, -0.9) {$B_2$};
        \node [style=right label] (58) at (2.4, -4.5) {$B_2$};
        \node [style=rectangle] (59) at (4.25, -4.25) { };
        \node [style=rectangle] (60) at (5.90, -0.95) { };
        \node [style=rectangle] (61) at (4.2, -3.75) { };
        \node [style=rectangle] (62) at (5.2, -1.45) { };
        \node [style=right label] (63) at (3.9, -4.5) {$C$};
        \node [style=right label] (64) at (4.75, -0.95) {$C$};
	\end{pgfonlayer}
	\begin{pgfonlayer}{edgelayer}
        \draw [qWire] (1) to (44);
        \draw [qWire] (4) to (45);
        \draw [qWire] (56) to (3);
        \draw [qWire] (55) to (2);
        \draw [qWire] (1) to (51);
        \draw [qWire] (3) to (52);
        \draw [qWire] (2) to (61);
        \draw [qWire] (5) to (62);
        \draw[] (4) to (12);
        \draw[] (3) to (13);
        \draw[] (5) to (14);
        \draw[] (4) to (15);
        \draw[] (5) to (16);
        \draw [qWire] (26) to (27);
        \draw [qWire] (29) to (30);
        \draw [qWire] (32) to (33);
        \draw [qWire] (35) to (36);
        \draw [qWire] (42) to (43);
        \draw [qWire] (48) to (49);
        \draw [qWire] (53) to (54);
        \draw [qWire] (59) to (60);
	\end{pgfonlayer}
\end{tikzpicture}}}
\end{equation}
where we represent the classical inputs and outputs by thin black wires while the quantum systems continue to be represented by thick purple wires. Note that, as mentioned in the main text, for the purpose of this paper, we make the conventional assumption that a setting or an outcome variable is encoded in a classical system carrying a trivial representation. 

Moreover, the $\$$ map preserves internal causal structures \emph{regardless} of the specific definition of causal-influence that is used. This is evident from the diagrammatic representation of $\$$ on a multipartite channel: 

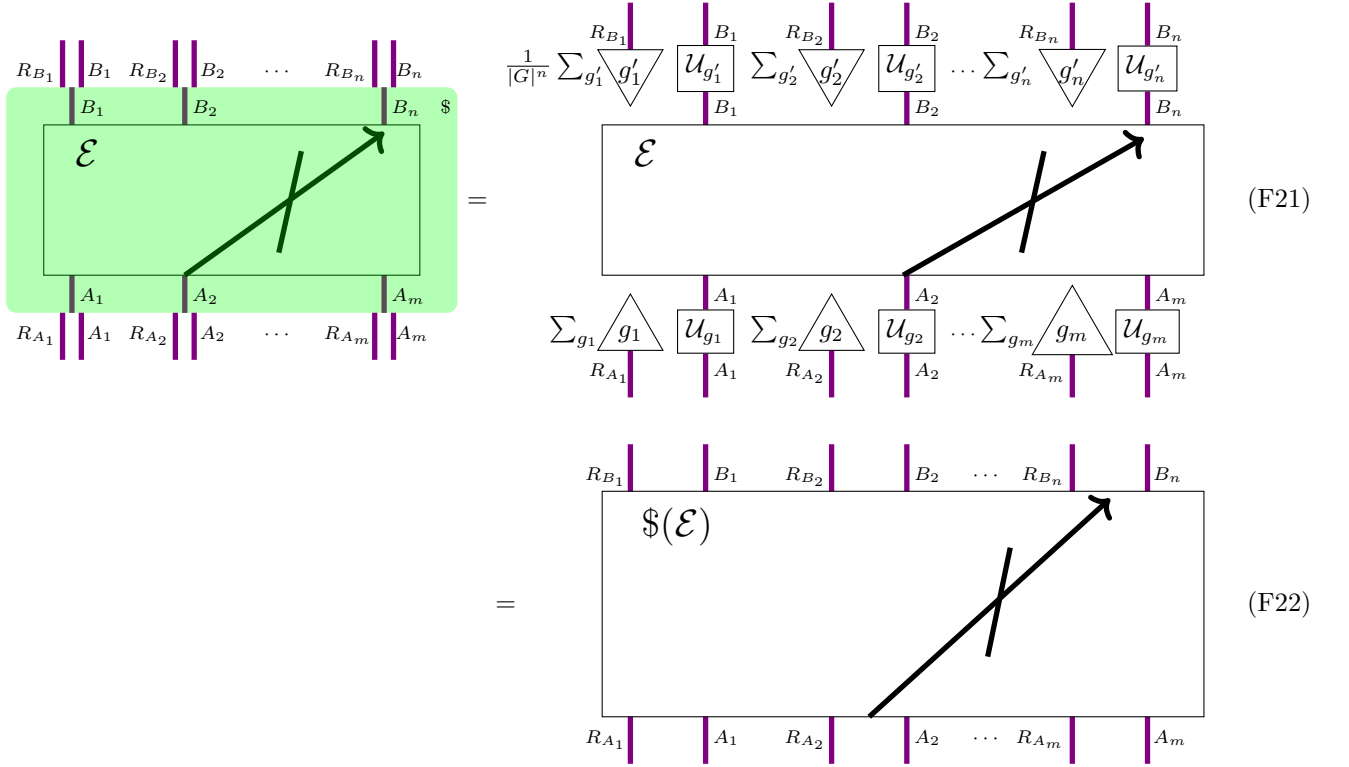
\begin{align}
  \begin{tikzpicture}
	\begin{pgfonlayer}{nodelayer}		    
        \node  [style=rectangle, minimum width=5cm, minimum height=2cm,draw=black] (1) at (-0, 0) {};
         \node [style=right label] (2) at (-5.8, -3.5) {$R_{A_1}$};
        \node [style=right label] (3) at (-3.9, -3.5) {$A_1$}; 
        \node [style=right label] (4) at (-2.8, -3.5) {$R_{A_2}$};
        \node [style=right label] (5) at (-0.9, -3.5) {$A_2$}; 
        \node [style=right label] (6) at (2.4, -3.5) {$R_{A_m}$};
        \node [style=right label] (7) at (4.3, -3.5) {$A_m$};
        \node [style=right label] (8) at (0.8, -3.5) {$\cdots$};
        \node [style=right label] (9) at (-5.8, 3.5) {$R_{B_1}$};
        \node [style=right label] (10) at (-3.9, 3.5) {$B_1$}; 
        \node [style=right label] (11) at (-2.8, 3.5) {$R_{B_2}$};
        \node [style=right label] (12) at (-0.9, 3.5) {$B_2$}; 
        \node [style=right label] (13) at (2.4, 3.5) {$R_{B_n}$};
        \node [style=right label] (14) at (4.3, 3.5) {$B_n$};
        \node [style=right label] (15) at (0.8, 3.5) {$\cdots$};
        \node [style=rectangle, minimum width=6cm, minimum height=3cm, fill=green, opacity=0.3, rounded corners] (16) at (0, 0) { };
        \node [style=rectangle] (21) at (-4.5, -2.75) { };
        \node [style=rectangle] (22) at (-4.5, -4.5) { };
        \node [style=rectangle] (23) at (-4, -2.75) { };
        \node [style=rectangle] (24) at (-4, -4.5) { };
        \node [style=rectangle] (25) at (-1.5, -2.75) { };
        \node [style=rectangle] (26) at (-1.5, -4.5) { };
        \node [style=rectangle] (27) at (-1, -2.75) { };
        \node [style=rectangle] (28) at (-1, -4.5) { };
        \node [style=rectangle] (29) at (3.8, -2.75) { };
        \node [style=rectangle] (30) at (3.8, -4.5) { };
        \node [style=rectangle] (31) at (4.3, -2.75) { };
        \node [style=rectangle] (32) at (4.3, -4.5) { };
        \node [style=rectangle] (33) at (-4.5, 2.75) { };
        \node [style=rectangle] (34) at (-4.5, 4.5) { };
        \node [style=rectangle] (35) at (-4, 2.75) { };
        \node [style=rectangle] (36) at (-4, 4.5) { };
        \node [style=rectangle] (37) at (-1.5, 2.75) { };
        \node [style=rectangle] (38) at (-1.5, 4.5) { };
        \node [style=rectangle] (39) at (-1, 2.75) { };
        \node [style=rectangle] (40) at (-1, 4.5) { };
        \node [style=rectangle] (41) at (3.8, 2.75) { };
        \node [style=rectangle] (42) at (3.8, 4.5) { };
        \node [style=rectangle] (43) at (4.3, 2.75) { };
        \node [style=rectangle] (44) at (4.3, 4.5) { };
        \node [style=rectangle] (45) at (-4.25, -3.25) { };
        \node [style=rectangle] (46) at (-4.25, -1.75) { };
        \node [style=rectangle] (47) at (-1.25, -3.25) { };
        \node [style=rectangle] (48) at (-1.25, -1.75) { };
        \node [style=rectangle] (49) at (4.05, -3.25) { };
        \node [style=rectangle] (50) at (4.05, -1.75) { };
        \node [style=rectangle] (51) at (-4.25, 3.25) { };
        \node [style=rectangle] (52) at (-4.25, 1.75) { };
        \node [style=rectangle] (53) at (-1.25, 3.25) { };
        \node [style=rectangle] (54) at (-1.25, 1.75) { };
        \node [style=rectangle] (55) at (4.05, 3.25) { };
        \node [style=rectangle] (56) at (4.05, 1.75) { };
        \node [style=right label] (57) at (-4.1, -2.5) {$A_1$}; 
        \node [style=right label] (58) at (-1.1, -2.5) {$A_2$};
        \node [style=right label] (59) at (4.2, -2.5) {$A_m$};
        \node [style=right label] (60) at (-4.1, 2.5) {$B_1$}; 
        \node [style=right label] (61) at (-1.1, 2.5) {$B_2$};
        \node [style=right label] (62) at (4.2, 2.5) {$B_n$};
        \node [style=right label, font=\Large] (63) at (-4.2, 1.1) {$\mathcal{E}$};
        \node [style=rectangle] (64) at (-1.25, -2.25) { };
        \node [style=rectangle] (65) at (4.05, 2.25) { };
        \node [style=right label] () at (5.5, 2.5) {$\$$};
	\end{pgfonlayer}
	\begin{pgfonlayer}{edgelayer}
        \draw [qWire] (21) to (22);
        \draw [qWire] (23) to (24);
        \draw [qWire] (25) to (26);
        \draw [qWire] (27) to (28);
        \draw [qWire] (29) to (30);
        \draw [qWire] (31) to (32);
        \draw [qWire] (33) to (34);
        \draw [qWire] (35) to (36);
        \draw [qWire] (37) to (38);
        \draw [qWire] (39) to (40);
        \draw [qWire] (41) to (42);
        \draw [qWire] (43) to (44);
        \draw [qWire] (45) to (46);
        \draw [qWire] (47) to (48);
        \draw [qWire] (49) to (50);
        \draw [qWire] (51) to (52);
        \draw [qWire] (53) to (54);
        \draw [qWire] (55) to (56);
        \draw[->, line width=2pt] (-1.25, -2) -- (4.05, 1.8);
        \draw[-, line width=2pt] (1.25, -1.4) -- (1.85, 1.3);
	\end{pgfonlayer}
\end{tikzpicture}=    \begin{tikzpicture}
	\begin{pgfonlayer}{nodelayer}		    
        \node  [style=rectangle, minimum width=8cm, minimum height=2cm,draw=black] (1) at (1, 0) {};
         \node [style=right label] (2) at (-7.4, -4.5) {$R_{A_1}$};
        \node [style=right label] (3) at (-4.1, -4.5) {$A_1$}; 
        \node [style=right label] (4) at (-2.2, -4.5) {$R_{A_2}$};
        \node [style=right label] (5) at (1.25, -4.5) {$A_2$}; 
        \node [style=right label] (6) at (4, -4.5) {$R_{A_m}$};
        \node [style=right label] (7) at (7.65, -4.5) {$A_m$};
        \node [style=right label] (8) at (2.2, -3.5) {$\cdots$};
        \node [style=right label] (9) at (-7.4, 4.5) {$R_{B_1}$};
        \node [style=right label] (10) at (-4.1, 4.5) {$B_1$}; 
        \node [style=right label] (11) at (-2.2, 4.5) {$R_{B_2}$};
        \node [style=right label] (12) at (1.25, 4.5) {$B_2$}; 
        \node [style=right label] (13) at (4, 4.5) {$R_{B_n}$};
        \node [style=right label] (14) at (7.65, 4.5) {$B_n$};
        \node [style=right label] (15) at (2.2, 3.5) {$\cdots$};
        \node [style=rectangle] (21) at (-4.5, -2.75) { };
        \node [style=rectangle] (22) at (-6.25, -5.5) { };
        \node [style=rectangle] (23) at (-4.25, -2.75) { };
        \node [style=rectangle] (24) at (-4.25, -5.5) { };
        \node [style=rectangle] (25) at (-1.5, -2.75) { };
        \node [style=rectangle] (26) at (-0.9, -5.5) { };
        \node [style=rectangle] (27) at (-1, -2.75) { };
        \node [style=rectangle] (28) at (1.1, -5.5) { };
        \node [style=rectangle] (29) at (3.8, -2.75) { };
        \node [style=rectangle] (30) at (5.5, -5.5) { };
        \node [style=rectangle] (31) at (4.3, -2.75) { };
        \node [style=rectangle] (32) at (7.5, -5.5) { };
        \node [style=rectangle] (33) at (-4.5, 2.75) { };
        \node [style=rectangle] (34) at (-4.25, 5.5) { };
        \node [style=rectangle] (35) at (-4.25, 1.75) { };
        \node [style=rectangle] (36) at (-4, 4.5) { };
        \node [style=rectangle] (37) at (-1.5, 2.75) { };
        \node [style=rectangle] (38) at (-0.9, 5.5) { };
        \node [style=rectangle] (39) at (-1, 2.75) { };
        \node [style=rectangle] (40) at (1.1, 5.5) { };
        \node [style=rectangle] (41) at (7.5, 1.75) { };
        \node [style=rectangle] (42) at (3.8, 4.5) { };
        \node [style=rectangle] (43) at (7.5, 5.5) { };
        \node [style=rectangle] (44) at (4.3, 4.5) { };
        \node [style=rectangle] (45) at (-4.25, -3.25) { };
        \node [style=rectangle] (46) at (-4.25, -1.75) { };
        \node [style=rectangle] (47) at (-1.25, -3.25) { };
        \node [style=rectangle] (48) at (1.1, -1.75) { };
        \node [style=rectangle] (49) at (4.05, -3.25) { };
        \node [style=rectangle] (50) at (7.5, -1.75) { };
        \node [style=rectangle] (51) at (-6.25, 5.5) { };
        \node [style=rectangle] (52) at (-4.25, 1.75) { };
        \node [style=rectangle] (53) at (-1.25, 3.25) { };
        \node [style=rectangle] (54) at (1.1, 1.75) { };
        \node [style=rectangle] (55) at (5.5, 5.5) { };
        \node [style=rectangle] (56) at (4.05, 1.75) { };
        \node [style=right label] (57) at (-4.1, -2.5) {$A_1$}; 
        \node [style=right label] (58) at (1.25, -2.5) {$A_2$};
        \node [style=right label] (59) at (7.65, -2.5) {$A_m$};
        \node [style=right label] (60) at (-4.1, 2.5) {$B_1$}; 
        \node [style=right label] (61) at (1.25, 2.5) {$B_2$};
        \node [style=right label] (62) at (7.65, 2.5) {$B_n$};
        \node [style=right label, font=\Large] (63) at (-6.2, 1.1) {$\mathcal{E}$};
        \node [style=rectangle] (64) at (-1.25, -2.25) { };
        \node [style=rectangle] (65) at (4.05, 2.25) { };
        \node  [style=rectangle, draw=black] (66) at (-4.25, -3.5) {$\mathcal{U}_{g_1}$};
        \node [style=Wsquare] (67) at (-6.25, -3.5) {$g_1$};
        \node [style=rectangle, minimum width=1.5cm, align=left] (68) at (-7.75, -3.5) {$\sum_{g_1}$};
        \node  [style=rectangle, draw=black] (68) at (1.1, -3.5) {$\mathcal{U}_{g_2}$};
        \node [style=Wsquare] (69) at (-0.9, -3.5) {$g_2$};
        \node [style=rectangle, minimum width=1.5cm, align=left] (70) at (-2.4, -3.5) {$\sum_{g_2}$};
        \node  [style=rectangle, draw=black] (71) at (7.5, -3.5) {$\mathcal{U}_{g_m}$};
        \node [style=Wsquare] (72) at (5.5, -3.5) {$g_m$};
        \node [style=rectangle, minimum width=1.5cm, align=left] (73) at (3.8, -3.5) {$\sum_{g_m}$};
        \node [style=rectangle, minimum width=1.5cm, align=left] (74) at (-8.25, 3.5) {$\frac{1}{|G|^n}\sum_{g'_1}$};
         \node [style=Wsquareadj] (75) at (-6.25, 3.5) {$g'_1$};
         \node  [style=rectangle, draw=black] (76) at (-4.25, 3.5) {$\mathcal{U}_{g'_1}$};
         \node [style=rectangle, minimum width=1.5cm, align=left] (77) at (-2.4, 3.5) {$\sum_{g'_2}$};
         \node [style=Wsquareadj] (78) at (-0.9, 3.5) {$g'_2$};
         \node  [style=rectangle, draw=black] (79) at (1.1, 3.5) {$\mathcal{U}_{g'_2}$};
         \node [style=rectangle, minimum width=1.5cm, align=left] (80) at (3.8, 3.5) {$\sum_{g'_n}$};
         \node [style=Wsquareadj] (81) at (5.5, 3.5) {$g'_n$};
         \node  [style=rectangle, draw=black] (82) at (7.5, 3.5) {$\mathcal{U}_{g'_n}$};
	\end{pgfonlayer}
	\begin{pgfonlayer}{edgelayer}
        \draw [qWire] (67) to (22);
        \draw [qWire] (69) to (26);
        \draw [qWire] (68) to (28);
        \draw [qWire] (72) to (30);
        \draw [qWire] (71) to (32);
        \draw [qWire] (76) to (34);
        \draw [qWire] (35) to (76);
        \draw [qWire] (78) to (38);
        \draw [qWire] (79) to (40);
        \draw [qWire] (41) to (82);
        \draw [qWire] (43) to (82);
        \draw [qWire] (68) to (48);
        \draw [qWire] (71) to (50);
        \draw [qWire] (51) to (75);
        \draw [qWire] (79) to (54);
        \draw [qWire] (55) to (81);
        \draw[->, line width=2pt] (1, -2) -- (7.4, 1.65);
        \draw[-, line width=2pt] (4.15, -1.4) -- (4.75, 1.3);
        \draw [qWire] (46) to (66);
        \draw [qWire] (24) to (66);
	\end{pgfonlayer}
\end{tikzpicture}\\ 
= \qquad \begin{tikzpicture}
	\begin{pgfonlayer}{nodelayer}		    
        \node  [style=rectangle, minimum width=8cm, minimum height=3cm,draw=black] (1) at (-0, 0) {};
         \node [style=right label] (2) at (-8.5, -3.5) {$R_{A_1}$};
        \node [style=right label] (3) at (-5.1, -3.5) {$A_1$}; 
        \node [style=right label] (4) at (-3.2, -3.5) {$R_{A_2}$};
        \node [style=right label] (5) at (0.25, -3.5) {$A_2$}; 
        \node [style=right label] (6) at (2.95, -3.5) {$R_{A_m}$};
        \node [style=right label] (7) at (6.6, -3.5) {$A_m$};
        \node [style=right label] (8) at (1.8, -3.5) {$\cdots$};
        \node [style=right label] (9) at (-8.5, 3.5) {$R_{B_1}$};
        \node [style=right label] (10) at (-5.1, 3.5) {$B_1$}; 
        \node [style=right label] (11) at (-3.2, 3.5) {$R_{B_2}$};
        \node [style=right label] (12) at (0.25, 3.5) {$B_2$}; 
        \node [style=right label] (13) at (3.1, 3.5) {$R_{B_n}$};
        \node [style=right label] (14) at (6.6, 3.5) {$B_n$};
        \node [style=right label] (15) at (1.8, 3.5) {$\cdots$};
        \node [style=rectangle] (21) at (-7.25, -2.75) { };
        \node [style=rectangle] (22) at (-7.25, -4.5) { };
        \node [style=rectangle] (23) at (-5.25, -2.75) { };
        \node [style=rectangle] (24) at (-5.25, -4.5) { };
        \node [style=rectangle] (25) at (-1.9, -2.75) { };
        \node [style=rectangle] (26) at (-1.9, -4.5) { };
        \node [style=rectangle] (27) at (0.1, -2.75) { };
        \node [style=rectangle] (28) at (0.1, -4.5) { };
        \node [style=rectangle] (29) at (6.5, -2.75) { };
        \node [style=rectangle] (30) at (6.5, -4.5) { };
        \node [style=rectangle] (31) at (4.5, -2.75) { };
        \node [style=rectangle] (32) at (4.5, -4.5) { };
        \node [style=rectangle] (33) at (-7.25, 2.75) { };
        \node [style=rectangle] (34) at (-7.25, 4.5) { };
        \node [style=rectangle] (35) at (-5.25, 2.75) { };
        \node [style=rectangle] (36) at (-5.25, 4.5) { };
        \node [style=rectangle] (37) at (-1.9, 2.75) { };
        \node [style=rectangle] (38) at (-1.9, 4.5) { };
        \node [style=rectangle] (39) at (0.1, 2.75) { };
        \node [style=rectangle] (40) at (0.1, 4.5) { };
        \node [style=rectangle] (41) at (6.5, 2.75) { };
        \node [style=rectangle] (42) at (6.5, 4.5) { };
        \node [style=rectangle] (43) at (4.5, 2.75) { };
        \node [style=rectangle] (44) at (4.5, 4.5) { };
        \node [style=right label, font=\Large] (63) at (-7, 2) {$\$(\mathcal{E}$)};
        \node [style=rectangle] (64) at (-1.25, -2.25) { };
        \node [style=rectangle] (65) at (4.05, 2.25) { };
	\end{pgfonlayer}
	\begin{pgfonlayer}{edgelayer}
        \draw [qWire] (21) to (22);
        \draw [qWire] (23) to (24);
        \draw [qWire] (25) to (26);
        \draw [qWire] (27) to (28);
        \draw [qWire] (29) to (30);
        \draw [qWire] (31) to (32);
        \draw [qWire] (33) to (34);
        \draw [qWire] (35) to (36);
        \draw [qWire] (37) to (38);
        \draw [qWire] (39) to (40);
        \draw [qWire] (41) to (42);
        \draw [qWire] (43) to (44);
        \draw[->, line width=2pt] (-0.9, -3) -- (5.5, 2.8);
        \draw[-, line width=2pt] (2.25, -1.4) -- (2.85, 1.5);
	\end{pgfonlayer}
\end{tikzpicture}
\end{align}

From the diagrams, it is clear that the $\$$ map does not add any new causal connections since each reference frame system associated to an input has no causal influence on any reference frame system associated to an output. In other words, if $S \nrightarrow S'$, then $R_S S \nrightarrow R_{S'} S'$, where $S$ and $S' $ can be any subsets of inputs and outputs.

We conclude that $\$$ is a causal-structure-preserving simulation.

\subsection{Coherent simulation}
\label{app:cohsim}

The {\em coherent simulation} map, which we denote by $\ER$, is defined analogously to $\$$, except that after injecting each reference system $R_{S_i}$ in its fiducial state $\ket{e}^{R_{S_i}}\bra{e}$ to each component system $R_{S_i}$, we apply coherent twirling (such as the one used in Ref.~\cite{hamette2020Quantum}) instead of incoherent twirling (a.k.a. $G$-twirling such as the one used in Ref.~\cite{bartlettReference2007}) to each $R_{S_i}S_i$ pair (independently from every other such pair). 

To define the $\ER$ map, we first define the superoperator:
\begin{align}
    {\cal V}_{gg'}(\cdot)\coloneqq U_g \cdot U_{g'}^{\dagger}.
\end{align}
Then, we can write the action of $\ER$ on a unipartite quantum operation $\mathcal{E}^{B|A}$ as:
\begin{align}
\label{eq:ERuni}
&    \ER( \mathcal{E}^{B|A})
 \coloneqq     \bigg( \frac{1}{|G|}  \sum_{ g_{B} {g'_{B}}}  \ketbra{g_B}{g'_B}^{R_B} \otimes  \mathcal{V}^{ B}_{ g_{B}{g'_{B}}} \bigg) \circ \mathcal{E}^{ B| A}  \circ \bigg(    \frac{1}{|G|}  \sum_{{ g_{A} {g'_{A}}}}   \tr_{R_{A}}\left( \ketbra{g_A}{g'_A}^{R_A} \cdot \right) 
 \otimes  (\mathcal{V}^{ A}_{ g_A{g'_{A}}})^{\dagger}  \bigg). \nonumber
\end{align}
The $\ER$ map preserves the property of being completely positive (CP) and also the property of being trace-preserving (TP).  This follows from the fact that the superoperators that act on the left and right of $\mathcal{E}^{B|A}$ in \cref{eq:ERuni} are CPTP. (The $\ER$ map also preserves the property of unitality.)

Furthermore, the image of any quantum operation under $\ER$ is a valid operation  in the $(G,{\cal U})$-twirled world since it is both left-invariant and right-invariant under the group action. That is $\forall g\in G$,
\begin{align}
  & (\mathcal{U}^{R_B}_g \otimes \mathcal{U}^{B}_g)  \circ \ER({\cal E}^{B|A})   \nonumber \\
  = &(\mathcal{U}^{R_B}_g \otimes \mathcal{U}^{B}_g)  \circ \left( \frac{1}{|G|} \sum_{{ g_B {g_B}'}}   \ketbra{g_B}{g'_B}^{R_B} \otimes  (\mathcal{V}^{ B}_{ g_B{g_B}'})^{\dagger}  \right) 
  \circ \mathcal{E}^{ B| A} \circ
  \left(   \frac{1}{|G|} \sum_{{ g_A {g_A}'}}  \tr_{R_{A}}\left( \ketbra{g_A}{g'_A}^{R_A} \cdot \right)  \otimes  (\mathcal{V}^{ A}_{ g_A{g_A}'})^{\dagger}   \right)
 \nonumber\\ 
=& \ER({\cal E}^{B|A}), 
\end{align}
and similarly
\begin{align}
  &\ER({\cal E}^{B|A}) \circ  (\mathcal{U}^{R_A}_g \otimes \mathcal{U}^{A}_g) = \ER({\cal E}^{B|A}).
\end{align}

If we specialize  $\mathcal{E}^{B|A}$ to the case of the superoperator that injects the state $\rho^B$, we get
\begin{align}
\ER(\rho^B) 
= & \frac{1}{|G|}\sum_{gg'}  \ketbra{g_B}{g'_B}^{R_B}  \otimes \left( \mathcal{V}^{B}_{gg'} \circ \rho^{B} \right).
\end{align}

If we specialize $\mathcal{E}^{B|A}$ to the case of the superoperator that contracts with the effect $E$, we get
\begin{align}
& \ER(\tr_A[E^A \cdot])  = \tr_{R_AA} \left[ \frac{1}{|G|}  \sum_{gg'} |g \rangle^{R_A} \langle g'|  \otimes ( E^{A} \circ (\mathcal{V}^{A}_{gg'})^{\dag} ) \cdot  \right]. \nonumber
\end{align}

Here, the unit effect is no longer mapped to the unit effect on the pair $R_AA$. Nevertheless, the statistics can still be reproduced since all states, quantum operations and effects are mapped by $\ER$ to processes in a subspace of the pair of system and its reference frame; in particular, the unit effect is mapped to the unit effect on that subspace (instead of the unit effect on the whole space), so a POVM is mapped to a POVM on that subspace. See Appendix \ref{app:rep} for the definition of the subspace. As such, we have 
\begin{align}
\label{eq:ERrepro}
\ER(\tr_B (E^B\;\cdot ))\circ \ER (\mathcal{E}^{B|A})\circ \ER (\rho^A) = \tr_B \left(E^B \mathcal{E}^{B|A}(\rho^A)\right).
\end{align}

For each multipartite quantum operation $\mathcal{E}^{{B}  | {A} }$, its image under $\ER$ is
\begin{align}
    \ER( \mathcal{E}^{B|A}) \coloneqq     \bigg( \frac{1}{|G|} \bigotimes_i \sum_{ g_{B_i} {g'_{B_i}}}   \ketbra{ g_{B_i}}{g'_{B_i}}^{R_{ B_i}} \otimes  \mathcal{V}^{ B_i}_{ g_{B_i}{g'_{B_i}}} \bigg) %
  \circ \mathcal{E}^{ B| A}\circ \bigg(    \frac{1}{|G|} \bigotimes_j \sum_{{ g_{A_j} {g'_{A_j}}}}  
\tr_{R_{A_i}}\left( \ketbra{ g_{A_j}}{g'_{A_j}}^{R_{ {A_j}}}\cdot \right) 
 \otimes  (\mathcal{V}^{ A_j}_{ g_A{g'_{A_j}}})^{\dagger}  \bigg). \label{eq:cohE}
\end{align}
When an input or an output system $S_i$ carries a trivial representation (e.g., when $S_i$ is classical), we can drop the reference frame system $R_{S_i}$.

Similar to the case for unipartite quantum operations, $\ER( \mathcal{E}^{ B|A})$ is a CPTP map whenever $\mathcal{E}^{ B|A}$ is CPTP and is trace-nonincreasing whenever $\mathcal{E}^{ B|A}$ is trace-nonincreasing. It is also straightforward to prove that $\ER(\mathcal{E}^{B|A})$ is a valid process in the ($G,\cal U$)-twirled quantum world. And the actions of $\ER$ on the superoperator describing the injection of a state or the
contraction with an effect are special cases of \cref{eq:cohE}.

The definition of $\ER(\mathcal{E}^{B|A})$ when $A$ and $B$ are multipartite systems ensures that $\ER$ commutes with parallel and sequential compositions. The proof is analogous to the one for \cref{eq:para} and \cref{eq:seq}. Together with \cref{eq:ERrepro}, this implies that it reproduces the statistics of any circuit. It also implies that the coherent simulation preserves the aspect of the causal structure described by the wiring of the circuit.  Furthermore, the $\ER$ map also preserves the aspect of the causal structure that is internal to the operations in the circuit, which can be seen by following the analogous argument for $\ER(\mathcal{E}^{B|A})$.  Consequently, $\ER$, like $\$$, provides a causal-structure-preserving simulation. 

\subsection{The local relational degrees of freedom used by the simulations}
\label{app:rep}

As mentioned in the main text, the basic ideas of these simulations is that, for each system $S$, we encode the degrees of freedom of $S$ that transform nontrivially under the action of the symmetry in {\em relational} degrees of freedom between $S$ and its associated reference frame system $R_S$. In particular, such relational degrees of freedom transform trivially under the collective actions of the symmetry on $S$ and $R_S$, but transform \emph{nontrivially} under $\{\mathcal{I}^{R_S} \otimes {\cal U}^{S}_g\}_{g \in G}$, which we term the {\em relational} group actions. 

Note that in a $(G,\mathcal{U})$-twirled world, the symmetry transformation $\mathcal{I}^{A} \otimes {\cal U}^{B}_g$ is equivalent to any transformation of the form  ${\cal U}^{A}_h \otimes {\cal U}^{A}_{gh}$ for some $ h \in G$ (for instance, it is equivalent to the transformation $\mathcal{U}_{g^{-1}}^{A} \otimes \mathcal{I}^{B}$).
This follows from the fact that in a twirled quantum world, every state $\rho_{AB}$ is invariant under the collective group action, i.e., $\forall g \in G: {\cal U}^{A}_{g} \otimes {\cal U}^{B}_{g} (\rho_{AB}) =\rho_{AB}$
which implies that
\begin{align}
&{\cal U}^{A}_h \otimes {\cal U}^{A}_{gh} (\rho_{AB})
= {\cal U}^{A}_h \otimes {\cal U}^{A}_{gh}  [{\cal U}^{A}_{h^{-1}} \otimes {\cal U}^{B}_{h^{-1}}] (\rho_{AB})
= {\cal I}^{A}  \otimes \mathcal{U}^{B}_{g} (\rho_{AB}).
\end{align}
It follows that any group action of the form $\{\mathcal{U}^{A}_h \otimes {\cal U}^{B}_{gh}\}_{g \in G}$ for some $h\in G$ suffices to implement the {\em relational} group action between $A$ and $B$.

It then follows that if $\rho^S$ is noninvariant under ${\cal U}_g^{S}$, then $\$(\rho^S)$ is noninvariant under $\mathcal{I}^{R_S}\otimes  {\cal U}_g^{S}$. The proof is  by contradiction. If $\$(\rho^S)$ were invariant under $\mathcal{I}^{R_S}\otimes{\cal U}_g^{S}$, we would have
\begin{align}
   \$(\rho^S) =&\mathcal{I}^{R_S}\otimes{\cal U}_g^{S}\circ \$(\rho^S) %
 = \mathcal{I}^{R_S}\otimes{\cal U}_g^{S} \left[ \frac{1}{|G|} \sum_{g'} {\cal U}^{R_S}_{g'}  \otimes {\cal U}^S_{g'}  \left(\ketbra{e}^{R_S} \otimes \rho^S \right) \right] 
 =  \frac{1}{|G|} \sum_{g'} {\cal U}^{R_S}_{g'} \left(\ketbra{e}^{R_S}\right) \otimes {\cal U}^S_{g} \left({\cal U}_{g'}^{S} (\rho^S) \right).
 \label{eq:eq}
\end{align}
Since $\{ {\cal U}^{R_S}_{g'} (\ketbra{e}^{R_S})\}_{g'}$ are linearly independent by definition, the only solution to \cref{eq:eq} is that
\begin{align}
 {\cal U}_{g'}^{S}(\rho^S)= {\cal U}_{g}^{S}\left({\cal U}_{g'}^{S}(\rho^S)\right), \forall g'\in G.
\end{align}
In the case where $g'=g$, this implies $\rho^S={\cal U}_g^{S} (\rho^S) $, contradicting the premise that $\rho^S$ is noninvariant under ${\cal U}_g^{S}$. 

The proof that the analogous claim holds for $\ER$ proceeds similarly. If $\ER(\rho^S)$ were invariant under $\mathcal{I}^{R_S}\otimes{\cal U}_g^{S}$, we would have
\begin{align}
   \ER(\rho^S) =&\mathcal{I}^{R_S}\otimes{\cal U}_g^{S}\circ \ER(\rho^S) 
 = \mathcal{I}^{R_S}\otimes{\cal U}_g^{S} \left[ \frac{1}{|G|} \sum_{g'g''} {\cal V}^{R_S}_{g'g''}  \otimes {\cal V}^S_{g'g''}  \left(\ketbra{e}^{R_S} \otimes \rho^S \right) \right] \nonumber \\
 = & \frac{1}{|G|} \sum_{g'g''} {\cal V}^{R_S}_{g'g''} \left(\ketbra{e}^{R_S}\right) \otimes {\cal U}_g^{S} \left({\cal V}^S_{g'g''} (\rho^S) \right),
 \label{eq:eq1}
\end{align}
Again, since $\{ {\cal V}^{R_S}_{g'g''} (\ketbra{e}^{R_S})\}_{g'g''}$ are linearly independent, the only solution to \cref{eq:eq1} is that
\begin{align}
 {\cal V}_{g'g''}^{S}(\rho^S)= {\cal U}_{g}^{S}\left({\cal V}_{g'g''}^{S}(\rho^S)\right), \forall g'g''\in G.
\label{eq:eq11}
\end{align}
In the case where $g'=g''=g$, we have ${\cal V}_{g'g''}={\cal U}_g$,  then \cref{eq:eq11} implies that $\rho^S={\cal U}_g^{S} (\rho^S) $, contradicting the premise that $\rho^S$ is noninvariant under ${\cal U}_g^{S}$.

Now, we comment on the nature of the relational degrees of freedom used by $\$$ and the nature of those
used by $\ER$. (The distinction between these two types of relational degrees of freedom was first discussed in Ref.~\cite{Bartlett_2009}.)

Consider the decomposition of the joint Hilbert space of a system $S$ and its reference frame $R_S$ with respect to the symmetry group $G$: 
\begin{align}
	\mathcal{H}^{R_SS}=\bigoplus_q \mathcal{M}_q^{R_SS} \otimes \mathcal{N}_q^{R_SS},
\end{align}
where each sector carries an inequivalent representation labeled by $q$, and for each $q$, $\mathcal{M}_q^{R_SS}$ carries an irreducible representation and $\mathcal{N}_q^{R_SS}$ is a multiplicity space, carrying a trivial representation. 

Then, the incoherent simulation $\$$ on a unipartite operator can be viewed~\cite{bartlettReference2007} as the sum of projectors into each sector $q$  together with randomization in the factor spaces carrying the irreducible representations; more precisely, 
\begin{equation}
    \$ (O^S) =\sum_q \left( \mathcal{D}_{\mathcal{M}_q}^{R_SS} \otimes \mathcal{I}_{\mathcal{N}_q}^{R_SS} \right) \left[ \Pi_{q}^{R_SS} \left(\ketbra{e}^{R_S} \otimes O^S \right)  \Pi_{q}^{R_SS} \right],
\end{equation}
where $\Pi_{q}$ is the projection into the sector $q$, $\mathcal{D}_{\mathcal{M}_q}$ denotes the trace-preserving operation that takes every operator on the Hilbert space $\mathcal{M}_q$ to a constant times the identity operator on that space, and $\mathcal{I}_{\mathcal{N}_q}$ denotes the identity map over operators in the space $\mathcal{N}_q$.

As such, the incoherent simulation $\$$ encodes information across all subspaces carrying different representations.

As for the coherent simulation $\ER$, it encodes information only in the subspace carrying the trivial representation~\cite{hamette2021Perspectiveneutral} (i.e., the subspace of the Hilbert space for which every vector  transforms trivially under the group action):
\begin{align}
    \ER (O^S) =
    & \Pi_0^{R_SS} \left(\ketbra{e} \otimes O \right) 
    \Pi_0^{R_SS}. \nonumber
\end{align}

\subsection{On the nonnecessity of correlated reference frames}

We end with a comment regarding the role of correlated reference frames in facilitating simulation, a topic discussed in the conclusions of the main text and in \cref{app:shareRF}. 
For simulation maps that associate a reference frame to each system in the circuit being simulated, there are different possibilities for how the reference frames 
might be correlated.  The $\$$ and $\ER$ maps both assume
a {\em lack} of  correlation among any of the reference frames.  An alternative would be to {\em allow} correlation among the reference frames that have some common ancestry in the causal structure (the map of \cref{footnote:yen} is of this type).  In this language, what is proven above
is that correlation among the reference frames is not needed for achieving the simulation.
Some prior work in the field of quantum reference frames {\em did} recognize this fact, but only in particular scenarios, such as a Bell scenario or a prepare-measure  scenario~\cite{bartlett2003Classical}. Other work~\cite{bartlettReference2007} proved the possibility of simulation in more general scenarios, but presumed a single reference frame that was accessible to all parties, which is equivalent to assuming correlation among the reference frames associated to the parties.  
What we have proven here, namely, that one can achieve a twirled world simulation of a quantum circuit having arbitrary causal structure while not requiring any correlation among the reference frames, was not noted in previous work.

\section{The failure of incoherent and coherent simulations in swirled quantum worlds}
\label{app:failure}

We begin with the proof of the impossibility of {\em incoherent} simulation in a swirled quantum world. Recall that the $\$_C$ map is the analogue of the $\$$ map but defined for a nonphysical symmetry. As mentioned in the main text, the $\$_C$ map {\em does} achieve a valid simulation if one restricts attention to a prepare-measure experiment on a {\em unipartite} system. (In this case, it is equivalent to the simulation proposed by Stueckelberg~\cite{Stueckelberg1960}.)  This is proven as follows. Because $\mathbb{Z}_2$ has order 2, it suffices to take a single qubit as the reference system.  Taking the reference frame state $|e\rangle$ to be the Pauli $Y$ eigenstate $|{+}y\rangle$, and noting that $\mathcal{C}(|{+}y\rangle \langle {+}y|)= |{-}y\rangle \langle {-}y|$, we find 
  \begin{align}
\label{eq:dolCrho}
   \$_C (\rho^A)  
   = & \frac{1}{2} (\ketbra{+y}^{R_A}\otimes\rho^{A}  
   + \ketbra{-y}^{R_A}\otimes\mathcal{C}(\rho^{A}),\\
      \$_C (\tr_A(E^A\;\cdot))  
   = & \ketbra{+y}^{R_A}\otimes \tr_A(E^A\;\cdot) %
   + \ketbra{-y}^{R_A}\otimes\tr_A(E^A\;\cdot) \circ {\mathcal{C}^{A}}^{\dagger}.
\end{align}  
It is straightforward to see that 
$\$_C (\tr_A(E^A\;\cdot)) \circ \$_C (\rho^A) = \tr_A(E^A \rho^A )$ and consequently that 
the statistics of any prepare-measure experiment on the unipartite system are reproduced under $\$_C$. 

Nonetheless, the $\$_C$ map  does not define a simulation in general experimental scenarios, such as a prepare-measure experiment on a {\em bipartite} system.
(Another example of a situation in which it fails to define a simulation is a prepare-transform-measure experiment on a unipartite system; this can be inferred from the failure in the bipartite scenario by leveraging the Choi isomorphism.) Here we give the explicit expressions for the image of $\$_C$ when acting on a bipartite state $\rho^{AB}$:
\begin{align}
\label{eq:dolCAB}
 \$_C (\rho^{AB}) = & \frac{1}{4} \big( 
 \ketbra{+y}{+y}^{R_A} \otimes {\cal I}^{A}  \otimes\ketbra{+y}{+y}^{R_B} \otimes {\cal I}^B
    +\ketbra{+y}{+y}^{R_A} \otimes {\cal I}^{A}  \otimes\ketbra{-y}{-y}^{R_B} \otimes {\cal C}^B
 \nonumber \\
 &
 +\ketbra{-y}^{R_A} \otimes {\cal C}^{A}  \otimes \ketbra{+y}{+y}^{R_B} \otimes {\cal I}^B
 + \ketbra{-y}^{R_A} \otimes {\cal C}^{A}  \otimes \ketbra{-y}{-y}^{R_B} \otimes {\cal C}^B\big)
\circ \rho^{AB}
\end{align}
The resulting matrix is block-diagonal with four blocks, each of which is given by one of the terms in the summation. The first and the last block are positive-semi-definite (PSD) matrices, but the middle two are not necessarily PSD, since partial conjugation (such as $ {\cal I}^{A} \otimes {\cal C}^{B}$) is not completely positive. Thus, $\$_C (\rho^{AB})$ may not be a valid state in RQT when $\rho^{AB}$ is entangled, and similarly for bipartite effects and channels. Indeed, there are choices of $\rho^{AB}$ such as the Bell state for which this is the case. 

We turn now to the proof of the impossibility of {\em coherent} simulation in a swirled quantum world.
We remind the reader that the $\ER_C$ map is the analogue of the $\ER$ map but defined for a nonphysical symmetry. The $\ER_C$ map performs worse even than the $\$_C$ map insofar as it fails to provide a simulation even in the case of a unipartite prepare-measure scenario. \blk This is because, in general, $\ER_C(\rho^A)$ for some state $\rho^A$ fails to even be a valid linear operator on ${\cal H}_A$. Specifically, its action on a unipartite state $\rho^A$ is: 
\begin{align}
    \ER_C (\rho^A) 
   \coloneqq  
    &\frac{1}{2} \left[  \ketbra{+y}{+y}^{R_A} \otimes \rho^A  + \ketbra{+y}{-y}^{R_A} \otimes (\rho^A C^A) 
    + \ketbra{-y}{+y}^{R_A} \otimes (C^A\rho^A)  +  \ketbra{-y}^{R_A} \otimes \overline{\rho}^A
    \right] 
    \nonumber \\
    = &\frac{1}{2} \big[  \ketbra{+y}{+y}^{R_A} \otimes \rho^A  %
    +( C^{R_A}\ketbra{-y}^{R_A} )\otimes (C^A \overline{\rho}^A )  %
    + (C^{R_A}\ketbra{+y}{+y}^{R_A} ) \otimes (C^A\rho^A)  
    +  \ketbra{-y}^{R_A} \otimes \overline{\rho}^A
    \big] 
    \nonumber \\
= & \frac{1}{2} \big[  (\mathbb{I}^{R_A}\otimes\mathbb{I}^A+ C^{R_A}\otimes C^A) 
(\ketbra{+y}{+y}^{R_A}\otimes\rho^A + \ketbra{-y}^{R_A}\otimes\overline{\rho}^A)\big]. \nonumber
\end{align}
where in the second equality we have used the property that $C^2=\mathbb{I}$, as proven in \cref{app:conju}. $\ER_C (\rho^A) $ is not necessarily a linear operator on ${\cal H}_A$ since $\mathbb{I}^{R_A}\otimes\mathbb{I}^{R_A}$ is linear while $C^{R_A} \otimes C^{A}$ is antilinear. 

To explicitly see that $\ER_C (\rho^A) $ is not a linear operator on ${\cal H}_A$ in general, we consider the concrete example of the state $\rho^A=\ketbra{0}^A$. The image of this state under $\ER_C$ is: 
\begin{align}
     &\ER_C (\ketbra{0}^A)  
     \frac{1}{2} \big[  (\mathbb{I}^{R_A}\otimes\mathbb{I}^A+ C^{R_A}\otimes C^A)(\mathbb{I}^{R_A}\otimes\ketbra{0}^A)\big].
\end{align}
We now check to see how this acts on the Hilbert space vectors $\ket{00}^{R_AA}$ and $i\ket{00}^{R_AA}$.
For the case of $\ket{00}^{R_AA}$, we find 
\begin{align}
    &\ER_C (\ketbra{0}^A) \ket{00}^{R_AA} 
    =
    \frac{1}{2}  (\mathbb{I}^{R_A}\otimes\mathbb{I}^A+ C^{R_A}\otimes C^A)\ket{00}^{R_AA} = \ket{00}^{R_AA},\\
    \end{align}
while for the case of $i\ket{00}^{R_AA}$, we find 
    \begin{align}
     &\ER_C (\ketbra{0}^A) (i\ket{00}^{R_AA}) 
     \frac{1}{2}  (\mathbb{I}^{R_A}\otimes\mathbb{I}^A+ C^{R_A}\otimes C^A) ( i\ket{00}^{R_AA} )  = 0.
\end{align}
Any operator on $\mathcal{H}_A$ that maps a vector on $\mathcal{H}_A$ to zero is noninvertible, hence not linear. 

We conclude that $\ER_C $ can map a Hermitian operator to an operator on that is not even linear, and therefore maps some quantum states and effects to operators that do not describe quantum states and effects. As such, it cannot define a simulation.

\end{widetext}

\section{Generalization from quantum theory to an arbitrary GPT}
\label{app:GPT}

The idea of symmetrizing a theory relative to the collective action of a symmetry can be generalized from quantum theory to general GPTs. Here, we only consider GPTs that satisfy the principle of tomographic locality. We defer a discussion of symmetrization within tomographically nonlocal GPTs to a future paper (just as was done in Ref.~\cite{centeno2024twirled}). 
Note that classical probability theory, quantum theory, boxworld~\cite{barrett2007information} and the toy theory of Ref.~\cite{Spekkens2007Evidence} are all examples of GPTs that are tomographically local, while some exotic variants of classical theories~\cite{d2020classicality,scandolo2019information,Chiribella2024} and of quantum theory~\cite{erba2024compositionrulequantumsystems} are not.
Below, $T$  will denote an arbitrary GPT that is tomographically local. 

We generalize the notions of twirling and swirling from quantum theory to a GPT $T$ as follows: twirling refers to symmetrization with respect to a physical symmetry, while swirling refers to symmetrization with respect to a nonphysical symmetry. 
(This harmonizes with the definition given in Ref.~\cite{centeno2024twirled}, where only twirling was considered.)
The symmetrized theory is termed a \emph{twirled $T$-world} or a \emph{swirled $T$-world} depending on whether the symmetry is physical or nonphysical. 

As proven in Ref.~\cite[Appendix D.2]{centeno2024twirled}, any twirled $T$-world is itself a valid GPT. Replacing the physical symmetry with a nonphysical symmetry  
in  this proof yields the proof that a swirled $T$-world is itself a valid GPT.

It turns out that the generalization from quantum theory to GPTs introduces the possibility of a distinction between two types of nonphysical symmetries, hence two types of swirled $T$-worlds, which we now describe. 

We begin by clarifying the notion of a physical operation. Intuitively, an operation is physical in theory $T$ if it is something that can be realized as a dynamical evolution map according to $T$. 

Just as in quantum theory, a necessary condition for an operation to be physical is that it must be \emph{logically possible} in $T$, that is, 
having this operation in $T$
does not lead to any logical inconsistency. (Recall from the main text that here \emph{logical inconsistency} means that one would obtain a number outside $[0,1]$ where one expects a probability.)

In quantum theory, being logically possible is not only a necessary condition for being physical, it is sufficient as well.
This follows from the fact that quantum theory satisfies the so-called \emph{no-restriction hypothesis}, which requires that in the theory, all logically possible operations must also be physically possible. 

While any logically impossible operation must be nonphysical in any GPT, in a GPT $T$ that {\em fails} to satisfy the no-restriction hypothesis (such as the toy theory of Ref.~\cite{Spekkens2007Evidence} and certain subtheories of quantum theory), a logically possible operation can also be nonphysical in $T$ simply because it is not included among the set of valid operations in $T$, even though supplementing the set with this operation would not lead to any logical inconsistencies.
Thus, for a GPT that fails to satisfy the no-restriction hypothesis, one can distinguish three distinct classes of operations:
(i) physical (hence necessarily logically possible), (ii) nonphysical but still logically possible, and (iii) not logically possible  (hence necessarily nonphysical). 

It follows that a given group of symmetry transformations (relative to which one can define a symmetrization map) will be one of three types: (i) consisting entirely of physical operations and termed a {\em physical symmetry}, (ii) including one or more operations that are nonphysical but still logically possible, termed a {\em weakly nonphysical symmetry}, and (iii) including one or more operations that are nonphysical by virtue of {\em failing} to be logically possible, termed a {\em strongly nonphysical symmetry}. Note that for the latter two types of symmetry transformations, i.e., the nonphysical ones, a transformation must still satisfy basic properties to qualify as a symmetry transformation, such as being reversible and preserving the inner product between states and effects.

A symmetrization process relative to a physical symmetry is, by definition, a twirling process.  Symmetrization processes relative to the two types of nonphysical symmetries lead to two types of swirling processes.  

\subsection{Connections to the unitary-nonunitary distinction and the complex-real distinction}
\label{app:eg}

Consider GPTs defined by modifying the standard Hilbert space formulation of quantum theory, such as subtheories of quantum theory and modifications thereof~\cite{Dmello_2024,Ansanelli2025GPT}. 
We now show explicitly that in such theories, the question of whether or not a symmetry is physical may not align with whether or not it is unitary, nor with whether or not the Hilbert space describing the pure states in the resulting symmetrized world is over the complex or the real field. To see this, consider the following two examples. 

The first example is a subtheory of quantum theory, where the states and effects are all and only those in quantum theory, but where the only valid operations are identity transformations, swaps, measure-and-reprepare channels, 
classically-controlled versions and convex combinations of these. We refer to this theory as PMQT for short (where PM stands for preparation and measurement). (In the language of Ref.~\cite{GPTembedding,rolino2024minimaloperationaltheoriesclassical}, this theory would be referred to as the minimal extension of the full prepare-measure fragment of quantum theory.) In PMQT, all nontrivial symmetries are nonphysical, including unitary symmetries; in particular, unitary symmetries are weakly nonphysical, while nonunitary symmetries are strongly nonphysical. Consequently, there are no twirled PMQT worlds, as symmetrization relative to any symmetry is always an instance of swirling. Moreover, the field on which the Hilbert space is defined in a swirled PMQT world can be either complex or real since a unitary symmetry can result in a complex-amplitude theory while a nonunitary symmetry necessarily results in a real-amplitude theory since as mentioned in the main text, any nonunitary swirling of quantum theory (and consequently its subtheories) will result in a subtheory of RQT.

The second example is a modification of quantum theory where the allowed states and effects exclude those with negative partial transpose, and the operations include all quantum operations \emph{and} complex conjugation operations. Here, complex conjugations, despite being antiunitary, can be included as a physical operation because no states or effects in the theory will be mapped to invalid states or effects even under a partial complex conjugation. We refer to this theory as {\em PPT-world} (where PPT stands for positive partial transpose). In  PPT-world, all symmetries are physical. 
Consequently, there is no possibility for a swirling process in a PPT-world, as symmetrization relative to any symmetry is always an instance of twirling. In particular, the real-amplitude PPT-world that results from symmetrizing relative to the time-reversal symmetry is a twirled PPT-world, instead of a swirled one. 

These two examples serve to illustrate that the distinction between unitary and nonunitary symmetries does not always align with the distinction between physical and nonphysical symmetries. In the first case (PMQT world), all nontrivial symmetries are non-physical, despite some being unitary. In the second case (PPT world), all symmetries are physical, even though some are nonunitary. 
Moreover, in these two examples, whenever the symmetry is nonunitary, the Hilbert space in the resulting symmetrized world can be defined over the real field. Thus, the distinction between physical and nonphysical symmetries also does not align with whether the Hilbert space over which pure states are defined in the resulting symmetrized world is over the complex or the real field.

\subsection{Symmetrized-nonsymmetrized gaps for a GPT $T$}

For any GPT $T$, the analogue of \cref{thm:twirledsim} holds:
\begin{theorem}
For all causal structures and all physical symmetries of $T$, there is no gap between the correlations realizable in $T$ and those realizable in the corresponding twirled $T$-world.
\label{thm:twirledsim_GPT}
\end{theorem}

The proof is simply that the incoherent simulation $\$$ can be straightforwardly generalized to any GPT $T$ and any symmetrization based on a physical symmetry, while still being causal-structure-preserving. We denote such a generalized map by $\$_T$. Explicitly, to generalize $\$$ to $\$_T$, in \cref{eq:dolEmul}, we replace $\mathcal{U}_g$ with $\mathcal{F}_g$, denoting the map in $T$ representing actions of the physical symmetry;
we replace $\{\ketbra{g}{g}^{R_S}\}_g$ with a set of GPT states, denoted $\{\omega_g^{R_S}\}_g $, such that $  \omega_{g}^{R_S} := \mathcal{F}_g (\omega_{e}^{R_S})$ and such that $ \omega_g^{R_S}$ is orthogonal to $  \omega_{g'}^{R_S}$ whenever $g\neq g'$; the trace operation is replaced by the inner product between GPT states and effects, denoted $\langle \cdot, \cdot \rangle$; and, $\mathcal{U}_g^{\dagger}$ is replaced by $\mathcal{F}_g^{-1}$. That is:
\begin{align}
\$_T(\mathcal{E}^{B|A}) \coloneqq &\left( \frac{1}{|G|^m} 
\bigotimes_{i} \sum_{{g}_i}\omega_{g_i}^{R_{B_i}}\otimes \mathcal{F}^{B_i }_{g_i}
\right) 
 \circ \mathcal{E}^{B|A} \nonumber \\
&\circ
 \left(
  \bigotimes_{j} \sum_{{g}_j} \langle\omega_{g_j}^{R_{A_j}},\cdot\rangle\otimes  {\mathcal{F}^{{A}_j}_{g_j}}^{\dag}
  \right).  
  \label{eq:ext}
\end{align}
It is then straightforward to verify that the proof that $\$$ is causal-structure-preserving in a twirled quantum world simulation of quantum theory also generalizes to show that $\$_T$ is causal-structure-preserving in a twirled $T$-world simulation of $T$.

Furthermore, $\$$ cannot be extended to be a simulation in any swirled $T$-world. The symmetry with respect to which the swirling is defined is nonphysical, meaning that there exists a $g\in G$ such that the representation of its action on a state $\mathcal{F}_g$ is not a valid transformation in the theory. It follows that $\mathcal{F}_g \circ \mathcal{I}$ is also not a valid transformation in the theory, since $\mathcal{F}_g \circ \mathcal{I}=\mathcal{F}_g$. However,  %
since the group action on the input and the output of an operation is uncorrelated, for a unipartite identity channel $\mathcal{I}^{A|A}$, there will be a term in its image of the form  $\big(\omega _g^{R_A'} \otimes \mathcal{F}_g^{A}\big) \circ \mathcal{I}^{A|A} \circ \big(\langle\omega_{e}^{R_{A}},\cdot\rangle \otimes \mathcal{I}^A\big)$, which equals $\omega _g^{R_A'} \otimes \langle\omega_{e}^{R_{A}},\cdot\rangle \otimes \mathcal{F}^A$, which is nonphysical. 
Thus, when the symmetry is nonphysical, such an extension does not define a simulation within the swirled world. 
This is analogous to the fact in \cref{app:failure} that $\$_C$ may map a bipartite state to a state that is nonvalid due to the uncorrelated group actions on different subsystems of the bipartite system.

Now we comment on the two examples of GPTs mentioned in \cref{app:eg}.

As mentioned earlier, there are no twirled PMQT worlds, since all symmetries are nonphysical, and so symmetrizing relative to them result only in swirled PMQT worlds. However, for any swirled PMQT world where the symmetry is unitary, we do {\em not} have a swirled-nonswirled gap. (This should be contrasted with the gap for full quantum theory swirled relative to time-reversal symmetry.) This is because---although the $\$$ map fails as a simulation, we can modify  $\$$ somewhat to obtain a causal-structure-preserving simulation. Specifically, we define a simulation in which any states and effects are still simulated by their images under the $\$$ map; as for any operation (which, recall, is by assumption a convex combination of the identity channel and measure-reprepare channels), we apply the $\$$ map to the measure-reprepare components, while mapping $\mathcal{I}^{B|A}$ simply to ${\mathcal I}^{R_BB|R_AA}$. Then, it is not hard to see that the resulting map is a simulation and is causal-structure-preserving. The fact that there is no swirled-nonswirled gap for any unitary symmetries in PMQT is closely related to the fact that any transformation of a unitary symmetry is logically possible, so the symmetry is weakly nonphysical instead of strongly nonphysical. 
This example shows that not all swirled-nonswirled pairs of GPTs admit of a swirled-nonswirled gap, especially in the case where the symmetry is weakly nonphysical. As such, we only conjecture for strongly nonphysical symmetries that one can always find a causal structure where there is swirled-nonswirled gap for any nonclassical GPT. (See \cref{cor:nonclassical} for why we restrict to nonclassical GPTs.)

For the second example, PPT-world, every symmetry induces a twirling, so there are no swirlings, as mentioned earlier. As such, even the time-reversal symmetry induces a twirling, and per \cref{thm:twirledsim_GPT}, the time-reversal-twirled PPT-world admits a causal-structure-preserving simulation of the PPT-world. Just as the time-reversal-swirled quantum world corresponds to real-amplitude quantum theory, the time-reversal-twirled PPT world can also be viewed as the \enquote{real-amplitude} version of the PPT world.  Since time-reversal symmetry is nonunitary, this example shows that the existence of a symmetrized-nonsymmetrized gap is not simply a function of whether the symmetrization is with respect to a unitary or nonunitary symmetry when considering arbitrary GPTs. Consequently, it also is not simply a function of the field over which the Hilbert space is defined (complex or real numbers) in the symmetrized theory.

\section{Shared reference frame states are always entangled in symmetrized worlds}
\label{app:shareRF}

In the following, we focus on twirled quantum worlds and shared reference frame states for a bipartite system, but all of our analysis can be straightforwardly generalized to any GPT, to both twirling and swirling, and to multipartite systems.

What does it mean for a bipartite state $\rho^{AB}$ to be useful as a shared reference frame? 
It means that it can encode some information about the relative orientation of local reference frames. 
For example, in the rotationally twirled spinor world~\cite{centeno2024twirled}, it should be able to encode some information about the relative orientation of Alice’s and Bob’s local Cartesian frames. 

To formalize this, consider a composite system $AB$ where $A$ is described relative to the first reference frame and $B$ is described relative to the second.
If we model a change to the relative orientation of the two local reference frames as a passive transformation on $AB$, then it is represented by an element of the group action $\{\mathcal{I}^{A} \otimes {\cal U}^{B}_g\}_{g \in G}$,\footnote{In the case where the symmetry is strongly nonphysical, for example, if it is nonunitary in quantum theory, then such a passive action should not be viewed as a physical operation, but merely a mathematical operation.} which we term the {\em relational} group action. For a bipartite state to be useful as a quantum token of a shared reference frame, it must be noninvariant under the relational group action. 
However, this is not a sufficient condition, since local actions on a multipartite state can change local properties as well as relational properties. Nevertheless, if a state is noninvariant under the relational group action {\em even after twirling} (which, recall from the main text that, implements the \emph{collective} group action), then one can be assured that the noninvariance of the state arises as a consequence of the relational group action changing relational properties of the state. Therefore, we have the following definition:
\begin{definition}
    A bipartite state acts as a shared reference frame state if and only if its image under the twirling map is not invariant under the relational group action.
\end{definition}
We say a state is a {\em perfect} shared reference frame state if the relational group action always takes its image under twirling to an orthogonal state.

It is worth noting that the definition of the relational group action given above (namely as  $\{\mathcal{I}^{A} \otimes {\cal U}^{B}_g\}_{g \in G}$) made a conventional choice, wherein the nontrivial unitary action occurs on system $B$ rather than $A$. \blk For a $(G,{\cal U})$-invariant state, the symmetry transformation $\mathcal{I}^{A} \otimes {\cal U}^{B}_g$ is equivalent to any transformation of the form  ${\cal U}^{A}_h \otimes {\cal U}^{A}_{gh}$ for some $ h \in G$.  
This follows from the fact that in a twirled quantum world, every state $\rho_{AB}$ is invariant under the collective group action, i.e., $\forall g \in G: {\cal U}^{A}_{g} \otimes {\cal U}^{B}_{g} (\rho_{AB}) =\rho_{AB}$
which implies that
\begin{align}
&{\cal U}^{A}_h \otimes {\cal U}^{B}_{gh} (\rho_{AB})
\nonumber\\
&= {\cal U}^{A}_h \otimes {\cal U}^{B}_{gh}  [{\cal U}^{A}_{h^{-1}} \otimes {\cal U}^{B}_{h^{-1}}] (\rho_{AB})\nonumber\\
&= {\cal I}^{A}  \otimes \mathcal{U}^{B}_{g} (\rho_{AB}).
\end{align}
It follows that there are many distinct but equivalent representations of the relational group action, as any set  $\{\mathcal{U}^{A}(h) \otimes {\cal U}^{B}(gh)\}_{g \in G}$ with $h\in G$ suffices to implement the {\em relational} group action. 

Now we prove 
\begin{restatable}{theorem}{share}\label{thm:share}
    In any twirled quantum world, a shared reference frame state is necessarily entangled.
\end{restatable}
\begin{proof}
We prove the contrapositive; that is, any separable state, i.e., any bipartite state that can be prepared by local operations and classical shared randomness in the twirled quantum world, does not act as a shared reference frame state. 

Any separable state in a twirled world is 
a mixture of products of invariant states, 
\begin{align}
    \rho^{R_1R_2}=\sum_i p_i \rho_i^{R_1} \otimes \rho_i^{R_2},
\end{align}
such that $p_i\in[0,1]$ and $\sum_i p_i=1$ and 
\begin{align}
  {\cal U}_g^{R_j} (\rho_j^{R_j}) = \rho_j^{R_j} \quad \forall g\in G, j=1,2.
\end{align}
Such a state is also invariant under the relational group action, that is,  $\forall g\in G$, 
\begin{align}
   ( {\cal I}\otimes  {\cal U}_g^{R_2} ) \rho^{R_1R_2}= \sum_i p_1  \rho_i^{R_1} \otimes  {\cal U}_g^{R_2} (\rho_i^{R_2})= \rho^{R_1R_2}, 
\end{align}
Thus, it cannot encode any information about the relational group action and consequently does not act as a shared reference frame state. 
\end{proof}

\section{Prospects for new causal structures admitting a swirled-nonswirled causal-compatibility gap}
\label{app:causal}

For the purposes of studying swirled-nonswirled gaps, it is sufficient to only consider causal structures where all observed systems are \emph{classical systems},  %
i.e., systems described by classical probability theory (CPT).\footnote{Causal structures where observed systems can be described by a GPT such as quantum theory have been studied in e.g., \cite{allen2017Quantum}.} As for latent systems (i.e., unobserved systems), we allow stipulations for some of them to be classical systems, while the rest are described by the GPT under consideration, referred to as \emph{GPT systems}. For example, the modification of the bilocality scenario considered in \cite{Renou_2021} includes one latent classical system, namely, the source of classical randomness shared between the pair of the quantum latent systems. Note, however, that in all explicit examples provided in this section, we allow all latent systems to be latent GPT systems so that in the figures illustrating these examples, we only need two types of wires indicating different system types: a thick purple wire indicates a GPT system, which is necessarily latent,  and a thin black wire indicates a classical system, which is necessarily observed.

Recall from \cref{app:preser} that in a circuit describing some experiment, each wire is associated with a system, and each box is associated with a particular operation. 
Given a causal structure, a circuit is said to be \emph{compatible} with it if, for any two systems $A$ and $B$ in the circuit, $A$ can influence $B$ in the circuit only if $A$ can influence $B$ in the causal structure. 
Then, a probability distribution among its observed variables (all of which are classical) is said to be \emph{$T$-realizable} in a causal structure (or $T$-compatible with it) if and only if there exists a circuit compatible with it that can yield that distribution when its nonlatent classical systems are described by the GPT $T$. We denote the swirled version of $T$ as $sT$. Then, a causal structure is said to admit of a $sT$-$T$ gap if the set of probability distributions that are $T$-realizable  (or its \emph{$T$-realizable correlations} for short) is a \emph{strict} superset of the set of probability distributions that are $sT$-realizable.

For simplicity, we will first describe our results for a subset of all possible causal structures, namely, those that do not have any nontrivial internal no-influence conditions (cf. \cref{app:preser}). As such, the gates in a circuit compatible with it are allowed to have {\em any} internal causal structure. We refer to this set of causal structures as the set with {\em unrestricted gates}, which is the most commonly studied type in the literature. The resulting class of causal structures is in fact a very broad class; indeed, this set encompasses all cases considered in the device-independent paradigm. 

Specifically, we first, in \cref{app:necessary}, provide necessary conditions for a causal structure with unrestricted gates to admit of a swirled-nonswirled causal compatibility gap for general GPTs. We then leverage this result, in \cref{app:lack}, to demonstrate that the lack of a shared reference frame state among all GPT systems, i.e., the presence of GPT systems that are independent or at most share classical randomness (cf. \cref{thm:share}), is not a sufficient condition for the existence of a swirled-nonswirled gap, even when the swirling is restricted to strongly nonphysical symmetries (cf. \cref{app:GPT}). At the end, in \cref{app:internal}, we make some comments on how our results generalize to causal structures wherein one makes nontrivial assumptions about the internal causal structure of the gates, i.e., gates with nontrivial no-influence conditions.

\subsection{Necessary conditions}
\label{app:necessary}

Now we present our necessary conditions for a given causal structure to admit of a swirled-nonswirled gap.

\subsubsection{District factorization}

The notion of district factorization has been previously defined in the study of classical causal inference (namely, one where all systems, including the latent ones are classical systems), which tells us when the problem of assessing compatibility of a distribution with a causal structure can be broken into the problem of assessing compatibility of each of a set of distributions with each of a set of substructures (or {\em fragments}) in a particular type of decomposition of that causal structure.
Here, we will generalize this notion to the case of causal structures where some latent systems can be described by an arbitrary GPT (but where the observed systems remain classical). With this generalization, the question of whether a causal structure has the potential for a swirled-nonswirled gap can sometimes be reduced to the question of whether any of its fragments admit of such a gap. 

The classical notion of district factorization is only defined for structures where latent systems do not have causal parents, i.e., where each is the output of a state preparation in the circuit rather than that of a transformation on another system. This is because in the classical case, it is known that every causal structure where a latent system has parents is observationally equivalent to another causal structure where no latent system has parents. (Here, observational equivalence means being compatible with the same set of distributions.) In the case where latent systems are described by quantum theory or some other nonclassical GPT, however, such observational equivalence may fail~\cite{quantum_inflation,Dani_intermediate}. Therefore, the generalization of district factorization we propose here allows for latent systems that have causal parents, while reducing to the traditional notion in the special case where none of the latent systems have parents. 

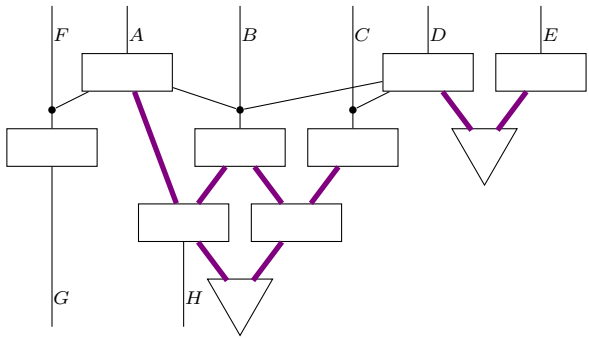
\begin{figure}[h!]
    \centering
\begin{tikzpicture}
        \node [style=Wsquareadj] (1) at (0,0) {%
        };
        \node  [style=proj2] (2) at (-1.5, 2) {%
        };
        \node  [style=proj2] (3) at (1.5, 2) {%
        };
         \node [style=rectangle] (4) at (-1.5, -1) { };
         \node [style=right label] (5) at (-1.5, 0) {$H$};
         \node  [style=proj2] (6) at (3, 4) { };
         \node  [style=proj2] (7) at (0, 4) { };
        \node  [style=proj2] (8) at (-3, 6) { };
        \node  [style=proj2] (9) at (-5, 4) { };
        \node [style=rectangle] (44) at (-5, 8) { };        
        \node [style=right label] (10) at (-5, 7) {$F$};
        \node [circle, fill, inner sep=1pt] (45) at (-5,5)  {};
        \node [style=rectangle] (11) at (-5, -1) { };
        \node [style=right label] (12) at (-5, 0) {$G$};
        \node  [style=proj2] (13) at (5, 6) { };
        \node  [style=proj2] (14) at (8, 6) { };
        \node [style=Wsquareadj] (15) at (6.5,4) {%
        };
        \node [style=rectangle] (16) at (-3, 8) { };
        \node [style=right label] (17) at (-3, 7) {$A$};
        \node [style=rectangle] (18) at (5, 8) { };
        \node [style=right label] (19) at (5, 7) {$D$};
        \node [style=rectangle] (20) at (0, 8) { };
        \node [style=right label] (21) at (0, 7) {$B$};
        \node [circle, fill, inner sep=1pt] (40) at (0,5)  {};
        \node [style=rectangle] (22) at (3, 8) { };
        \node [style=right label] (23) at (3, 7) {$C$};
        \node [circle, fill, inner sep=1pt] (41) at (3,5)  {};
        \node [style=rectangle] (24) at (8, 8) { };
        \node [style=right label] (25) at (8, 7) {$E$};
        \draw [] (2) to (4);
        \draw [qWire] (1) to (2);
        \draw [qWire] (1) to (3);
        \draw [qWire] (6) to (3);
        \draw [qWire] (7) to (3);
        \draw [qWire] (7) to (2);
        \draw [qWire] (8) to (2);
        \draw [] (8) to (40);
        \draw [] (8) to (45);
        \draw [] (11) to (9);
        \draw [] (13) to (40);
        \draw [] (13) to (41);
        \draw [qWire] (15) to (13);
        \draw [qWire] (15) to (14);
        \draw [] (16) to (8);
        \draw [] (13) to (18);
        \draw [] (20) to (7);
        \draw [] (22) to (6);
        \draw [] (24) to (14);
        \draw [] (9) to (44);
\end{tikzpicture}   
    \caption{An example of a causal structure. A thick purple wire indicates a latent GPT system. A thin black wire indicates an observed classical system. A black dot denotes the copy operation of the classical system. The copies of the same classical system share the same name.}
    \label{fig:bigcircuit}
\end{figure}

\begin{figure}[h!]
    \centering
\begin{tikzpicture}
        \node [style=Wsquareadj] (1) at (0,0) {%
        };
        \node  [style=proj2] (2) at (-1.5, 2) {%
        };
        \node  [style=proj2] (3) at (1.5, 2) {%
        };
         \node [style=rectangle] (4) at (-1.5, -1) { };
         \node [style=right label] (5) at (-1.5, 0) {$H$};
         \node  [style=proj2] (6) at (3, 4) { };
         \node  [style=proj2] (7) at (0, 4) { };
        \node  [style=proj2] (8) at (-3, 6) { };
        \node  [style=proj2] (9) at (-6, 4) { };
        \node [style=rectangle] (44) at (-6, 8) { };        
        \node [style=right label] (10) at (-6, 7) {$F$};
        \node [style=rectangle] (11) at (-6, -1) { };
        \node [style=right label] (12) at (-6, 0) {$G$};
        \node  [style=proj2] (13) at (6, 6) { };
        \node  [style=proj2] (14) at (9, 6) { };
        \node [style=Wsquareadj] (15) at (7.5,4) { };
        \node [style=rectangle] (16) at (-3, 8) { };
        \node [style=right label] (17) at (-3, 7) {$A$};
        \node [style=rectangle] (46) at (-3, -1) { };
        \node [style=right label] (47) at (-3, 0) {$F$};
        \node [style=rectangle] (18) at (6, 8) { };
        \node [style=right label] (19) at (6, 7) {$D$};
        \node [style=rectangle] (50) at (6, -1) { };
        \node [style=right label] (51) at (6, 0) {$C$};
        \node [style=rectangle] (52) at (5.3, -1) { };
        \node [style=rectangle] (53) at (5.3, 5.7) {};
        \node [style=right label] (54) at (5.3, 0) {$B$};
        \node [style=rectangle] (20) at (0, 8) { };
        \node [style=right label] (21) at (0, 7) {$B$};
        \node [circle, fill, inner sep=1pt] (40) at (0,5)  {};
        \node [style=rectangle] (22) at (3, 8) { };
        \node [style=right label] (23) at (3, 7) {$C$};
        \node [style=rectangle] (24) at (9, 8) { };
        \node [style=right label] (25) at (9, 7) {$E$};
        \draw [] (2) to (4);
        \draw [qWire] (1) to (2);
        \draw [qWire] (1) to (3);
        \draw [qWire] (6) to (3);
        \draw [qWire] (7) to (3);
        \draw [qWire] (7) to (2);
        \draw [qWire] (8) to (2);
        \draw [] (8) to (40);
        \draw [] (11) to (9);
        \draw [] (13) to (50);
        \draw [qWire] (15) to (13);
        \draw [qWire] (15) to (14);
        \draw [] (16) to (8);
        \draw [] (13) to (18);
        \draw [] (20) to (7);
        \draw [] (22) to (6);
        \draw [] (24) to (14);
        \draw [] (9) to (44);
        \draw [] (8) to (46);
        \draw [] (52) to (53);
\end{tikzpicture}   
    \caption{All districts of the circuit of \cref{fig:bigcircuit}.}
    \label{fig:bigcircuitdistricts}
\end{figure}
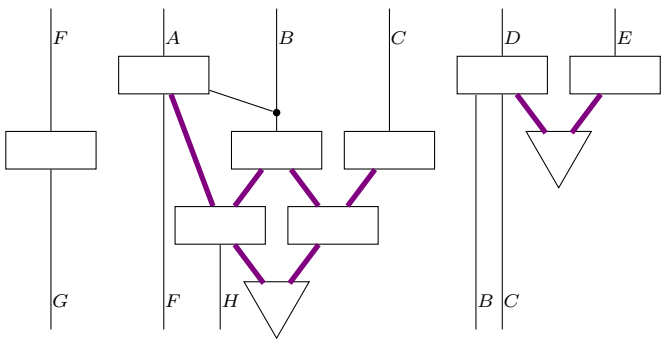

The procedure of district factorization corresponds to a surgery on the circuit wherein we cut certain classical wires. First, we cut all \emph{observed} classical wires in the circuit and identify all fragments of the circuit that become topologically independent from the rest of the circuit. It is important that we do \emph{not} cut any latent wire, even if it is classical. Then, for each fragment, we reconnect all the observed classical wires \emph{within} the fragment; the resulting fragment is called a \emph{district} of the circuit.\footnote{In the original definition of district factorization given by Ref.~\cite{Evans_2018} for classical causal structures, the term \emph{district} only refers to the output classical variables in each fragment instead of the fragment itself. 
We are consequently adopting the term for a slightly different purpose here.} For example, for the 
causal structure depicted in \cref{fig:bigcircuit}, the districts appearing in the district factorization thereof are depicted in \cref{fig:bigcircuitdistricts}. 
In the process of obtaining these districts, the wire $B$ that feeds into the box that has output $A$ was  initially cut, but subsequently reconnected due to being \emph{within} one district. All of the other observed classical wires are cut and \emph{not} reconnected, because they connect different districts.

Given a causal structure, denote its set of open output wires by $\bf O$, and its set of open input wires by $\bf I$. Moreover, for each district $k$, denote the set of its output wires as  
${\bf O}_k$, and denote the set of its input wires as ${\bf I}_k$. A probability distribution $P({\bf O}|{\bf I})$ that is compatible with the original causal structure must admit of a factorization of the form: 
\begin{align}\label{eq:district}
    P({\bf O}|{\bf I})=\Pi_k P({\bf O}_k|{\bf I}_k),
\end{align} 
where each $P({\bf O}_k|{\bf I}_k)$ is a conditional distribution that is compatible with district $k$.

The proof is a straightforward analogue of the proof for the classical case~\cite{Evans_2018}, where we substitute the classical Markov compatibility condition with the generalized Markov condition~\cite{henson2014Theoryindependent}.

Since district factorization is valid for any choice of the GPT $T$ describing the latent systems, the following is a necessary \emph{and sufficient} condition for a causal structure to admit of a $sT$-$T$ gap: 
\begin{proposition}
\label{pro:district}
    A causal structure admits of an $sT$-$T$ gap if and only if at least one of its districts admits of an $sT$-$T$ gap.
\end{proposition}
Thus, we can determine if a causal structure admits of a $sT$-$T$ gap by checking each of its districts.

\subsubsection{Check each district}
A necessary condition one can check on each district is the following.
\begin{proposition}
\label{pro:sub}
A causal structure admits of a $sT$-$T$ gap only if it admits of a CPT-$T$ gap, where CPT stands for classical probability theory.
\end{proposition}
The proof is as follows. First, note that classical probability theory is a subtheory of any GPT\footnote{If one adopts a more general definition of a GPT in which CPT need not be a subtheory, then for the purpose of discussing correlations realizable within a causal structure, the scope of GPTs must be restricted to those that embed CPT. Also note that, since we only consider collective presentations of a symmetry, if $T$ embeds CPT, then its symmetrized world necessarily embeds CPT as well.~\cite{ying2024twirling2}}---otherwise, we could not have classical systems labeling measurement settings and outcomes in the causal structure---so any distribution that is CPT-realizable is also $sT$-realizable in a given causal structure. If there is no CPT-$T$ gap in a causal structure, implying that all distributions that are $T$-realizable are also CPT-realizable, then all $T$-realizable distributions must also be $sT$-realizable, i.e., there is no $sT$-$T$ gap.

A corollary of this proposition is that 
\begin{corollary}
\label{cor:nonclassical}
    A GPT $T$ must be nonclassical to have the possibility to have a $sT$-$T$ gap in some causal structures.
\end{corollary}
Here, nonclassical theories refer to those that are not subtheories~\cite{GPTembedding} of CPT. If $T$ is a subtheory of CPT, then, there cannot be a CPT-$T$ gap in any causal structure (recall that CPT is always a subtheory of any GPT), and hence \cref{cor:nonclassical}. For example, the toy theory of \cite{Spekkens2007Evidence} is a subtheory of CPT~\cite{GPTembedding} and so there is no possibility of finding a symmetrized-nonsymmetrized gap with any symmetry for this toy theory. 

\begin{figure}[h!]
    \centering
\begin{tikzpicture}
        \node [style=Wsquareadj] (1) at (-2, -3) {};
        \node [style=Wsquareadj] (2) at (2, -3) {};
        \node [style=right label,font=\normalsize ] (19) at (-2, -4.6) {$X$};
        \node [style=right label,font=\normalsize ] (19) at (2, -4.6) {$Y$};
        \node [style=rectangle] (4) at (-2, -5.5) { };
        \node [style=rectangle] (5) at (2, -5.5) { };
        \node  [style=proj2] (3) at (0, 0) {}; 
        \node [style=rectangle] (13) at (0, 2.25) { };
        \node [style=right label,font=\normalsize ] (19) at (0, -2) {$Z$};
        \node [style=rectangle] (10) at (0, -3) { };
        \node [style=right label,font=\normalsize ] (20) at (0,1.5) {$A$};
        \draw [qWire] (1) to (3);
        \draw [qWire] (2) to (3);
        \draw[] (3) to (13);
        \draw[] (3) to (10);
        \draw[] (1) to (4);
        \draw[] (2) to (5);
\end{tikzpicture} 
    \caption{The causal structure of the PBR scenario.}
    \label{fig:PBR}
\end{figure}
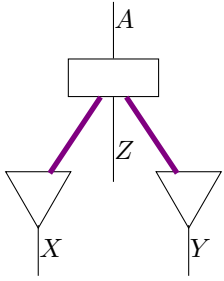

There is a class of causal structures, termed \emph{algebraic}~\cite{Khanna_classifying} (which were previously termed \enquote{interesting}~\cite{henson2014Theoryindependent}), that are known to not admit of a CPT-$T$ gap for any GPT $T$. Specifically, a causal structure is \emph{algebraic} if it does not impose any inequality constraints on the probability distributions that can be classically realized (i.e., it only imposes \emph{equality} constraints). Since all such equality constraints can be shown to hold regardless of the theory $T$ that describes the latent systems ~\cite{henson2014Theoryindependent}\footnote{Strictly speaking, Ref.~\cite{henson2014Theoryindependent} only proved that conditional independence constraints are theory-independent. However, equality constraints of the other possible type, namely, the nested Markov constraints~\cite{Shpitser2014}, are also theory-independent~\cite{privateElie}.}, only nonalgebraic causal structures have the potential for a CPT-$T$ gap. Thus, we have the following corollary:
\begin{corollary}
\label{cor:algebraic}
     A causal structure admits of a $sT$-$T$ gap only if it is nonalgebraic.
\end{corollary}
The fact that the causal structure of the PBR scenario~\cite{Pusey2012PBR}  (\cref{fig:PBR}) does not admit of a RQT-QT gap, which was shown in \cite[Supp. Mat.]{Renou_2021}, is a special case of \cref{cor:algebraic} because it is an algebraic causal structure.

We now turn to the second necessary condition. 

\begin{proposition}
\label{pro:ind}
Consider a causal structure that has only one district. It admits of a $sT$-$T$ gap only if it does not allow for the preparation of a shared reference frame state among all latent GPT systems.
\end{proposition}
Recall from \cref{app:shareRF} that causal structures that do not allow for the preparation of a shared reference frame state include both the ones that have GPT latent independent systems and those that have latent GPT systems that only share classical randomness.
The proof of this proposition is to use the result of \cite[Prop. V.1]{weilenmann2025partial} (which can be straightforwardly extended to all swirled worlds) to infer that when the causal structure allows for a shared reference state among all systems, then it cannot admit of a swirled-nonswirled gap.

For example, in any multipartite Bell scenario, all latent GPT systems can share a reference frame state as they are prepared by the same source, and thus, it does not admit a RQT-QT causal compatibility gap, even though it admits of a CPT-QT causal compatibility gap, as shown in \cite{mckague2009Simulating}. On the other hand, the bilocality scenario (even when classical randomness is added among the two sources) and natural generalizations thereof termed {\em star networks} {\em do} have latent GPT systems that cannot share a reference frame state, and have been shown in \cite{sarkar2025Gap} to admit of RQT-QT gap.

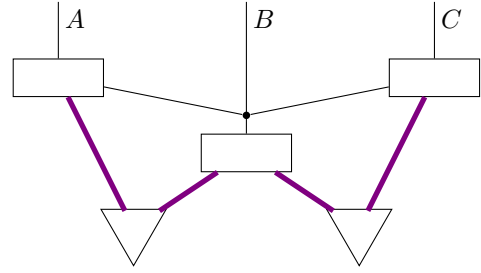
\begin{figure}[h!]
    \centering
\begin{tikzpicture}
        \node [style=Wsquareadj] (1) at (-3, -4) {};
        \node [style=Wsquareadj] (2) at (3, -4) {};
        \node  [style=proj2] (3) at (0, -2) {}; 
        \node  [style=proj2] (4) at (-5, 0) {};
        \node  [style=proj2] (5) at (5, 0) {};
        \node [style=rectangle] (12) at (-5, 2.25) { };
        \node [style=rectangle] (13) at (0, 2.25) { };
        \node [style=rectangle] (14) at (5, 2.25) { };
        \node [style=right label,font=\normalsize ] (19) at (-4.85, 1.5) {$A$};
        \node [style=right label,font=\normalsize ] (20) at (0.15,1.5) {$B$};
        \node [circle, fill, inner sep=1pt] (30) at (0,-1)  {};
        \node [style=right label,font=\normalsize ] (21) at (5.15, 1.5) {$C$};
        \draw [qWire] (1) to (3);
        \draw [qWire] (2) to (3);
        \draw [qWire] (1) to (4);
        \draw [qWire] (2) to (5);
        \draw[] (4) to (30);
        \draw[] (5) to (30);
        \draw[] (4) to (12);
        \draw[] (3) to (13);
        \draw[] (5) to (14);
\end{tikzpicture} 
    \caption{The causal structure of the Evans scenario.}
    \label{fig:evans}
\end{figure}

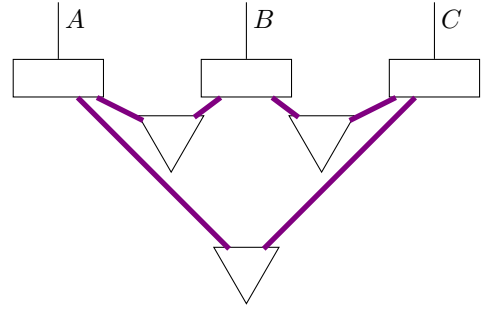
\begin{figure}[h!]
    \centering
\begin{tikzpicture}
        \node [style=Wsquareadj] (1) at (-2, -1.5) {};
        \node [style=Wsquareadj] (2) at (2, -1.5) {};
        \node [style=Wsquareadj] (30) at (0, -5) {};
        \node  [style=proj2] (3) at (0, 0) {}; 
        \node  [style=proj2] (4) at (-5, 0) {};
        \node  [style=proj2] (5) at (5, 0) {};
        \node [style=rectangle] (12) at (-5, 2.25) { };
        \node [style=rectangle] (13) at (0, 2.25) { };
        \node [style=rectangle] (14) at (5, 2.25) { };
        \node [style=right label,font=\normalsize ] (19) at (-4.85, 1.5) {$A$};
        \node [style=right label,font=\normalsize ] (20) at (0.15,1.5) {$B$};
        \node [style=right label,font=\normalsize ] (21) at (5.15, 1.5) {$C$};
        \draw [qWire] (1) to (3);
        \draw [qWire] (2) to (3);
        \draw [qWire] (30) to (4);
        \draw [qWire] (30) to (5);
        \draw [qWire] (1) to (4);
        \draw [qWire] (2) to (5);
        \draw[] (4) to (12);
        \draw[] (3) to (13);
        \draw[] (5) to (14);
\end{tikzpicture} 
    \caption{The causal structure of the triangle scenario.}
    \label{fig:triangle}
\end{figure}

Among well-known causal structures that have not been studied for RQT-QT gaps, the Evans' scenario~\cite{Evans_2015} (depicted  in \cref{fig:evans}) and the Triangle scenario (depicted  in \cref{fig:triangle}) are two that may have the potential for a swirled-nonswirled gap, since they both exhibit a CPT-QT gap and cannot prepare a share reference frame for all latent GPT systems. 

\subsection{On the necessity and sufficiency of lacking correlated reference frames}
\label{app:lack}

From the above discussion, it is now clear that 
\begin{corollary}
Not allowing for the preparation of a shared reference frame state among all latent GPT systems is not a sufficient condition for a causal structure to admit of a swirled-nonswirled gap (even when the swirling is restricted to strongly nonphysical symmetries).
\end{corollary}

\begin{figure}[h!]
    \centering
\begin{tikzpicture}
        \node [style=Wsquareadj] (1) at (-3, -4) {};
        \node [style=Wsquareadj] (2) at (3, -4) {};
        \node  [style=proj2] (3) at (0, 0.5) {}; 
        \node  [style=proj2] (4) at (-5, -2) {};
        \node  [style=proj2] (5) at (5, -2) {};
        \node [style=rectangle] (12) at (-5, 2.25) { };
        \node [circle, fill, inner sep=1pt] (30) at (-5,-1)  {};
         \node [circle, fill, inner sep=1pt] (31) at (5,-1)  {};
        \node [style=rectangle] (13) at (0, 2.25) { };
        \node [style=rectangle] (14) at (5, 2.25) { };
        \node [style=right label,font=\normalsize ] (19) at (-4.85, 1.5) {$A$};
        \node [style=right label,font=\normalsize ] (20) at (0.15,1.5) {$B$};
        \node [style=right label,font=\normalsize ] (21) at (5.15, 1.5) {$C$};
        \draw [qWire] (1) to (3);
        \draw [qWire] (2) to (3);
        \draw [qWire] (1) to (4);
        \draw [qWire] (2) to (5);
        \draw[] (30) to (3);
        \draw[] (31) to (3);
        \draw[] (4) to (12);
        \draw[] (3) to (13);
        \draw[] (5) to (14);
\end{tikzpicture} 
    \caption{An example of a causal structure that has independent sources but is algebraic, and thus does not admit of a $sT$-$T$ gap. }
    \label{fig:evansmod}
\end{figure}
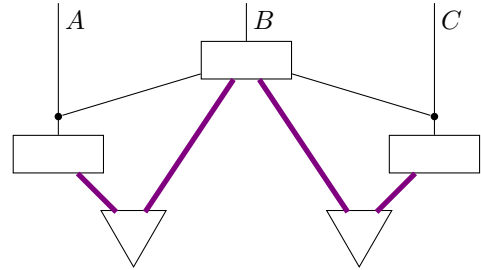

We have already given an example of a causal structure---one that does not allow for the preparation of a share reference frame state among all latent GPT systems---that does not admit of a swirled-nonswirled gap, namely, the causal structure of the PBR scenario in \cref{fig:PBR}. The reason is that such a causal structure is algebraic and by \cref{cor:algebraic}, no algebraic causal structure admits of such a gap. We give another example of a causal structure that has independent sources but is algebraic in \cref{fig:evansmod}.

 Furthermore, we also have that
\begin{corollary}
    Even if a causal structure admits of a CPT-T gap, not allowing for the preparation of a shared reference frame state among all latent GPT systems is still not a sufficient condition for it to admit of a swirled-nonswirled gap (even when the swirling is restricted to strongly nonphysical symmetries).
\end{corollary}

This is true because it is possible to have a causal structure that has a CPT-$T$ gap and independent latent GPT systems, but where these latent GPT systems belongs to a different district.  Because as such, as long as there are no independent latent GPT systems within each district, \cref{pro:ind} implies that each district fails to admit of an $sT$-$T$ gap, and then by \cref{pro:district}, it follows that the original causal structure does not admit of a $sT$-$T$ gap either. 

The causal structure of \cref{fig:Dbell} provides an example: it admits of a CPT-$T$ gap \emph{and} have independent latent GPT systems, but, as shown in \cref{fig:Dbell_dist}, each of its two districts has the causal structure of a Bell scenario, which does not admit of a $sT$-$T$ gap. 
We thus conclude by \cref{pro:district} that the causal structure in \cref{fig:Dbell} does not admit of an $sT$-$T$ gap either.

It is an open question whether lacking the possibility of preparing a shared reference frame state among all latent GPT systems within a single district is a \emph{sufficient} condition for a $sT$-$T$ gap.  For instance, as noted above, we do not know if the causal structure of the Evans scenario (\cref{fig:evans}) or that of the triangle scenario (\cref{fig:triangle}) admits of such a gap.

\begin{figure}[h!]
    \centering
\begin{tikzpicture}
        \node [style=Wsquareadj] (1) at (-4, -3) {};
        \node [style=Wsquareadj] (2) at (6, -6) {};
        \node  [style=proj2] (3) at (0, -4) {}; 
        \node  [style=proj2] (4) at (2, -1) {};
        \node  [style=proj2] (5) at (6, 2) {};
        \node  [style=proj2] (6) at (-4, 6) {};
        \node [style=rectangle] (12) at (0, -6) { };
        \node [style=rectangle] (13) at (-4, 9) { };
        \node [style=right label,font=\normalsize ] (19) at (0, -5) {$A$};
        \node [style=right label,font=\normalsize ] (20) at (0,8) {$B$};
        \node [style=right label,font=\normalsize ] (21) at (2, 8) {$C$};
        \node [style=right label,font=\normalsize ] (21) at (6, 8) {$D$};
        \node [style=right label,font=\normalsize ] (21) at (-4, 8) {$E$};
        \node [circle, fill, inner sep=1pt] (31) at (0,-3)  {};
        \node [circle, fill, inner sep=1pt] (32) at (2,0.5)  {};
        \node [style=rectangle] (34) at (0, 9) { };
        \node [style=rectangle] (35) at (2, 9) { };
         \node [style=rectangle] (36) at (6, 9) { };
         \node [circle, fill, inner sep=1pt] (37) at (6,3)  {};
        \draw [qWire] (1) to (6);
        \draw [qWire] (2) to (3);
        \draw [qWire] (1) to (4);
        \draw [qWire] (2) to (5);
        \draw[] (31) to (4);
        \draw[] (4) to (35);
        \draw[] (37) to (6);
        \draw[] (3) to (12);
        \draw[] (6) to (13);
        \draw[] (3) to (34);
         \draw[] (5) to (32);
         \draw[] (5) to (36);
\end{tikzpicture} 
    \caption{An example of a causal structure that is non-algebraic and has two independent sources, but nevertheless \emph{cannot} admit of a $sT$-$T$ gap.}
    \label{fig:Dbell}
\end{figure}
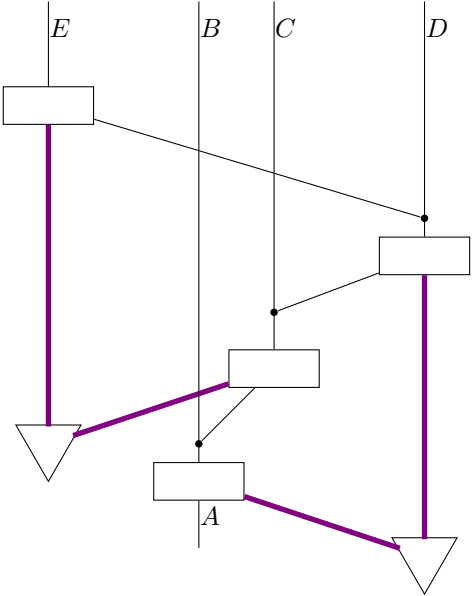

\begin{figure}
    \centering
\resizebox{7cm}{!}{\begin{tikzpicture}
        \node [style=Wsquareadj] (1) at (-10, -3) {};
        \node [style=Wsquareadj] (2) at (6, -6) {};
        \node  [style=proj2] (3) at (0, -4) {}; 
        \node  [style=proj2] (4) at (-4, -1) {};
        \node  [style=proj2] (5) at (6, 2) {};
        \node  [style=proj2] (6) at (-10, 6) {};
        \node [style=rectangle] (12) at (0, -9) { };
        \node [style=rectangle] (13) at (-10, 9) { };
        \node [style=right label,font=\normalsize ] (19) at (0, -8) {$A$};
        \node [style=right label,font=\normalsize ] (20) at (0,8) {$B$};
        \node [style=right label,font=\normalsize ] (21) at (-4, 8) {$C$};
        \node [style=right label,font=\normalsize ] (21) at (6, 8) {$D$};
        \node [style=right label,font=\normalsize ] (21) at (-10, 8) {$E$};
        \node [style=rectangle] (34) at (0, 9) { };
        \node [style=rectangle] (35) at (-4, 9) { };
        \node [style=rectangle] (36) at (6, 9) { };
        \node [style=rectangle] (37) at (7, -9) { };
        \node [style=right label,font=\normalsize ] (38) at (7, -8) {$C$};
        \node [style=right label,font=\normalsize ] (38) at (-4, -8) {$B$};
         \node [style=rectangle] (39) at (-4, -9) { };
        \node [style=rectangle] (60) at (7, 1.75) { };
        \node [style=rectangle] (40) at (-11, -9) { };
        \node [style=right label,font=\normalsize ] (38) at (-11, -8) {$D$};
        \node [style=rectangle] (61) at (-11, 5.75) { };
        \draw [qWire] (1) to (6);
        \draw [qWire] (2) to (3);
        \draw [qWire] (1) to (4);
        \draw [qWire] (2) to (5);
        \draw[] (4) to (35);
        \draw[] (3) to (12);
        \draw[] (6) to (13);
        \draw[] (3) to (34);
         \draw[] (5) to (36);
        \draw[] (60) to (37);
        \draw[] (4) to (39);
        \draw[] (40) to (61);
\end{tikzpicture} }
    \caption{Two districts of the causal structure of \cref{fig:Dbell}.}
    \label{fig:Dbell_dist}
\end{figure}
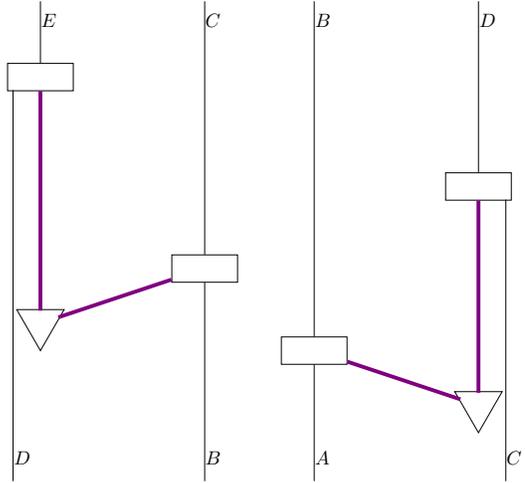

\subsection{Causal structures having gates with nontrivial no-influence conditions}
\label{app:internal}

When we consider causal structures outside the set with unrestricted gates, i.e., causal structures that have nontrivial internal no-influence conditions, most of the results presented in \cref{app:necessary} still hold, except for possibly \cref{cor:algebraic} concerning nonalgebraic causal structures and \cref{pro:ind} concerning the lack of a shared reference frame state, as we will explain below. 

For \cref{pro:district} concerning district factorization, when a causal structure could be outside the set of unrestricted gates, one can still identify the districts using the same procedure and \cref{eq:district} still holds (since the proof for \cref{eq:district} only relies on no-influence conditions external to gates). We do not claim that this is the most fine-grained notion of district factorization that is possible since it does not make use of the internal causal structures. Nevertheless, \cref{pro:district} is valid for a general causal structure. 

Similarly, since neither the proof of \cref{pro:sub} nor the definition of whether a causal structure admits of a CPT-$T$ gap rely on the absence of nontrivial no-influence conditions, \cref{pro:sub} is also valid for a general causal structure. 

As for \cref{cor:algebraic} and \cref{pro:ind}, we first need to introduce some concepts related to faithful decompositions of a causal structure, where we decompose a causal structure with nontrivial internal no-influence conditions into a causal structure without nontrivial no-influence conditions. The notion of faithful decompositions has been previously studied in the literature (e.g., in \cite{lorenz2021Causala}) for classical and quantum circuits where all gates are restricted to reversible gates. Here, we will discuss decompositions for GPT causal structures and we do not restrict our notion of faithful decompositions to reversible gates only, since the question we are interested in concerns the set of realizable correlations in a causal structure, which includes all correlations that can be realized in some circuit compatible with it.

Specifically, a causal structure $\alpha$ is said to be \emph{faithful} to another causal structure $\beta$ if, for any systems $A$ and $B$ in $\beta$, both also exist in $\alpha$, and $A$ can influence $B$ in $\alpha$ if and only if $A$ can influence $B$ in $\beta$.  Furthermore, a causal structure $\beta$ is said to admit of a \emph{faithful decomposition within $T$}, denoted $\alpha$, if $\alpha$ is a causal structure that belongs to the set with unrestricted gates, is faithful to $\beta$, and has the same set of $T$-realizable correlations as $\beta$. When a causal structure does not have any nontrivial no-influence conditions, evidently, it is a faithful decomposition of itself within any $T$. Furthermore, while any causal structure admits of a faithful decomposition within CPT, it is currently unclear if, within a nonclassical GPT $T$ such as quantum theory (QT), any causal structure admits of a faithful decomposition.\footnote{Recall that here we do not have any restrictions on the gates in the decomposition. If the gates in the decomposition are required to be reversible, it has been proven that not all causal structures admit of a faithful decomposition within CPT, nor do all admit of a faithful decomposition within QT~\cite{lorenz2021Causala}.}

If a causal structure admits of a faithful decomposition within $sT$ and $T$, then the question of $sT$-$T$ gap can be answered by checking if the faithful decomposition admits of a gap. That is
\begin{lemma}
    When a causal structure admits of a faithful decomposition, denoted $\alpha$, within $sT$ and $T$, it admits of a $sT$-$T$ gap if and only if $\alpha$ admits of a $sT$-$T$ gap.
\end{lemma}
When a causal structure does not admit of a faithful decomposition within $T$ but does admit of a faithful decomposition within $sT$, we can still have a sufficient condition for it to admit of a $sT$-$T$ gap. That is, 
\begin{lemma}
    When a causal structure admits of a faithful decomposition within $sT$, denoted $\alpha$, it admits of a $sT$-$T$ gap as long as $\alpha$ admits of a $sT$-$T$ gap.
\end{lemma}
This is because the set of $sT$-realizable correlations in the causal structure is the same as that in $\alpha$, while its set of $T$-realizable correlations must be a superset of that in $\alpha$. To see the latter, it suffices to realize that if $\alpha$ is also a faithful decomposition of the causal structure within $T$, then its set of $T$-realizable correlations is the same as that in $\alpha$, and if $\alpha$ is not a faithful decomposition within $T$, then there may exist a $T$-realizable correlation in the causal structure that is not realizable in $\alpha$. For the same reason, if the causal structure does not admit a faithful decomposition within $sT$, then, even if it does admit of a faithful decomposition $\alpha$ within $T$ and even if $\alpha$ admits of a $sT$-$T$ gap, it does not mean that the causal structure itself admits of a $sT$-$T$ gap. Note that although $sT$ is a subtheory of $T$, the fact that the causal structure admits of a faithful decomposition within $T$ does not mean that it also admits of a faithful decomposition within $sT$, since it is conceivable that for certain gates in $sT$, its faithful decomposition needs gates that are in $T$ but not $sT$.

Now, let us turn back to \cref{cor:algebraic} and  \cref{pro:ind}. As for \cref{cor:algebraic}, although the definition of a causal structure being algebraic or not does not rely on it being in the set with unrestricted gates, and, as we will explain soon, one can in fact utilize tools for the set with unrestricted gates to study the algebraic or nonalgebraic properties of a causal structure outside the set, it is unfortunately no longer clear if an algebraic causal structure outside the set with unrestricted gates necessarily cannot admit of a CPT-$T$ gap.

Let us first see how to determine whether a general causal structure is algebraic, since all previous works on the classification of algebraic and non-algebraic causal structures, such as~\cite{henson2014Theoryindependent,Pienaar2017,Khanna_classifying,Ansanelli_Observational}, focus on the set with unrestricted gates. To do so, we make use of the following two facts. First, any causal structure admits of a faithful decomposition within CPT.
Second, by definition, whether a causal structure is algebraic or not is determined by the set of CPT-realizable correlations in it. From these two facts, we have that
\begin{lemma}
    A causal structure is algebraic if and only if its faithful decomposition within $CPT$ is algebraic.
\end{lemma}
Thus, to determine whether a general causal structure is algebraic, we can first find its faithful decomposition within $CPT$ and then apply the tools from the existing literature to check whether that decomposition is algebraic.

Unlike for a causal structure from the set with unrestricted gates, it is currently unclear whether for a general causal structure, being algebraic implies an absence of a CPT-$T$ gap. This is because the set of $T$-realizable correlations in the causal structure can be a strict superset of the set of $T$-realizable correlations in its CPT faithful decomposition. Consequently, even if the faithful decomposition within CPT does not admit a CPT-$T$ gap, the original causal structure could still admit one. 

Therefore, to claim that all algebraic causal structures (including those with nontrivial no-influence conditions) cannot admit a CPT-$T$ gap for any $T$, one would need to generalize previous proofs---established only for the set with unrestricted gates---showing that all equality constraints on the set of CPT-realizable correlations cannot be violated by the set of $T$-realizable correlations. We leave it to future work to determine whether those proofs can indeed be generalized. If so, then \cref{cor:algebraic} is valid for a general causal structure.

\begin{figure}[h!]
    \centering
    \begin{subfigure}{0.22\textwidth}
    \centering
\begin{tikzpicture}
        \node [style=Wsquareadj] (1) at (-1.5, -4.5) {};
        \node [draw=black,rectangle, minimum height=1.3cm,  minimum width=1.4cm] (3) at (0, -1.2) {}; 
        \node  [style=proj2] (4) at (-3, -2.5) {};
        \node  [style=proj2] (5) at (3, 2) {};
        \draw[->, line width=1pt] (-0.5, -2.3) -- (0.93, -0.25);
        \draw[-, line width=1pt] (0, -1) -- (0.3, -1.6);
        \node [style=rectangle] (13) at (0, 2.25) { };
        \node [style=rectangle] (12) at (-3, -0.25) { };
        \node [style=rectangle] (14) at (3, 4.25) { };
        \node [style=rectangle] (16) at (-3, -5.25) { };
        \node [style=rectangle] (17) at (3, -1) { };
        \node [style=right label,font=\normalsize ] (19) at (-3.85, -1) {$A$};
        \node [style=right label,font=\normalsize ] (19) at (-3.85, -4) {$X$};
        \node [style=right label,font=\normalsize ] (20) at (0,1.5) {$B$};
        \node [style=right label,font=\normalsize ] (21) at (3.15, 3.5) {$C$};
        \node [style=right label,font=\normalsize ] (21) at (3.15, 0.5) {$Z$};
        \draw [qWire] (1) to (3);
        \draw [qWire] (1) to (4);
        \draw [qWire] (3) to (5);
        \draw[] (4) to (12);
        \draw[] (4) to (16);
        \draw[] (3) to (13);
        \draw[] (5) to (14);
        \draw[] (5) to (17);
\end{tikzpicture} 
\caption{ }
\label{fig:CJbilo}
\end{subfigure}
    \begin{subfigure}{0.2\textwidth}
    \centering
\begin{tikzpicture}
        \node [style=Wsquareadj] (1) at (-1.5, -4.5) {};
        \node [style=Wsquareadj] (2) at (1, -2) {};
        \node  [style=proj2] (3) at (0, 0) {}; 
        \node  [style=proj2] (4) at (-3, -2.5) {};
        \node  [style=proj2] (5) at (3, 2) {};
        \node [style=rectangle] (13) at (0, 2.25) { };
        \node [style=rectangle] (12) at (-3, -0.25) { };
        \node [style=rectangle] (14) at (3, 4.25) { };
        \node [style=rectangle] (16) at (-3, -5.25) { };
        \node [style=rectangle] (17) at (3, -1) { };
        \node [style=right label,font=\normalsize ] (19) at (-3.85, -1) {$A$};
        \node [style=right label,font=\normalsize ] (19) at (-3.85, -4) {$X$};
        \node [style=right label,font=\normalsize ] (20) at (0,1.5) {$B$};
        \node [style=right label,font=\normalsize ] (21) at (3.15, 3.5) {$C$};
        \node [style=right label,font=\normalsize ] (21) at (3.15, 0.5) {$Z$};
        \draw [qWire] (1) to (3);
        \draw [qWire] (2) to (3);
        \draw [qWire] (1) to (4);
        \draw [qWire] (2) to (5);
        \draw[] (4) to (12);
        \draw[] (4) to (16);
        \draw[] (3) to (13);
        \draw[] (5) to (14);
        \draw[] (5) to (17);
\end{tikzpicture} 
\caption{ }
\label{fig:bilo}
\end{subfigure}
    \caption{Comparison of (a) a causal structure with nontrivial no-influence conditions and (b) the causal structure of the bilocality scenario.}
    \label{fig:internal}
\end{figure}
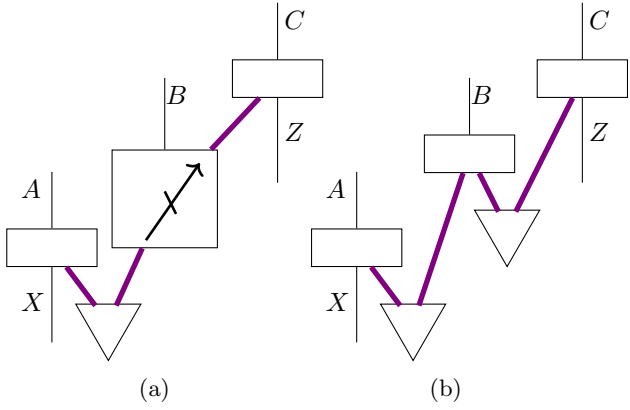

Now we turn to \cref{pro:ind}. \cref{pro:ind} does \emph{not} directly apply to causal structures outside the set of unrestricted gates; that is, even if a causal structure allows for a shared reference frame state among all systems, it may still admit of a swirled-nonswirled gap when it has nontrivial no-influence conditions. For example, consider the causal structure in \cref{fig:CJbilo}, where we have a nontrivial no-influence condition in the middle gate. Without this nontrivial no-influence condition, this causal structure cannot admit of a swirled-nonswirled gap since it is algebraic. However, {\em with} this nontrivial no-influence condition, it is no longer algebraic and in fact, it admits of an RQT-QT gap. This is because it corresponds to the bilocality scenario, as shown in \cref{fig:bilo}. (This follows from the fact that the operation in the middle gate has only one input and the fact that such an operation always admits of a faithful causal decomposition even if it is not classical.)

Admittedly, the example in \cref{fig:CJbilo} is a case where independent latent GPT systems do appear in its faithful decomposition. So one might conjecture that not allowing for a shared reference frame state among all systems is still a necessary condition for a causal structure with one district to admit of a swirled-nonswirled gap, in the sense that when the causal structure admits of a faithful decomposition within $sT$ and $T$, not allowing for a shared reference frame state among all latent GPT systems must be a feature of its faithful decomposition. We leave this question to future research.

\end{document}